\documentclass[11pt]{article}
\usepackage{jheppub}

\usepackage{amsmath}
\usepackage{stackrel}
\usepackage{mathtools} 
\usepackage{ulem} 
\usepackage{subcaption} 
\usepackage{multirow} 
\usepackage{color} 
\usepackage{tablefootnote} 
\usepackage{pifont} 

\def\backtick{\char18}
\usepackage{listings}
\graphicspath{
  {.}
  {./figures}
  {./figures/onestep}
  {./figures/twostep}
  {./figures/semisoft}
  {./figures/Tc}
  {./figures/triptych}
  {./figures/schematic-mass-hierarchy-plots}
}

\definecolor{mymagenta}{rgb}{1,0,1}

\newcommand\MSbar{$\overline{\rm MS}$}

\newcommand{\rmii}[1]{{\mbox{\tiny\rm{#1}}}}

\newcommand{\sumint}[1]{{\hbox{$\sum$}\!\!\!\!\!\!\!\int\,}_{\!\!\!\!\raise-0.9ex\hbox{$\scriptstyle{#1}$}}}
\newcommand{\triptychwidth}{0.93\textwidth}

\newcommand{\cmark}{\ding{51}}%
\newcommand{\xmark}{\ding{55}}%

\newcommand{\Tint}[1]{{\hbox{$\sum$}\!\!\!\!\!\!\!\int\,}_{\!\!\!\!\raise-0.9ex\hbox{$\scriptstyle{#1}$}}}
\newcommand{\Tinti}[1]{{{\Sigma}\!\!\!\!\raise0.3ex\hbox{$\int$}_\rmii{${#1}$}}}
\newcommand{\Tintip}[1]{{{\Sigma'}\!\!\!\!\!\raise0.3ex\hbox{$\int$}_\rmii{${#1}$}}}

\newcommand{\veps}{\varepsilon}

\newcommand{\scale}{\Lambda_{4}}
\newcommand{\scaleft}{\Lambda_{3}}

\title{Perturbative effective field theory expansions for cosmological phase transitions}

\author[a]{Oliver Gould}
\author[b,c,d]{and Tuomas V.~I.~Tenkanen}

\affiliation[a]{School of Physics and Astronomy, University of Nottingham, Nottingham NG7 2RD, United Kingdom}
\affiliation[b]{Nordita, KTH Royal Institute of Technology and Stockholm University, Roslagstullsbacken 23, SE-106 91 Stockholm, Sweden}
\affiliation[c]{Tsung-Dao Lee Institute \& School of Physics and Astronomy, Shanghai Jiao Tong University, Shanghai 200240, China}
\affiliation[d]{Shanghai Key Laboratory for Particle Physics and Cosmology, Key Laboratory for Particle Astrophysics and Cosmology (MOE), Shanghai Jiao Tong University, Shanghai 200240, China}

\emailAdd{oliver.gould@nottingham.ac.uk}
\emailAdd{tuomas.tenkanen@su.se}

\preprint{NORDITA 2023-037}

\abstract{
Guided by previous non-perturbative lattice simulations of a two-step electroweak phase transition, we reformulate the perturbative analysis of equilibrium thermodynamics for generic cosmological phase transitions in terms of effective field theory (EFT) expansions.
Based on thermal scale hierarchies, we argue that the scale of many interesting phase transitions is in-between the soft and ultrasoft energy scales, which have been the focus of studies utilising high-temperature dimensional reduction.
The corresponding EFT expansions provide a handle to control the perturbative expansion, and allow us to avoid spurious infrared divergences, imaginary parts, gauge dependence and renormalisation scale dependence that have plagued previous studies.
As a direct application, we present a novel approach to two-step electroweak phase transitions, by constructing separate effective descriptions for two consecutive transitions.
Our approach provides simple expressions for effective potentials separately in different phases, a numerically inexpensive method to determine thermodynamics,   and significantly improves agreement with the non-perturbative lattice simulations.
}

\begin{document}

\maketitle

\section{Introduction}
\label{sec:intro}

Gravitational waves from a cosmological phase transition could provide a window to directly observe the very early universe, preceding the birth of the cosmic microwave background as well as Big Bang nucleosynthesis. This would offer a probe of the fundamental constituents of matter and their interactions which is complementary to particle colliders.

In recent years, studies of the electroweak phase transition (EWPT) have sparked a lot of interest, motivated by the possibility of explaining the baryon asymmetry of the universe \cite{Kuzmin:1985mm, Bodeker:2020ghk}, and also of generating a stochastic gravitational wave (GW) background \cite{Caprini:2019egz} observable by LISA-generation experiments \cite{LISA:2017pwj, TianQin:2015yph, Ruan:2018tsw}. In the Standard Model, electroweak symmetry breaking occurs via a smooth crossover \cite{Kajantie:1996mn, DOnofrio:2015gop}, so a possible first-order electroweak phase transition requires the existence of new physics beyond the Standard Model (BSM). The search for BSM physics that could alter the thermal history of electroweak symmetry breaking provides a target and challenge for future collider experiments \cite{Ramsey-Musolf:2019lsf}. Of particular interest are multi-step phase transitions in the presence of multiple Higgs-like fields, where the transition to the EW phase can be preceded by another phase at a higher temperature \cite{Weinberg:1974hy}. In this work, we discuss a two-step EWPT \cite{Land:1992sm}.

In determining the thermodynamic properties of a BSM theory, thermal enhancements of infrared (IR) physics play an important role. In practice, this means that perturbative computations require thermal resummations \cite{Dolan:1973qd, Kapusta:1979fh, Parwani:1991gq, Arnold:1992rz}. In the imaginary-time formalism of high-temperature quantum field theory \cite{Matsubara:1955ws}, the most elegant solution to organise these resummations is by means of effective field theory \cite{Kajantie:1995dw, Braaten:1995cm, Braaten:1995jr}. In such a computation, a dimensionally reduced effective theory is constructed for the IR sensitive zero Matsubara modes, while all non-zero Matsubara modes are integrated out and their effects are captured in the effective parameters of the EFT. Physically, this accounts for thermal screening, whereby the hard thermal scale modifies the dynamics of the softer IR physics that drives the EWPT.

The most infrared modes of the magnetic gauge bosons become strongly coupled at high temperatures \cite{Linde:1980ts}, leading to non-perturbative effects on the thermodynamics which require use of Monte-Carlo lattice simulations \cite{Farakos:1994xh, Kajantie:1995kf, Kajantie:1996qd}. For decades this non-perturbative physics at the ultrasoft scale has caused worry, calling into question the applicability and validity of perturbative determinations of thermodynamics. In the work at hand, we argue that these worries have been somewhat misplaced and argue that in fact first-order phase transitions take place above the ultrasoft scale. Indeed, it has long been known \cite{Arnold:1992rz, Moore:2000jw} that there is a scale in between the soft and ultrasoft scales, the latter of which have been the focus of studies utilising high-temperature dimensional reduction. A proper treatment of physics at this in-between \textit{supersoft scale} requires a chain of EFTs that we construct in this work. In this approach, these non-perturbative effects are typically subleading for first-order phase transitions, so that both the leading order thermodynamics and several corrections can be obtained by a purely perturbative expansion in powers of a small expansion parameter. More recently, supersoft-scale EFTs have been studied in \cite{Ekstedt:2020abj, Gould:2021ccf, Hirvonen:2021zej, Ekstedt:2021kyx, Ekstedt:2022ceo, Ekstedt:2022zro} (see also \cite{Gould:2019qek, Lofgren:2021ogg, Gould:2022ran, Lofgren:2023sep}).

By direct comparison to previous non-perturbative lattice simulations, we demonstrate that EFTs at the supersoft scale describe thermodynamics with striking accuracy. Further information on the validity of a perturbation approach can be extracted from the expansion itself, by considering renormalisation group (RG) invariance, gauge invariance and the convergence of successive terms. For the former, it has been shown in \cite{Croon:2020cgk, Gould:2021oba} that perturbative computations below two-loop order suffer from large intrinsic uncertainties in terms of sensitivity to RG scales, which reflect the magnitude of missing perturbative corrections. Throughout the paper, we denote RG scales as $\scale$ and $\scaleft$, in the full four-dimensional parent theory and in the dimensionally reduced EFT respectively.

The remainder of this article is organised as follows. In Section~\ref{sec:motivation} we motivate our analysis by reviewing previous lattice results for a two-step phase transition in the real-triplet extended Standard Model, as well as shortcomings of previous perturbative analyses. In Section~\ref{sec:setup} we discuss thermal scale hierarchies for generic first-order transitions. Based on these hierarchies, in Section~\ref{sec:xpansion} we introduce corresponding effective field theories, paying particular attention to a scale in between the soft and ultrasoft scales, that we dub the supersoft scale. In Section~\ref{sec:methods} we present what we refer to as strict perturbative expansions for the thermodynamics of phase transitions discussed in the preceding section. In Section~\ref{sec:results} we provide a concrete application to the real-triplet extended SM, presenting numerical results while relegating a number of technical details throughout the article to appendices. In Section~\ref{sec:discussion} we summarise and discuss our results.

\subsection{Motivation}
\label{sec:motivation}

For a two-step electroweak phase transition, the current state-of-the-art determination of the equilibrium thermodynamics is provided by the non-perturbative lattice simulations of Ref.~\cite{Niemi:2020hto}, concretely performed for a real-triplet extended SM \cite{Patel:2012pi}. In this work, we keep our discussion generic, and applicable to a wide variety of cosmological phase transitions, yet for the numerical analysis turn to the real-triplet extended SM. In this model, the scalar sector comprises the Higgs doublet $\phi$ and a real triplet scalar $\Sigma^a$, where $a=1,2,3$ is an SU(2) adjoint index. We follow the conventions of \cite{Niemi:2020hto} and define the scalar part of the Lagrangian in 4d Euclidean space as%
\footnote{
We make an exception for the mass parameters $\mu_\phi^2$ and $\mu_\Sigma^2$ for which we use the opposite sign compared to \cite{Niemi:2018asa,Niemi:2020hto}
}
\begin{align} \label{eq:SigmaSM_Lagrangian}
\mathcal{L}(\phi,\Sigma) =& (D_\mu \phi)^\dagger (D_\mu \phi) + \mu_\phi^2 \phi^\dagger \phi + \lambda (\phi^\dagger\phi)^2 \nonumber \\
& +\frac{1}{2} (D_\mu \Sigma^a)^2  + \frac12 \mu_\Sigma^2 \Sigma^a \Sigma^a + \frac{b_4}{4} (\Sigma^a\Sigma^a)^2 \nonumber \\ 
& + \frac{a_2}{2} \phi^\dagger\phi \Sigma^a \Sigma^a,
\end{align} 
where the definitions for covariant derivatives are standard, and can be found in \cite{Niemi:2018asa}. This model admits a two-step EW phase transition, where the system undergoes a first phase transition to the triplet phase at some high temperature, after which the system undergoes a second transition to the EW phase. Phase transitions in this model were first studied in perturbation theory in Ref.~\cite{Patel:2012pi} and Ref.~\cite{Niemi:2018asa} performed the dimensional reduction from the hard to the soft scale 3d EFT for this model. Non-perturbative lattice simulations of the 3d EFT were presented in \cite{Niemi:2020hto}, together with the perturbative computation of the two-loop thermal effective potential.

In our numerical analysis, we study the two benchmark points of \cite{Niemi:2020hto}. These points are defined as
\begin{align}
\text{BM1: } \quad (M_\Sigma, a_2,b_4) &= (160 \text{ GeV},1.1,0.25), \\
\text{BM2: } \quad (M_\Sigma, a_2,b_4) &= (255 \text{ GeV},2.3,0.25),
\end{align}
where $M_\Sigma$ is the physical triplet pole mass, and $a_2$ and $b_4$ are \MSbar\ parameters at the input renormalisation scale $\scale=M_Z$ equal to the Z-boson pole mass. Both of these points exhibit a two-step EW phase transition. According to the lattice study of \cite{Niemi:2020hto}, in BM1 the first (higher temperature) transition to the triplet phase is a crossover, whereas in BM2 the first transition is first order. The second transition is of first order in both benchmark points.

\begin{figure}
\centering
\begin{subfigure}{0.48\textwidth}
    \centering
    \includegraphics[width=\textwidth]{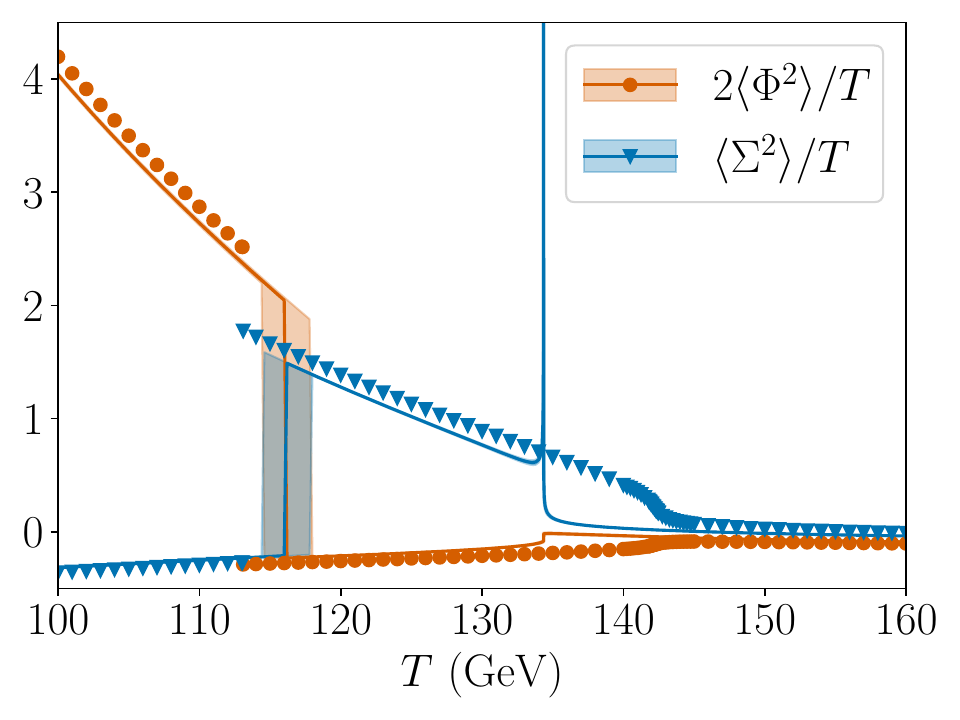}
    \caption{BM1}
    \label{fig:BM1_soft_mixed}
\end{subfigure}
\hfill
\begin{subfigure}{0.48\textwidth}
    \centering
    \includegraphics[width=\textwidth]{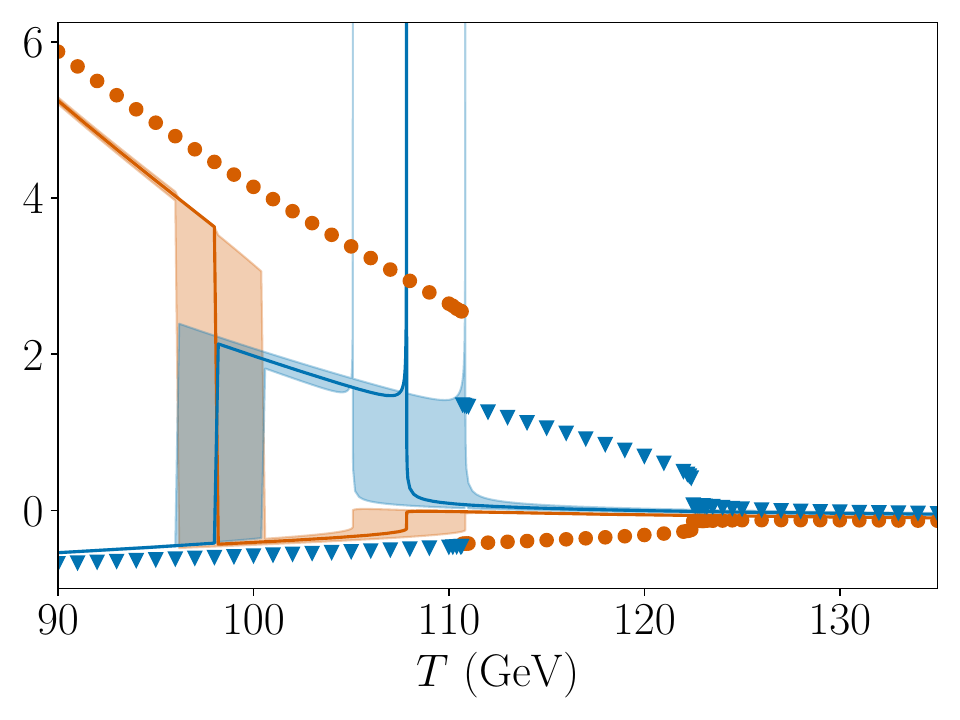} 
    \caption{BM2}
    \label{fig:BM2_soft_mixed}
\end{subfigure}
\caption{
Quadratic scalar condensates as functions of temperature $T$, in analogy to Fig.~2 of Ref.~\cite{Niemi:2020hto}. Circular and triangular markers depict lattice results, while solid lines show the two-loop perturbative counterpart following the approach of Ref.~\cite{Niemi:2020hto}. We have added coloured bands which show the RG-scale dependence of the perturbative calculation, as the scale within the EFT is varied over the set $\scaleft \in \{0.5 T, T, 2T\}$. In this work at hand, we will fix the relatively poor agreement between the perturbative and lattice results apparent here, including the spurious divergence in the perturbative results for the triplet condensate.
}
\label{fig:soft_strict_direct}
\end{figure}

In Ref.~\cite{Niemi:2020hto} a comparison to a state-of-the-art perturbative calculation was provided, utilising the two-loop order effective potential, computed within the dimensionally reduced 3d EFT. The computation utilised the $\hbar$-expansion of the effective potential \cite{Fukuda:1975di} -- where the potential is perturbatively expanded around its leading order minima, order-by-order -- thereby ensuring order-by-order gauge invariance \cite{Fukuda:1975di, Nielsen:1975fs, Laine:1994zq}. However, it suffered from IR divergences related to the determination of the critical temperature of the first transition to the triplet phase. Such divergences were reported already in Ref.~\cite{Laine:1994zq}: in the $\hbar$-expansion of the effective potential, the leading order potential does not have a first, but a second-order phase transition, and the critical temperature at leading order is identified with the temperature where the effective mass parameter of the scalar undergoing the transition vanishes. At two-loop order, i.e.\ $\mathcal{O}(\hbar^2)$, such a vanishing mass parameter inflicts an IR divergence on the scalar condensate (defined below). This is illustrated in Fig.~\ref{fig:soft_strict_direct}, adapted from Fig.~2 of Ref.~\cite{Niemi:2020hto}.
The condensates shown there are equal to derivatives of the free energy density (or equivalently the effective potential) with respect to 3d EFT parameters, 
\begin{align}
\label{eq:cond-phi}
\langle \phi^\dagger \phi \rangle &\equiv \frac{\partial  V_{\text{eff}}}{\partial  \mu^2_{\phi,3}}, \\
\label{eq:cond-sigma}
\left\langle \text{Tr} \Sigma^2 \right\rangle &\equiv 2 \frac{\partial   V_{\text{eff}}}{\partial  \mu^2_{\Sigma,3}}.
\end{align}
These relations follow simply from the path integral definition of the free-energy density, or effective potential \cite{Farakos:1994xh}. A discontinuity in these condensates signals a first-order phase transition. This is because the first temperature derivative of the pressure $p=-T V_\text{eff}$, or free energy density, can be written in terms of the condensates,
\begin{align}
\Delta p'(T_\text{c}) &= -T \frac{d}{dT} \Delta V_{\text{eff}}(\kappa_i) = -\sum_i T \frac{d\kappa_i}{dT} \frac{\partial \Delta V_{\text{eff}} }{\partial \kappa_i} \nonumber \\
&= \eta(\mu^2_{\phi,3}) \Delta \langle \phi^\dagger \phi \rangle + \eta(\lambda_{\phi,3}) \Delta \langle (\phi^\dagger \phi)^2 \rangle  + \ldots \; ,
\end{align}
where $\Delta$ denotes the difference between two phases, the prime denotes a temperature derivative, the $\eta$-functions are defined as $\eta(\kappa_i) \equiv T d\kappa_i/dT$, and $\kappa_i$ runs over all 3d EFT parameters. The $\eta$-functions depend only on the ultraviolet (UV) thermal scale, and hence are smooth, continuous functions of temperature. It is the condensates that have discontinuities at phase transitions (or kinks for higher-order transitions).

In perturbation theory, jumps in the condensates are related to jumps in the position of the global minimum of the effective potential as a function of temperature.%
\footnote{Note, that the minima depend on the values of EFT parameters, so one first evaluates the potential in the corresponding minimum, and then differentiates with respect to the EFT parameters to determine the condensates.} 
At leading order, the square roots of the quadratic condensates, Eqs.~\eqref{eq:cond-phi} and \eqref{eq:cond-sigma}, agree with the minima of the effective potential. However, this relationship breaks down beyond leading order, and while the condensates are manifestly gauge invariant, this is not the case for the minima of the effective potential \cite{Nielsen:1975fs, Andreassen:2014eha}. A further benefit of the condensates is that they can be computed directly on the lattice as volume-averaged expectation values \cite{Farakos:1994xh}, unlike the minima of the effective potential \cite{Elitzur:1975im}. As defined above, based on derivatives with respect to \MSbar\ parameters, the condensates are RG dependent, inheriting their RG dependence from that of the \MSbar\ parameters themselves. However, this RG dependence is simple and known exactly, due to the superrenormalisability of the 3d EFT. It can be subtracted off to define the following RG invariant combination
\begin{align}
{\langle \phi^\dagger \phi \rangle}_{\text{RG}} &\equiv \langle \phi^\dagger \phi \rangle - \beta_{\langle \phi^\dagger \phi \rangle} \ln\Big( \frac{\scaleft}{T} \Big),
\end{align}
where 
\begin{align}
\beta_{\langle \phi^\dagger \phi \rangle}
= \frac{1}{(4\pi)^2}(3g_{3}^{2}+g_{3}^{'2}),
\end{align}
and likewise for the triplet condensate with $\beta_{\left\langle \text{Tr} \Sigma^2 \right\rangle} = 12g_{3}^{2}/(4\pi)^2$. From now on we will omit the RG subscript, but when we refer to condensates, we mean the above RG-invariant condensates. Note that these beta functions are exact, and independent of the phase, due to the superrenormalisability of the 3d EFT \cite{Farakos:1994xh}. On the lattice, this combination is exactly RG invariant, whereas in any finite order perturbative calculation some RG dependence will remain, as a consequence of missing higher loop terms. By varying the RG scale, we gain some estimate of the size of these missing higher loop terms.

In this work, we resolve the failures of perturbation theory visible in Fig.~\ref{fig:soft_strict_direct} by providing a consistent setup for the radiatively-generated first transition, as inspired by Ref.~\cite{Ekstedt:2022zro}. In this setup, a physically correct picture of a radiatively-generated transition is provided by consistent power counting, where the barrier separating symmetric and broken phases exists already at leading order, due to integrating out heavy degrees of freedom that generate the barrier. For the second transition, the hierarchies of scale can change, necessitating the construction of separate EFTs for the two transitions, occurring at distinctive thermal scales. This novel construction can be used to avoid the spurious IR divergence of the triplet condensate at symmetry breaking, and provide sound predictions for the critical temperature and strength of the first transition. We find that a complete and gauge-invariant resolution of the failures of perturbation theory also requires perturbatively expanding the critical temperature, following ideas presented in Refs.~\cite{Laine:1994zq, Gould:2022ran}. In order to scrutinize the accuracy of our purely perturbative computation, we compare our results to the non-perturbative lattice simulations of Ref.~\cite{Niemi:2020hto}. In addition, we provide a thorough comparison to some alternative perturbative methods -- such as direct, gauge-dependent minimisation of the effective potential -- and discuss their reliability and accuracy, despite their obvious theoretical blemishes.

\section{Thermal scale hierarchies}
\label{sec:setup}

In weakly coupled quantum field theories, scale hierarchies are a necessary prerequisite for a thermal phase transition. The argument goes as follows. Thermal effects arise through loop diagrams, which are subleading in the vanilla loop expansion. Yet, for there to be a phase transition, these thermal effects must change the effective dynamics of the transitioning field at leading order (LO). This \textit{requires} a breakdown of the vanilla loop expansion, because the subleading order must match the leading order.%
\footnote{A partial way out of this argument is if a model is close to a phase transition already at zero temperature, then only small thermal corrections are needed to undergo the transition. However, this setup requires a hierarchy of scales already at tree-level, whereby the potential difference between the minima is parametrically small compared with the curvature of the potential.}
Finally, for equilibrium physics, which is time independent and hence absent a light cone, the only kinematic enhancements possible are simple scale hierarchies. In fact, as we will see, at phase transitions in weakly coupled theories there are typically multiple scale hierarchies.

Scale hierarchies wreak havoc with the loop expansion. Any large ratio of UV to IR energy scales $\Lambda_\text{UV}/\Lambda_\text{IR}$ can multiply loop corrections, enhancing them relative to their naive loop counting. EFT provides a systematic means to account for such enhancements. To construct a reliable perturbative expansion to describe a given energy scale $\Lambda$, one must first integrate out all energy scales which are parametrically larger. In studying thermal phase transitions, the first step therefore is to identify the energy scale of the transitioning field.

In constructing the EFT for the transitioning field, one integrates out heavy degrees of freedom step by step. Each heavy degree of freedom that is integrated out modifies the effective infrared dynamics, including the effective mass of the transitioning field. How then do we know when to stop integrating modes out? After constructing the EFT for energy scales $\Lambda$ and below, if the mass of the transitioning field remains of order $\Lambda$ through the transition, then one can be sure that all contributions which are enhanced by a ratio of scales have been captured. On the other hand, if there is an apparent second-order phase transition, then the effective mass of the transitioning field goes to zero, and any other massive degrees of freedom will become relatively heavier than the transitioning field. Thus, new hierarchies of scale arise, and the fields which remain of mass $\Lambda$ through the transition must be integrated out.

The same conclusions can be reached from a rather different perspective. In general, a successful perturbative expansion requires a LO approximation which is relatively close to the complete result. If the LO approximation is qualitatively different from the complete result, then perturbation theory will fail, and may exhibit all manner of weird and wonderful pathologies. In the study of phase transitions, such pathologies arise when the LO approximation fails to get the order of the phase transition right.

At nonzero temperature, infrared modes (with energy  $E\ll T$) of bosonic fields become highly occupied, and their collective effective coupling is enhanced. For strong first-order phase transitions, the transitioning field remains gapped, and weak-coupling (i.e.~mean-field) expansions can work rather well, as long as the effective coupling is small. For weaker transitions, the bosonic field undergoing the transition is relatively lighter, so the convergence of the expansion is slower, until for transitions of second order or higher, the effective coupling is large, and there is therefore no weak-coupling expansion. One must then resort to other approaches, such as bootstrapping \cite{El-Showk:2012cjh}, lattice Monte-Carlo \cite{Farakos:1994xh, Rummukainen:1998as}, weak-strong dualities \cite{Sberveglieri:2020eko}, or the $\epsilon$ expansion \cite{Wilson:1973jj, Arnold:1993bq}.

Combining the observations of the previous two paragraphs, we reach the following conclusion: when studying phase transitions using a weak-coupling expansion, one should \textit{always} start with a LO approximation in which the transition is of first order. This is the best that one can do with perturbation theory. If the transition is indeed of first order, then the perturbative expansion will converge well. On the other hand, if the transition is really of second order or higher, then perturbation theory will fail, but there is anyway no way around it.

In this section, we present a generic setup for the perturbative analysis of thermal phase transitions. We will use EFT to construct LO approximations  in which a given phase transition is of first order. This will provide us with the best possible starting position for perturbation theory and gives results which are gauge invariant, real and free from spurious infrared divergences. It also improves agreement with the lattice, even when the transition is not of first order.   

\subsection*{Thermal scale hierarchies}

The starting point of our computation is the EFT picture for thermal phase transitions \cite{Ginsparg:1980ef, Appelquist:1981vg, Kajantie:1995dw, Braaten:1995cm}. This starts from the assumption that we are at high temperatures compared to relevant mass scales, $T \gg m$, and is based on the following chain of scale hierarchies
\begin{align} \label{eq:thermal_scale_hierarchies}
\underbrace{\vphantom{\frac{g^{\frac{1}{2}}}{4\pi}}\pi T}_{\text{hard scale}}
\gg
\underbrace{\left(\frac{g}{4\pi}\right)^{\frac{1}{2}} \pi T}_{\text{semisoft scale}}
\gg
\underbrace{\left(\frac{g}{4\pi}\right)^{1} \pi T}_{\text{soft scale}}
\gg
\underbrace{\left(\frac{g}{4\pi}\right)^{\frac{3}{2}} \pi T}_{\text{supersoft scale}}
\gg
\underbrace{\left(\frac{g}{4\pi}\right)^2 \pi T}_{\text{ultrasoft scale}}
\;,
\end{align}
in terms of a weak coupling $g \ll 1$, and the temperature $T$. The factors of $\pi$ arise from Matsubara modes ($\pi T$) and loop integrals ($\frac{g}{4\pi}$). However, from here on we shall omit the factors of 4 related to loop integrals, as they are often compensated by group theory factors in Feynman diagrams multiplying the loop integral. The hard, semisoft, soft and supersoft scales are perturbative; for these scales the effective expansion parameters are small,  $\veps_{\text{eff}} \ll 1$. Indeed, there are separate expansion parameters for each energy scale, with $\veps_\text{hard}\sim (g/\pi)^2$, and the expansion parameters for softer scales are larger, indicating slower convergence. We dedicate the next section to discuss how EFT expansions arise for the semisoft, soft and supersoft scales. Energy scales higher than the hard scale are exponentially (Boltzmann) suppressed. At the other extreme, energy scales at or below the ultrasoft scale are non-perturbative, as the effective expansion parameter therein is of order unity \cite{Linde:1980ts}. However, as we argue below, the dynamics of strong first-order thermal phase transitions generally takes place at either the soft or supersoft scales. Only for very weak transitions can the dynamics take place at the ultrasoft scale. In principle, other energy scales between the hard and ultrasoft scales may arise, though we have not encountered them.

In applying the above power counting to a given model, the parameter $g$ should be chosen such that $\veps_{\text{hard}} \sim g^2/(\pi)^2$ determines the convergence of the loop expansion for the hard scale. Thus, for models with a single dimensionless coupling, $g$ can be identified with a 3-point coupling, or $g^2$ with a 4-point coupling. For models with multiple couplings $g$ should be identified with the \textit{largest} relevant coupling, as this is what limits the convergence of the loop expansion.

\paragraph{Hard scale}
The temperature sets the most UV scale for thermal fluctuations, as energies above this are Boltzmann suppressed. At high temperatures, these hard scale fluctuations dominate the free energy density. For equilibrium physics, Matsubara's imaginary time formalism reveals that $n \pi T$ sets the energy scale of thermal-scale fluctuations, where $n$ is an even integer for bosonic fields, or an odd integer for fermionic fields. All modes except the (bosonic) zero Matsubara mode $n=0$ therefore have energies of at least the hard scale. Fields with masses above the hard scale can be integrated out as at zero temperature \cite{Laine:2000kv, Brauner:2016fla, Hirvonen:2022jba}. For the hard scale fluctuations, each successive loop is suppressed by $\veps_{\text{hard}} \sim g^2/(\pi)^2$ compared to previous one.

\paragraph{Soft scale}
The effective dynamics of softer modes is screened by hard scale fluctuations. At one-loop order this screening induces an effective mass of order $g T$. Bosonic zero Matsubara modes are therefore generically of the soft scale, unless there is some mechanism for the partial or full cancellation of one-loop screening. The EFT construction between the hard scale and the soft scale is the well-known high-temperature dimensional reduction \cite{Kajantie:1995dw, Braaten:1995cm}.   Technically, this amounts to integrating out nonzero Matsubara modes with masses of order $\pi T$, and constructing the EFT for the three-dimensional zero Matsubara modes of all lighter bosonic fields. The temporal components of gauge fields acquire a thermal Debye mass due to the heat bath breaking Lorentz invariance \cite{Kapusta:2006pm, Laine:2016hma}. These modes always live at the soft scale. Their squared Debye masses are solely generated by screening of the hard scale, so are a sum of positive definite terms each of order $(g T)^2$. Heavy bosonic zero-modes, with masses comparable to the hard scale $\pi T$, are integrated out along with the nonzero Matsubara modes \cite{Laine:2000kv, Brauner:2016fla, Niemi:2018asa, Gould:2019qek, Lofgren:2023sep}. Within the soft scale EFT, each successive loop is suppressed by $\veps_{\text{soft}} \sim g/\pi$ compared to previous one, at least when the interaction corresponding to $g$ is present at the soft scale.

\paragraph{Supersoft scale}
The thermal effective masses of Lorentz scalar fields can be parametrically smaller than the soft scale. If the quadratic mass parameter of a scalar field is negative at zero temperature $\mu^2 < 0$, then there can be cancellations between this and the positive hard thermal contributions to the effective mass,
\begin{align} \label{eq:supersoft_cancellation}
\mu^2_3 \approx \mu^2 + c g^2 T^2 \ll g^2 T^2,
\end{align}
where the subscript 3 is used to denote the effective mass of the field in the 3d EFT and $c > 0$ is an $\mathcal{O}(1)$ numerical coefficient.%
\footnote{In fact, it is possible to have $c < 0$ in multi-scalar theories with negative cross-couplings, in which case there can be inverse symmetry breaking \cite{Weinberg:1974hy, Jansen:1998rj}. In that case there is a phase transition for $\mu^2>0$, and it can also take place at the supersoft scale.} For broad classes of transitions, the transitioning field becomes lighter than the soft scale at the critical temperature. The dominant subleading corrections to its effective squared mass come from integrating out soft-scale fields, and are of order $\mathcal{O}(g^3T^2/\pi)$. Thus, barring additional parametric cancellations, the mass of the transitioning field lives at the supersoft scale $g^{3/2} T/\sqrt{\pi}$. This is the case for symmetry-breaking first-order phase transitions \cite{Arnold:1992rz, Ekstedt:2020abj}. The supersoft scale is in fact even more widely applicable to thermal first-order phase transitions, as we find below.

The construction of an EFT for the supersoft scale was introduced in Ref.~\cite{Gould:2021ccf}. The transition of a supersoft field can modify the masses of soft-scale fields at leading order. In this case, the effective Lagrangian at the supersoft-scale will have non-polynomial dependence on the transitioning field. It is nevertheless local, as the hierarchy of scales ensures the derivative expansion holds. This is akin to the EFT of inflation \cite{Burgess:2020tbq}. Within the supersoft scale EFTs we consider, the supersoft scale fields have parametrically small couplings at zero temperature, and each successive loop is suppressed by $\veps_{\text{super}} \sim (g/\pi)^{3/2}$ compared to previous one.

\paragraph{Semisoft scale}
This scale lies between the hard and soft scales. It can arise naturally for very strong first-order phase transitions: if the jump in a scalar background field becomes as large as $\sqrt{\pi} T / \sqrt{g}$, then it can impart a mass of order $\sqrt{\pi g} T$ on other fields through the Higgs mechanism. Below we find that this situation occurs in $Z_2$-symmetric multi-field models where there are two successive first-order phase transitions, and where there is sufficient supercooling between them. For such setups, when integrating out the semisoft scale fluctuations, each new perturbative order is suppressed by $\veps_{\text{semi}} \sim \sqrt{g/\pi} $ compared to previous one, though multiple orders in this expansion appear at each loop order (suitably resummed).

\paragraph{Ultrasoft scale}
This scale and below are nonperturbative. In gauge theories, the spacelike Ward identities ensure that the spatial components of gauge bosons do not receive a thermal mass correction within perturbation theory. In the absence of a Higgs mechanism, they therefore remain massless until the ultrasoft scale, where they receive a nonperturbative thermal mass \cite{Linde:1980ts}. Lorentz scalar fields can also become ultrasoft in the near vicinity of a second-order phase transition. However, unlike for gauge fields, fine tuning is typically required for a scalar to be this light. The contribution of the ultrasoft scale to the free energy density is of order $T (g^2 T / \pi)^3$, and hence subdominant to the contributions of the higher energy scales.

\section{Effective field theory expansions}
\label{sec:xpansion} 

We begin by discussing the thermal effective potential, or the background-field-dependent free energy density of the plasma. From this the phase transitions of a model can be determined. We discuss the computation of the effective potential for a transition taking place at the soft or supersoft scale. The scale inducing such a transition must be heavier than the transitioning field, hence it can be the hard, soft or semisoft scale. 

A starting point of our discussion is the tree-level potential for the soft scale 3d EFT, in terms of real background fields $v$%
\footnote{
Here, we indicate the size of the potential in terms of the mass term, quadratic in $v$. In the vicinity of a phase transition, terms with other powers of $v$ are of comparable size. Note that the mass dimension of $v$ is 1/2, following from the canonical normalisation of a scalar field in 3d. The relation to the corresponding 4d field is $v^2 \sim v_\text{4d}^2 / T$.
}
\begin{align} \label{eq:V_soft_tree_counting}
V^{\text{soft}}_{\text{tree}}
&\sim g^2T^2 v^2( 1 + \veps_{\text{hard}}  + \mathcal{O}(\veps^2_{\text{hard}})).
\end{align}
By tree-level we mean that no loop diagrams from within the soft scale EFT are included. However loop diagrams from the hard scale \textit{are} included in $V^{\text{soft}}_{\text{tree}}$, and are captured in the parameters of the EFT. This is reflected in the expansion in $\veps_\text{hard}$ on the right hand side, where we will assume that the first two orders have been calculated. The calculation of the NLO term was pioneered in Refs.~\cite{Kajantie:1995dw, Braaten:1995cm, Braaten:1995jr} and is a mainstay of high-temperature dimensional reduction. It has now been automated for generic models \cite{Ekstedt:2022bff}.

Based on the scale hierarchy between inducing and transitioning scales, we present EFT expansions of the effective potential. In such expansions, perturbation theory is organised in terms of power counting with respect to dimensionless quantities within the 3d description. In Sec.~\ref{sec:methods}, we then discuss the computation of thermodynamics using these EFT expansions. 

The validity of our perturbative EFT expansions will depend on the magnitude of the background field $v^2$. For very weak transitions, a field with effective coupling $g_3^2\sim g^2 T$ and mass $\sim g_3|v|$ will become nonperturbative unless $v^2 \gg g^2 T / \pi^2$. On the other hand, for very strong transitions, the thermal scale hierarchy will break down altogether unless $v^2 \ll \pi^2 T / g^2$. These conditions ensure that a particle of mass $g_3|v|$ is much heavier than the ultrasoft scale, and much lighter than the thermal scale $\pi T$. Together they determine the relatively wide range of transition strengths which we can describe perturbatively,
\begin{align} \label{eq:vev_range}
\frac{g^2}{\pi^2} \ll \frac{v^2}{T} \ll \frac{\pi^2}{g^2}.
\end{align}
In the context of electroweak baryogenesis, the geometric midpoint $v^2 \sim T$ is favoured in simple BSM models which produce the observed baryon asymmetry \cite{Kuzmin:1985mm, Morrissey:2012db}. The midpoint also often makes a convenient choice for power counting, but in what follows we will discuss both weaker and stronger transitions.

\subsection{Transition for a soft field}
\label{sec:soft-EFT}

As argued above, we wish to construct a perturbative expansion which predicts a first-order phase transition at leading order, which in this case means at tree-level within the soft-scale EFT. This implies a certain structure for the tree-level soft-scale potential: there should be at least two coexisting local minima, separated by a potential barrier. In the absence of such a tree-level barrier, we argue that there are no soft-scale phase transitions that can be described reliably within perturbation theory. Assuming the theory is weakly coupled at zero temperature, we can then conclude that \textit{either} there are no transitions at all, \textit{or} the transition takes place further into the IR.

The perturbative expansion of the soft scale effective potential can be expressed in terms of a formal expansion in $\veps_{\text{soft}}$ which is a dimensionless ratio of EFT parameters. Such an expansion parameter inherits its scaling $\veps_{\text{soft}} \sim g/\pi$  from the original theory, but can be treated as an independent expansion parameter in the following sense. Within the EFT, perturbation theory can be organised as an expansion in $\veps_{\text{soft}}$, and such a computation can be used to find critical values for 3d parameters, and the condensates as functions of the 3d parameters. Then, one wants to relate these to the temperature and the original parameters of the parent 4d theory, and this is done in an expansion in $\veps_{\text{hard}} \sim g^2/\pi^2$ in dimensional reduction. Indeed, there are two different expansions, one related to UV physics at the hard scale, and another to IR physics at the soft scale. The effective potential at the soft-scale admits the formal expansion
\begin{align}
V^{\text{soft}}_{\text{eff}} &= \underbrace{V_{0}^{\text{soft}}}_{\sim g^2 T^3} + \underbrace{V_{1}^{\text{soft}} }_{\sim \veps_{\text{soft}} V_{0}^{\text{soft}}} + \underbrace{V_{2}^{\text{soft}} }_{\sim \veps^{2}_{\text{soft}} V_{0}^{\text{soft} }} +\ \mathcal{O}(\veps^3_{\text{soft}} V_{0}^{\text{soft}}), 
\end{align}
where we have introduced $V_{0}^{\text{soft}} \equiv V^{\text{soft}}_{\text{LO}}$ to simplify notation, and denote higher order corrections with increasing numeral in the subscript.
Here the scaling of the LO potential is indicated in terms of the original weak expansion parameter of the parent theory, and we have assumed $v \sim T$ for simplicity. The magnitudes of higher order corrections are given with respect to LO in terms of $\veps_{\text{soft}} \sim g/\pi$.

\begin{figure}
\centering
\includegraphics[width=0.7\textwidth]{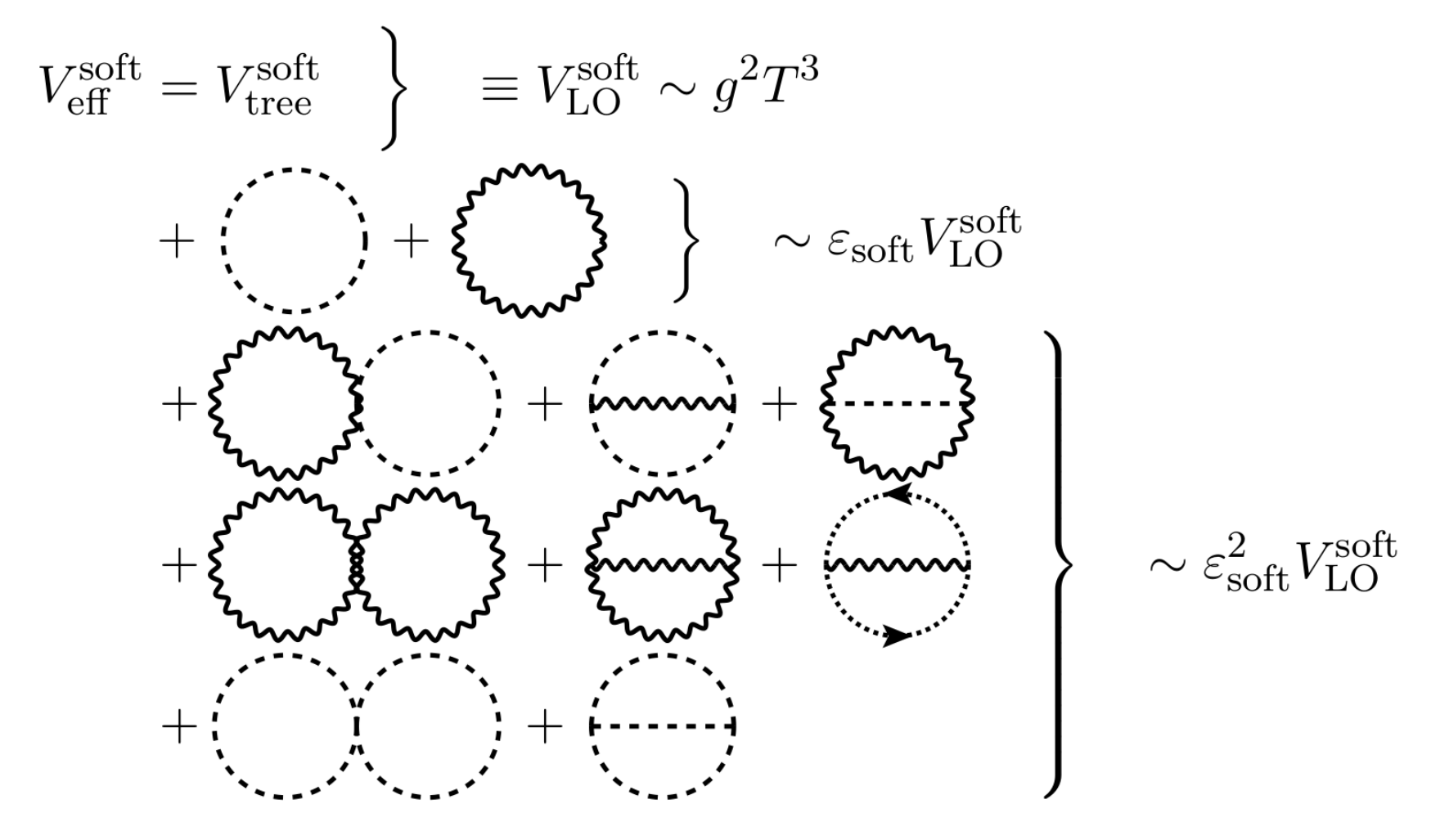}
\caption{
Schematic diagrammatic expansion of the soft scale effective potential up to N2LO. Expansion in $\veps_{\text{soft}} \sim \frac{g}{\pi}$ aligns with the loop expansion within the EFT, each new loop order introducing one power of $\veps_{\text{soft}}$.
}
\label{fig:soft-veff}
\end{figure}

Diagrammatically, the expansion in $\veps_{\text{soft}}$ aligns with the loop expansion within the EFT. The computation of the effective potential up to two loops is straightforward, and has recently been automated for general models \cite{Ekstedt:2022bff}. It is illustrated in Fig.~\ref{fig:soft-veff}.
Here dashed lines represent all scalars, wiggly lines gauge fields and dotted lines ghosts. The next term, of order $\mathcal{O}(\veps^3_{\text{soft}} V_{0}^{\text{soft}})$, arises at three-loops. This is the last order which is computable in perturbation theory in theories with non-Abelian gauge fields \cite{Linde:1980ts}. For later convenience, we denote next-to-next-to leading order as N2LO and higher orders with increasing numeral.

Possible realisations of a soft-scale EFT showing a first-order phase transition at tree-level include a real scalar with cubic and quartic interactions \cite{Gould:2021dzl}, a real scalar with quartic and sextic interactions \cite{Croon:2020cgk}, and multi-scalar models where there is a transition between different broken phases \cite{Gould:2021oba}. The tree-level potential of the cubic-quartic model, here written in terms of a real background $v$, reads
\begin{align}
\label{eq:cubic_quartic_potential}
V^{\text{soft}}_{\text{tree,cubic}} &= \frac{1}{2}m^2_3 v^2 + \frac{1}{3} \kappa_3 v^3 + \frac{1}{4} \lambda_{3} v^4,
\end{align}
where the linear term has been removed by a shift $v \to v + \text{const}$.%
\footnote{Note that because a scalar field in 3d has mass dimension 1/2, the cubic $\kappa_3$ and quartic $\lambda_{3}$ couplings have mass dimension 3/2 and 1 respectively.}
For $\kappa_3^2 > 4 \lambda_3 m_3^2$, this potential admits two minima separated by a maximum, and these two minima have the same height when $\kappa_3^2 = (9/2) \lambda_3 m_3^2$.

Starting with the potential of Eq.~\eqref{eq:cubic_quartic_potential} as our LO approximation, let us consider loop corrections. These arise both from the soft scale (within the EFT) and from the hard scale (the construction of the EFT). The loop expansion parameters within the soft scale EFT are
\begin{align}
\veps_{\text{soft}} 
\sim
\frac{\lambda_3}{(4\pi) m_3}
\; , \;
\frac{\kappa_3^2}{(4\pi) m_3^3}.
\end{align}
The powers of $\kappa_3$ and $\lambda_3$ follow from standard graph-theoretic identities \cite{Peskin:1995ev}, the inverse powers of mass arise from loop integrals and can be determined from dimensional analysis, and the factors of $(4\pi)$ follow from the angular integrals arising in 3d loop integrals. These should be compared with the loop expansion parameters arising within the corresponding 4d theory (with analogous parameters dropping subscripts 3),
\begin{align}
\veps_{\text{hard}} 
\sim
\frac{\lambda}{(4\pi)^2}
\; , \;
\frac{\kappa^2}{(4\pi)^2 m^2},
\end{align}
which determine corrections arising from the hard scale.

In the vicinity of the critical temperature, the joint requirements that there are two coexisting minima, and that their potential energies are approximately equal, imply that all three terms in the potential are of the same order, so that $\kappa_3^2 \sim \lambda_3 m_3^2$, and hence that the convergence of the loop expansion within the EFT is determined by a single expansion parameter $\veps_{\text{soft}}  \sim \lambda_3/(4\pi m_3) \sim \kappa_3^2 / (4\pi m_3 ^3)$ \cite{Gould:2021dzl}. Using $\lambda_3 \sim g^2 T$ and $m_3 \sim g T$ together implies $\veps_{\text{soft}} \sim g/(4\pi)$. The loop expansion within the EFT therefore converges more slowly than the loop expansion used in constructing the EFT. In addition, the expansion within the EFT diverges for a second order phase transition $m_3 \to 0$, though in the approach to this point the field becomes lighter than the soft scale.

For the $Z_2$-symmetric model with quartic and sextic interactions, the potential reads
\begin{align}
V^{\text{soft}}_{\text{tree,sextic}} &= \frac{1}{2}m^2_3 v^2 + \frac{1}{4}\lambda_3 v^4 + \frac{1}{8}c_{6,3} v^6,
\end{align}
where $\lambda_3 < 0$. The general conclusions about the perturbative expansion in the cubic-quartic model carry over to this case. In the vicinity of the critical temperature the loop expansion parameter within this EFT is $\veps_{\text{soft}} \sim \lambda_3/(4\pi m_3) \sim \sqrt{c_{6,3}}/(4\pi)$ \cite{Croon:2020cgk,Camargo-Molina:2021zgz,Ekstedt:2021kyx}. For $\lambda_3 \sim g^2 T$, the soft expansion parameter is again of order $\veps_{\text{soft}} \sim g / (4\pi)$.

Our third example is provided by a phase transition with two scalar fields participating in the transition, such that the transition happens between the different broken minima of the potential. For simplicity, we assume here a $Z_2$-symmetric model where scalars are charged under a gauge group with a gauge coupling $g$. Concretely we discuss the following tree-level potential with two background fields $x$ and $y$
\begin{align}
\label{eq:two-field-potential}
V^{\text{soft}}_{\text{tree,x,y}} &=
\frac{1}{2} \mu^2_{x,3} x^2
+ \frac{1}{2} \mu^2_{y,3} y^2
+ \frac{1}{4} \lambda_{x,3} x^4
+ \frac{1}{4} \lambda_{y,3} y^4
+ \frac{1}{4} \lambda_{xy,3} x^2 y^2.
\end{align}
For power counting, we assume that the scalar masses lie at the soft scale, $\mu^2_{x,3}, \mu^2_{y,3} \sim (gT)^2$, as well as possible gauge field and Debye masses $m_W, m_D \sim g T$, and that the couplings are all equally perturbative $\lambda_{x,3}, \lambda_{y,3}, \lambda_{xy,3} \sim g^2 T$. The loop expansion parameter within the EFT is then $\veps_{\text{soft}} \sim \lambda_{x,3} / (4\pi |\mu_{x,3}|) \sim g / (4\pi)$.

Depending on the signs of the mass terms, there is a symmetric minimum where both background fields vanish $(x, y) = (0, 0)$, and broken minima at $(x_0, 0)$ and $(0, y_0)$, where 
\begin{align}
x_0 = & \sqrt{\frac{-\mu^2_{x,3}}{\lambda_{x,3}}} , &
y_0 = & \sqrt{\frac{-\mu^2_{y,3}}{\lambda_{y,3}}} .
\end{align}
For certain choices of parameters, the symmetry breaking pattern $(0, 0) \to (x_0, 0) \to (0, y_0)$ is possible. The second step can be described reliably by this soft-scale EFT as, in field space, the two broken minima are separated by a barrier, so the transition is of first-order within the EFT.

\subsection{Transition for a supersoft field}
\label{sec:supersoft-EFT}

We continue by discussing the generic setup for a supersoft scale EFT of a single scalar field with a \textit{tree-level} potential barrier. By tree-level we mean that the potential does not include any loop corrections from the supersoft scale, yet note that it can still be non-polynomial due to contributions from the soft scale fields that have been integrated out. Indeed, the barrier is typically generated by the soft scale. Our following discussion is inspired by \cite{Arnold:1992rz, Weinberg:1992ds, Metaxas:1995ab, Metaxas:2000cw, Garny:2012cg, Ekstedt:2020abj, Gould:2021ccf, Hirvonen:2021zej} that discuss one-step phase transitions with radiative barriers. For a two-step EWPT, this setup describes the first transition to an intermediate phase before the transition to the EW phase.

Our discussion here is schematic: we assume a single supersoft scalar field $\phi$, with real background field $v$, that couples to a gauge field, with 3d gauge coupling $g_3$. The one-loop effective potential at the \textit{soft scale} reads \cite{Farakos:1994kx} 
\begin{align}
V^{\text{soft}}_{\text{1-loop}} &\simeq \frac{1}{2} \mu^2_3 v^2 + \frac{1}{4} \lambda_3 v^4 - \frac{1}{12\pi} \Big(6 m^3_V + M^3 \Big). 
\end{align}
The first two terms are tree-level terms in the soft scale EFT, and the last term is a one-loop contribution where $m_V^2 \simeq g_3^2 v^2/4$ is the gauge boson mass eigenvalue and $M$ represents the scalar mass eigenvalues $M^2 \simeq \mu^2_3 + 3 \lambda_3 v^2$ or $M^2 \simeq m^2_D + h_3 v^2$.%
\footnote{
We have used numerical coefficients here which correspond to an SU(2) gauge theory with fundamental Higgs, and for simplicity have dropped Goldstone contributions. However, barring the precise values of these numerical coefficients, the discussion here applies to more general gauge groups.
}
Here $m_D$ is the Debye mass for the temporal component of a gauge field, and $h_3$ its coupling to $\phi$.

For there to be a first-order transition, the potential should have more than one minimum separated by a barrier. At tree-level in the soft scale EFT this is not possible, since there is only one minimum at any given temperature: when the mass parameter $\mu^2_3$ is positive, the only minimum is at $v_{\text{sym}}=0$, and when it is negative the minimum is instead at $v_{\text{broken}}=\sqrt{-\mu^2_3 / \lambda_3}$. At $\mu^2_3 = 0$ there is a second-order transition. However, the one-loop gauge boson contribution is cubic $\propto |v|^3$ and can provide a barrier between minima. Similarly, contributions from the temporal components of gauge fields contribute to the barrier -- as well as other soft scalar fields in the case of multiple scalars -- but in order to simplify the presentation we do not discuss their contribution further in this section, but assume that the 3d gauge field is the only soft field. For a transition to be of first order, the one-loop gauge boson term should be of the same parametric order as the tree-level terms of the potential, i.e.~
\begin{align} 
\label{eq:equating_orders}
\mu^2_3 v^2 \sim \lambda_3 v^4 \sim \frac{g^3_3 v^3}{\pi}.
\end{align}
Additional loop corrections from gauge bosons are suppressed relatively by $g_3^2/(\pi g_3 |v|)$, and those from the scalar undergoing the transition are suppressed by $\lambda_3/(\pi M_3)$. To keep track of perturbative corrections, we introduce the dimensionless power counting parameter $\veps_{\text{soft}} \sim g_3/(\pi |v|)$, which counts soft-scale loops and satisfies $0<\veps_{\text{soft}} \ll 1$ when perturbative corrections are small. Expressing all three parameters in units of $v$, Eq.~\eqref{eq:equating_orders} then implies that
\begin{align}
\frac{g_3^2}{\pi^2} \sim \veps_{\text{soft}}^2 v^2,
\qquad \qquad
\frac{\mu_3^2}{\pi^2} \sim \veps_{\text{soft}}^3 v^4,
\qquad \qquad 
\frac{\lambda_3}{\pi^2} \sim \veps_{\text{soft}}^3 v^2.
\end{align}
Here, and in what follows, we use $\sim$ to denote that two quantities have the asymptotic scaling as $\veps_{\text{soft}}\to 0_+$. Note that this power counting is equivalent to that of Refs.~\cite{Arnold:1992rz, Ekstedt:2022zro} when $\veps_{\text{soft}} \sim g/(4\pi)$, and often the ratio $\veps_{\text{soft}} \equiv \lambda_3/g^2_3$ is used as an expansion parameter and denoted by $x$ in previous literature, e.g.~\cite{Kajantie:1996qd}. Note in particular that with this counting, the one-loop scalar terms are of order $\simeq M^3/\pi  \sim \mu_3^3/\pi \sim \pi^{2}\veps_{\text{soft}}^{9/2}v^3 $ and hence do not contribute at leading order to the potential 
\begin{align}
V^{\text{supersoft}}_{\text{EFT,LO}} = \frac{1}{2} \mu^2_3 v^2 + \frac{1}{4} \lambda_3 v^4 - \frac{1}{16\pi} g_3^3|v|^3
\end{align}
Here we have indicated that this is the LO, or tree-level, potential of the \textit{supersoft} scale EFT. The mass of the transitioning field is
\begin{align}
\widetilde{M}^2 \equiv \frac{d^2}{dv^2} V^{\text{supersoft}}_{\text{EFT,LO}} = \mu^2_3 + 3 \lambda_3 v^2 + \Pi \sim \Big(\frac{g^{\frac{3}{2}}}{\sqrt{\pi}} T \Big)^2, 
\end{align}
i.e.~at the supersoft scale. Here $\Pi \equiv - \frac{3}{8\pi} g^3_3 |v|$ is the resummed contribution from the soft gauge field with mass $m_W^2 \simeq g_3^2 v^2/4 \sim (g T)^2 $. The resummation arises from integrating out the soft fields, and it decorates the supersoft scalar propagator with one-loop insertions of the soft field. The broken minimum of the LO potential reads
\begin{align}   
\label{eq:v_broken}
v_{\text{broken}} = \frac{3 g_3^3}{32\pi\lambda_3} \left(1 + \sqrt{1 - \frac{1024 \pi^2 \mu^2_3 \lambda_3}{9 g_3^6}}\right).
\end{align}
This minimum is separated from the symmetric minimum at $v_{\text{sym}} = 0$ by a barrier for temperatures such that the effective mass term lies in the range $0 < \mu_3^2 < 9 g_3^6 / (1024 \pi^2 \lambda_3)$. We comment that should we include contributions of other soft fields, the expression for the broken minimum becomes readily much more complicated analytically. In addition, we point out that the supersoft EFT is constructed in the broken phase, or for sufficiently large background fields, where the gauge field is indeed soft and can be integrated out.

Next, we consider higher orders in $\veps_\text{soft}$. The NLO corrections to the effective potential, suppressed by one power of $\veps_\text{soft}$, are given by two-loop digrams of purely soft modes. This can be formally performed by matching the effective potentials of the soft and supersoft EFTs, treating supersoft masses and momenta in strict perturbation theory, cf.~\cite{Burgess:2020tbq, Gould:2021ccf, Hirvonen:2021zej, Hirvonen:2022jba}. In this approach, the propagator of the supersoft field is treated as massless for the matching computation~\cite{Braaten:1995cm}, so that pure supersoft diagrams vanish identically, and only soft-scale contributions from mixed scalar/gauge diagrams are included. The outcome of this formal procedure is the potential for the supersoft EFT, and the procedure is illustrated in Fig.~\ref{fig:supersoft-veff}. In Appendix \ref{sec:supersoft-matching}, we present explicit computations up to N2LO for some relevant example models. 
\begin{figure}
\centering
\includegraphics[width=0.8\textwidth]{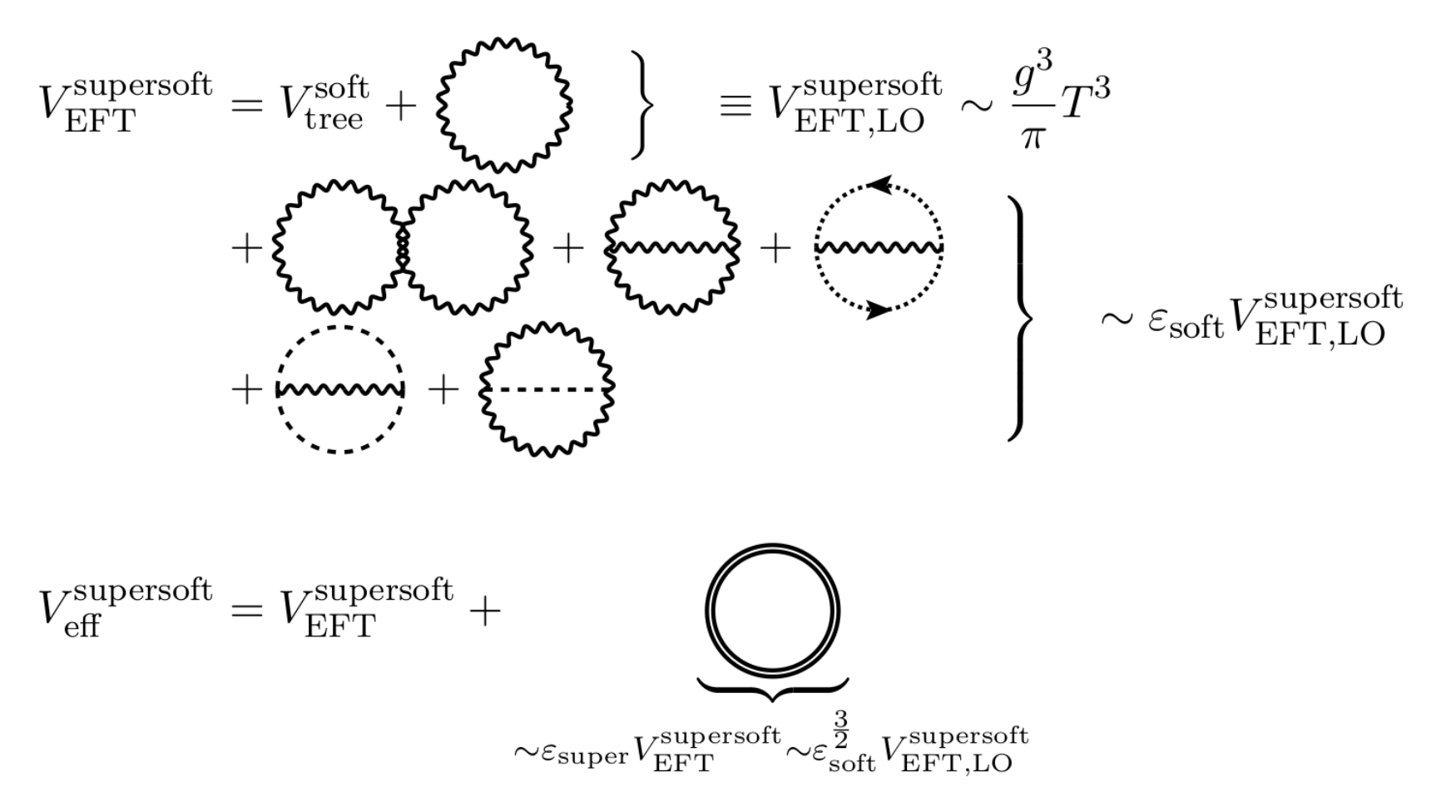}
\caption{
Schematic diagrammatic expansion of the supersoft scale effective potential. Computationally, this splits into two: EFT matching between soft and supersoft scales, and loop corrections within the supersoft theory. The perturbative expansion is misaligned with the loop expansion, as one-loop contributions of soft fields contribute at LO and lead to resummation of the supersoft field (denoted by the solid double line). 
}
\label{fig:supersoft-veff}
\end{figure}

In the construction of the effective potential for the supersoft scale, there are multiple expansions. First, there is the expansion related to integrating out the hard scale, which we will here take for granted. Second, there is the expansion related to integrating out the soft scale, and finally there is the loop expansion within the supersoft scale EFT. The parameters for these latter two expansions are related as $\veps_{\text{super}} \sim \veps^{3/2}_{\text{soft}}$, so that $\veps_{\text{super}} \sim (g/\pi)^{3/2}$. One-loop diagrams within the supersoft theory contribute at $\mathcal{O}(\veps_{\text{super}})$ relative to LO and have the simple form
\begin{align}
V^{\text{supersoft}}_{\text{N2LO}} &= -\frac{\widetilde{M}^3}{12\pi},
\end{align} 
in terms of the resummed mass $\widetilde{M}$ depicted with the double line in Fig.~\ref{fig:supersoft-veff}. Note that at the order we work, and because we are only interested in observables for homogeneous background fields, we do not need to include the effect of the momentum dependent field normalisation in matching, c.f.\ e.g.\ \cite{Hirvonen:2021zej}.

The N3LO contributions to the potential, or $\mathcal{O}(\veps^2_{\text{soft}})$ relative to LO, are given by three-loop soft-scale diagrams. This is followed by terms of order $\veps_{\text{soft}}\veps_{\text{supersoft}} \sim \veps_\text{soft}^{5/2}$ relative to LO, which are given by the resummation of the NLO corrections to the mass within the supersoft one-loop diagram \cite{Hirvonen:2022jba}. We do not compute either of these contributions, leaving them for future work. N4LO is the highest order that can be computed perturbatively, since at the next order one encounters Linde's Infrared Problem \cite{Linde:1980ts}, where all loop topologies of the ultrasoft scale contribute at the same order in powers of couplings. These contributions are suppressed relative to the LO potential by $\veps^3_{\text{soft}} \sim (g/\pi)^3$, in terms of the weak coupling of the original theory.

\subsection{A two step phase transition}

Let us return to the example model with two background fields $(x, y)$, and the potential of Eq.~\eqref{eq:two-field-potential}. We consider the interesting case where there is a two step transition with the first step $(0, 0) \to (x_0, 0)$ taking place at the critical temperature $T_{c,1}$ followed by a second step $(x_0, 0) \to (0, y_0)$ at a lower temperature $T_{c,2}$. Here $x_0$ and $y_0$ are generic nonzero background expectation values for $x$ and $y$.

The thermodynamics of such two-step phase transitions depends on the relative magnitudes of the couplings and masses. As argued above, for perturbation theory to work, we need to find EFTs for the transitioning fields in which these transitions appear first order. We find there are (at least) two natural options for the power counting relations, which we outline below. 

\subsubsection{The first step}

The first transition appears to be of second order within the soft-scale EFT. So, the transitioning field $x$ becomes parametrically lighter than the soft scale.

Let us start by assuming that all couplings are equally perturbative $\lambda_{x,3} \sim \lambda_{y,3} \sim \lambda_{xy,3} \sim g^2 T$, following the discussion after Eq.~\eqref{eq:two-field-potential}. Then, integrating out the soft scale fields around this transition, one finds that the largest possible discontinuity in the background field is $x_0^2 \sim g^2 T /(4 \pi)$. At this point the effective mass of the transitioning field is at the non-perturbative ultrasoft scale, and the transition is either very weak or a crossover and perturbation theory is not viable. This conclusion crucially relies on the absence of hierarchies between the couplings.

The first step $(0, 0) \to (x_0, 0)$ can be strongly first order if $\lambda_{x,3}$ is parametrically smaller than some other couplings to the $x$ field. For example, if the power counting for the portal coupling is unchanged $\lambda_{xy,3} \sim g^2 T$ but the self coupling scales as $\lambda_{x,3} \sim g^3 T / \pi$, then the $x$ field is supersoft at the first transition, and the analysis of Sec.~\ref{sec:supersoft-EFT} applies directly. For the first transition, the leading order potential then reads
\begin{align} \label{eq:V_supersoft_x}
V^{\text{supersoft}}_{\text{LO}}(x,0) &\simeq
\frac{1}{2} \mu^2_{x,3} x^2
+ \frac{1}{4} \lambda_{x,3} x^4
- \frac{1}{12\pi} \Big( 6 (m^2_V)^{\frac{3}{2}} + (m^2_y)^{\frac{3}{2}}   \Big),
\end{align}
where $m^2_y \simeq \mu^2_{y,3} + \frac{1}{2} \lambda_{xy,3} x^2$, and we have included contributions from a vector boson with mass $m^2_V = g^2_3 x^2/4$.

We assume that only the $x$ field becomes supersoft as it transitions, with other fields remaining soft. The balance of the three terms in the potential then implies that $x^2 \sim T$. In the supersoft scale EFT, the mass of the transitioning field is resummed 
\begin{align}
\widetilde{m}^2_x = \frac{d^2 V^{\text{supersoft}}_{\text{LO}} }{dx^2} = \mu^2_{x,3} + 3 \lambda_{x,3} x^2 - 3 \frac{g^3_3}{8\pi} |x| - \frac{\lambda_{xy,3}}{2\pi} \frac{\mu^2_{y,3} + \lambda_{xy,3} x^2}{\sqrt{\mu^2_{y,3} + \frac{1}{2} \lambda_{xy,3} x^2}}.
\end{align}
We emphasize that the effective couplings of the soft scale EFT do not vary much with respect to temperature, apart from their dimensional scaling, i.e.\ ratios $\lambda_{x,3}/T$, $\lambda_{y,3}/T$ and $g^2_3/T$ are approximately constant, whereas scalar mass parameters, $\mu^2_{x,3}$ and $\mu^2_{y,3}$, can vary non-trivially with temperature, in particular they can go through zero.

\begin{figure}
\centering
\includegraphics[width=0.48\textwidth]{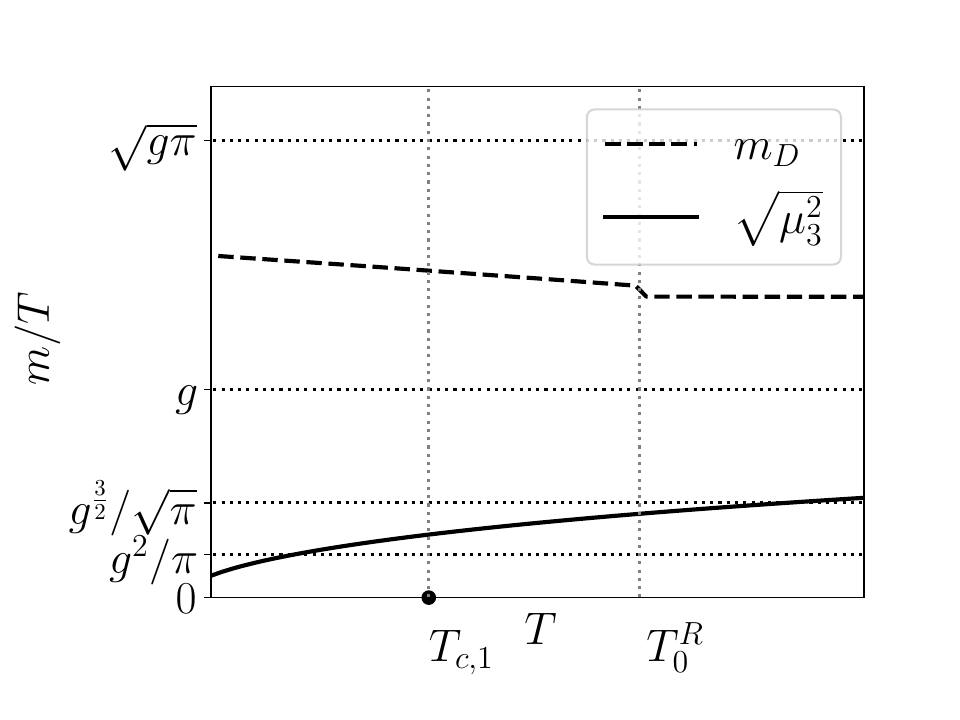} 
\includegraphics[width=0.48\textwidth]{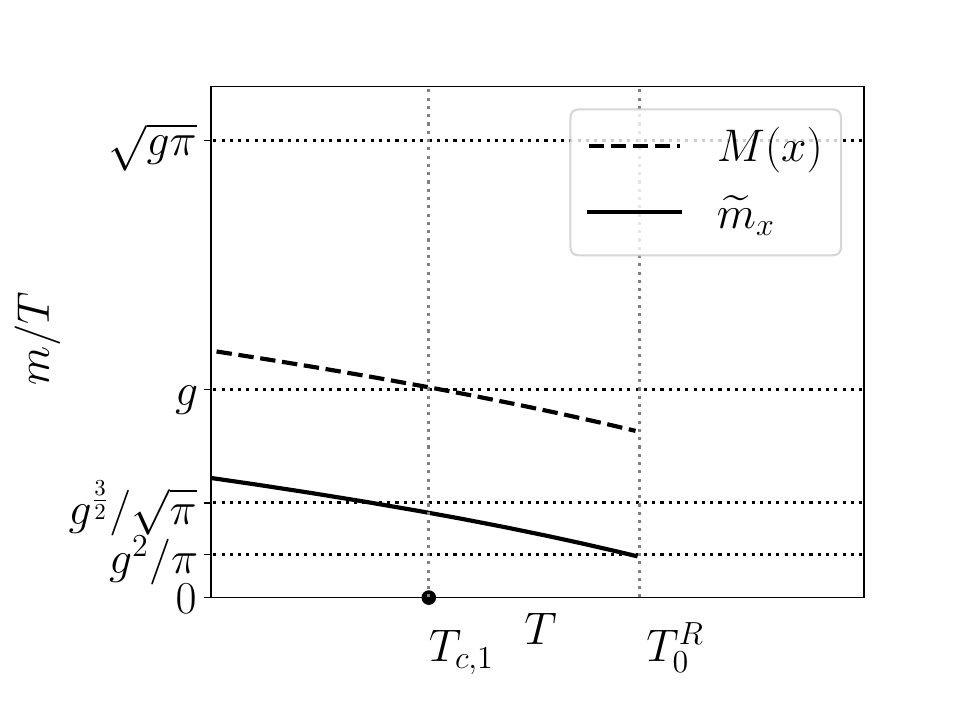} 
\caption{
Schematic evolution of masses as a function of temperature for a transition $(0,0)\to (x_0,0)$ in our example model of Eq.~\eqref{eq:two-field-potential}. A local minimum in the $x$-direction exists for temperatures $T<T_0^R$, and for $T<T_{c,1}$ it is the global minimum. Left: mass parameters in the soft scale 3d EFT. For the Debye mass, the contribution from the background field $x$ is included, leading to a small increase for $T<T_0^R$. Right: mass eigenvalues in the $x$-phase, where $\widetilde{m}_x$ denotes the resummed mass for the $x$-field, while $M(x)$ depicts the scaling of soft scale masses in the $x$-phase (for the $y$-field and the vector boson). The key features are: (i) scalar mass parameters are small near the transition (depicted by $\mu^2_3$ for both $x$- and $y$-fields), whereas the Debye mass is noticeably larger (note that the Debye mass is enhanced by group theoretic factors); (ii) for the final mass eigenvalues, there is a clear hierarchy between the light transitioning field and other fields. Note that such hierarchies might not be so clearly manifest at any given parameter point, yet this plot is inspired by our numerical application in Sec.~\ref{sec:results}.
}
\label{fig:masses-vs-temperature}
\end{figure}

Fig.~\ref{fig:masses-vs-temperature} shows a schematic plot of the temperature dependence of the effective masses in the vicinity of the first transition.
The left panel shows a generic scalar mass parameter, which grows with temperature. Shown in the right panel, however, when contributions from the background field $x$ are included, the scalar masses become decreasing functions of temperature below $T<T_0^R$ for which the $x$-minimum exists. The right panel depicts a generic soft mass $M$ and the resummed $x$-field mass in the $x$-phase of the supersoft EFT. Additionally in the left panel the Debye mass is shown, together with contributions from the background field. The Debye mass is typically significantly heavier than scalar masses. If the transition is weak, so that the correction to the Debye mass due to the background field is relatively small, it should be possible to integrate out the corresponding field following Ref.~\cite{Kajantie:1995dw}.

\subsubsection{Two consecutive supersoft scale transitions}
\label{sec:2-supersoft-EFTs}

Since 3d effective couplings do not vary significantly with temperature, we can assume they satisfy the same formal power counting relations for the second step $(x_0, 0) \to (0, y_0)$, as for the first. Furthermore, we also assume that $\lambda_{y,3} \sim \lambda_{x,3}$.%
\footnote{
In practice -- for strong electroweak phase transitions in models with relatively heavy BSM fields, and with a generic gauge coupling $g_4$-- the portal coupling $\lambda_{xy}$ is often the largest coupling, and there is a hierarchy $\lambda_{y} \sim \lambda_x < g^2_4 < \lambda_{xy}$ \cite{Caprini:2015zlo, Kainulainen:2019kyp, Niemi:2020hto, Niemi:2021qvp}. This suggests to organise all power countings with respect to $\lambda_{xy}$ instead of the gauge coupling.
}
Given this, what is the scale for the second transition?

The first possibility is that there is no significant supercooling between the critical temperatures, and while the masses of both $x$- and $y$-fields increase with increasing background field, they are still supersoft at $T_{c,2}$, in the $x$- and $y$-phases respectively (note that the $x$-field is soft in the $y$-phase, and vice versa). Alternatively, given enough supercooling down from $T_{c,1}$, the masses of transitioning fields could grow to become soft $\sim(g T)^2 $.

First, let us assume that the second transition to the $y$-phase occurs without much supercooling. Then the $x$-phase free-energy should still be computed in the supersoft scale EFT which described the first step, Eq.~\eqref{eq:V_supersoft_x}. What about the $y$-phase free energy?

Given the assumptions that $\lambda_{y,3} \sim \lambda_{x,3} \ll \lambda_{xy,3} \sim g_3^2$, the mass hierarchies in the $y$-phase are mirror those in the $x$-phase. The $y$-field is supersoft in the $y$-phase, and all else is soft. At leading order, in the vicinity of $T_{c,2}$ the $y$-phase potential therefore reads 
\begin{align} \label{eq:V_supersoft_y}
V^{\text{supersoft}}_{\text{LO}}(0,y) &\simeq
\frac{1}{2} \mu^2_{y,3} y^2
+ \frac{1}{4} \lambda_{y,3} y^4
- \frac{1}{12} \Big( 6 (m^2_V)^{\frac{3}{2}} + (m^2_x)^{\frac{3}{2}}   \Big),
\end{align}
where  $m^2_V = g^2_3 y^2 / 4 \sim (g T)^2$ and $m^2_x \simeq \mu^2_{x,3} + \frac{1}{2} \lambda_{xy,3} y^2 \sim (g T)^2$, i.e.\ the gauge field and $x$-field are soft in the $y$-phase. The critical temperature $T_{c,2}$ is determined from the condition that the free-energies of the phases are equal, with each being computed in separate a supersoft scale EFT, Eqs.~\eqref{eq:V_supersoft_x} and \eqref{eq:V_supersoft_y}.

\subsubsection{Transition for a soft field, induced by semisoft scale}
\label{sec:semisoft-to-soft-EFT}

\begin{figure}
\centering
\includegraphics[width=0.48\textwidth]{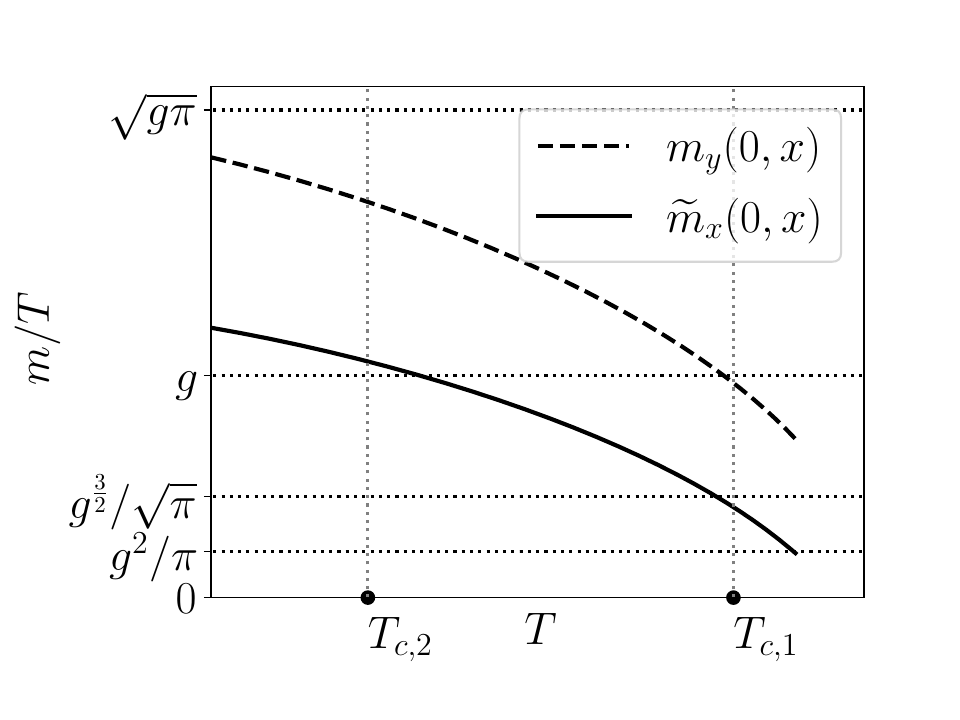} 
\includegraphics[width=0.48\textwidth]{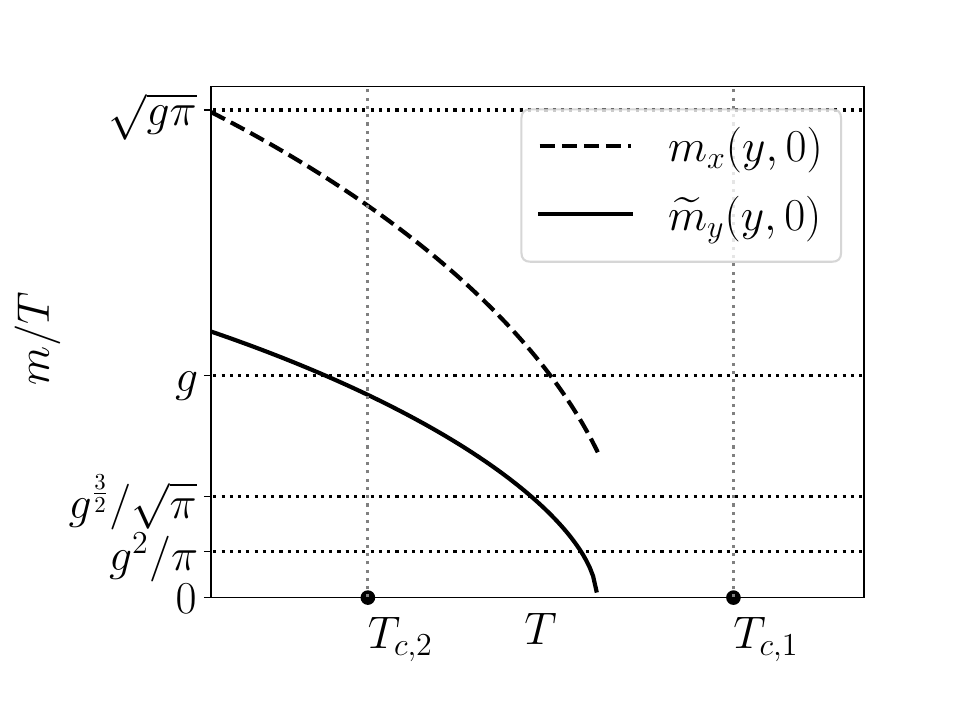} 
\caption{
Left (right): schematic visualisation of $x$-phase ($y$-phase) masses in the range between the critical temperatures of the two transitions.  Given enough supercooling between the transitions, the mass of the $x$-field can grow from supersoft to soft, and masses of other fields can grow from soft to semisoft. In this case a separate soft EFT should be constructed with the inducing scale being semisoft. Alternatively, should $T_{c,2}$ be sufficiently close to $T_{c,1}$, the transitioning $x$-field mass is still parametrically light and the supersoft EFT should be used. In this case also the mass of the transitioning $y$-field would be supersoft in the $y$-phase and a separate supersoft EFT can be constructed, in the $y$-phase. 
}
\label{fig:2step-masses}
\end{figure}

\begin{figure}
\centering
\begin{subfigure}{0.48\textwidth}
    \centering
    \includegraphics[width=\textwidth]{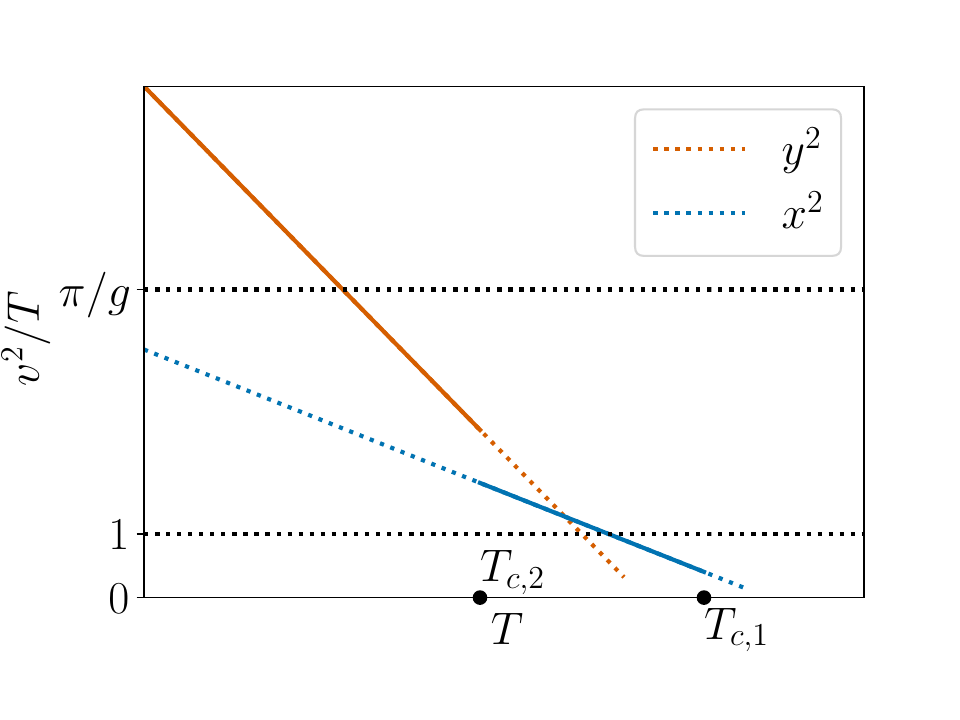}
    \label{fig:minima}
\end{subfigure}
\hfill
\begin{subfigure}{0.48\textwidth}
    \centering
    \includegraphics[width=\textwidth]{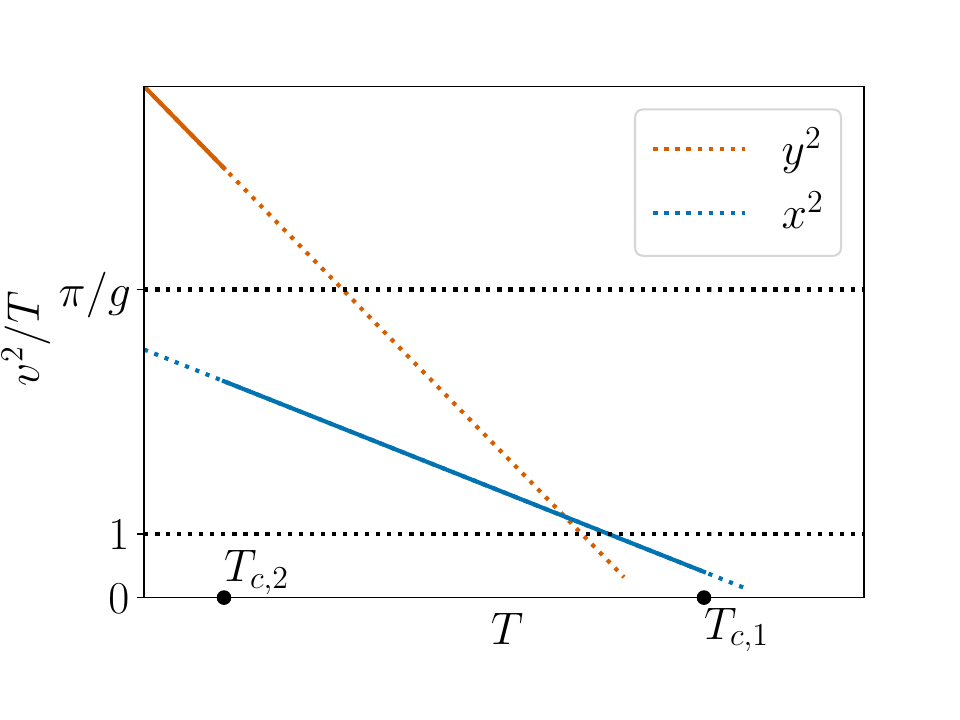} 
    \label{fig:minima-supercooling}
\end{subfigure}
\caption{
Schematic evolution of background fields for a two step transition. Dotted lines indicate metastable local minima, and solid lines stable global minima. Left: the second transition occurs without much supercooling after the first transition. For both transitions $\frac{\Delta v^2}{T_c} \sim 1$ and supersoft EFTs are constructed. Right: there is significant supercooling between the transitions, and the second transition is parametrically stronger $\frac{\Delta v^2}{T_{c,2}} \sim \frac{\pi}{g} \gg 1$. In this case, soft scale EFTs are constructed for the second transition by integrating out semisoft fields. 
}
\label{fig:vevs}
\end{figure}

On the other hand, if there is enough supercooling between $T_{c,1}$ and $T_{c,2}$, the masses of the transitioning fields can grow to the soft scale at the second transition. This is illustrated schematically in Fig.~\ref{fig:2step-masses}. 
In this case, background fields at minima are parametrically larger, and the leading order potential agrees with the standard tree-level potential. In the $x$ phase, this is
\begin{align}
V^{\text{soft}}_{\text{LO}}(x,0) &\simeq
\frac{1}{2} \mu^2_{x,3} x^2
+ \frac{1}{4} \lambda_{x,3} x^4,
\end{align}
which implies that at the broken minimum $x^2 \sim \frac{\mu^2_{x,3}}{\lambda_{x,3}} \sim \frac{\pi}{g}T$. Similarly, for the $y$-phase the background field satisfies $y^2 \sim \frac{\mu^2_{y,3}}{\lambda_{y,3}} \sim \frac{\pi}{g}T$. This is schematically illustrated in Fig.~\ref{fig:vevs}.
In the $x$-phase, the large background field of $x$ enhances and dominates the mass of $y$, and vice versa for the mass of $x$ in the $y$-phase,
\begin{align}
m^2_y(x,0) \simeq \mu^2_{y,3} + \frac{1}{2} \lambda_{xy,3} x^2 \sim \lambda_{xy,3} x^2  \sim  (\sqrt{g \pi} T)^2,\\
m^2_x(0,y) \simeq \mu^2_{x,3} + \frac{1}{2} \lambda_{xy,3} y^2 \sim \lambda_{xy,3} y^2   \sim (\sqrt{g \pi} T)^2.
\end{align}
These masses are \textit{semisoft}; the semisoft scale is the geometric mean between the soft and hard scales. Large background fields push the masses of other fields to the semisoft scale, e.g.~the gauge field has mass $m^2_V \sim (g_3 x)^2 \sim (\sqrt{g \pi} T)^2 $ in the $x$-phase and similarly in the $y$-phase.

This new scale hierarchy allows us to construct a soft scale EFT, where we integrate out the semisoft scale. The minima in the broken phase of these EFTs scale as $\frac{x^2}{T} \sim  \frac{\pi}{g} \gg 1$ (and likewise for the $y$-field) and describe an extremely strong transition. The critical temperature $T_{c,2}$ is determined from the condition that the free-energies of both phases are equal, where each is computed within a \textit{separate} soft-scale EFT, with fields at the semisoft scale integrated out. Note, that in the field space of two fields, there is a barrier at leading order that separates the phases.

\begin{figure}
\centering
\includegraphics[width=0.8\textwidth]{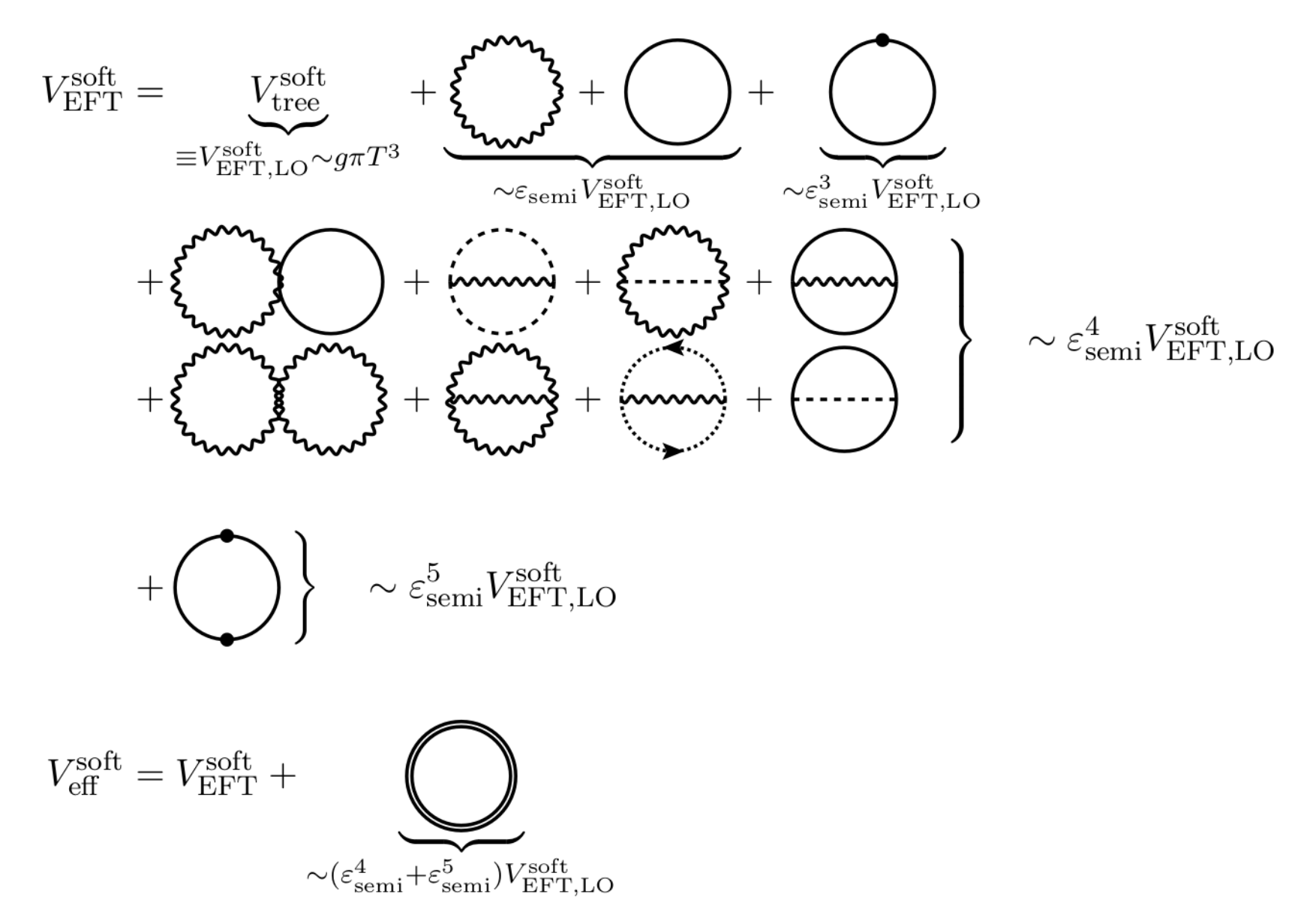}
\caption{
Schematic diagrammatic expansion of semisoft to soft scale matching and soft scale effective potential with $\veps_{\text{semi}} \sim \sqrt{g / \pi}$. Dashed line is soft scalar, solid line semisoft scalar. Soft mass insertions are denoted by two-point vertices depicted by a dot. Consequently, dashed propagator is massless, while solid line has semisoft mass $M^2 \sim a_{2,3} v^2$. In particular, note that $\mathcal{O}(\veps^2_{\text{semi}})$ with respect to LO does not appear.
}
\label{fig:semisoft-to-soft-veff}
\end{figure}

At leading order the size of the field-dependent effective potential can be read off from Eq.~\eqref{eq:V_soft_tree_counting} with $v^2 \sim \pi T / g$, leading to $V^{\text{soft}}_{\text{eff}} \sim g \pi T^3$. Counting powers of couplings, and ratios of scales, the perturbative expansion in each phase takes the form
\begin{align}
V^{\text{soft}}_{\text{eff}} &= g \pi T^3 \sum_{n=0}^{5} V_n \veps^n_{\text{semi}} + \mathcal{O}(\veps^6_{\text{semi}} V_0),
\end{align}
where $\veps_{\text{semi}} \sim \sqrt{g / \pi}$ and we have used the shorthand notation $V_0 \equiv V^{\text{soft}}_{\text{LO}}$ and truncated the series to fifth order, i.e.~N5LO. A diagrammatic rundown of these corrections is depicted in Fig.~\ref{fig:semisoft-to-soft-veff}. Computation beyond that requires soft mass insertions at two-loop order, which would provide the result at N6LO, and 3-loop semi-soft scale diagrams \cite{Rajantie:1996np} are required for N7LO.

The computation of the effective potential splits into two: first, matching from the semisoft to soft scale, and then soft scale contributions. In the matching, all loop momenta are formally semisoft \cite{Hirvonen:2022jba}, and one expands propagators in soft mass parameters, \textit{before} integration. For example, in the $y$-phase the one-loop $x$-field bubble diagram is expanded as%
\footnote{
We define the integral measure in the standard way for \MSbar\ regularisation, as $\int_p \equiv \Big( \frac{e^\gamma \mu^2_3}{4\pi} \Big)^{\epsilon} \int \frac{d^dp}{(2\pi)^d}$ in $d=3-2\epsilon$ dimensions, where $\gamma$ is the Euler-Macheroni constant.
}
\begin{align}
\label{eq:mass-insertion}
\frac{1}{2} \int_p \ln(p^2 + \underbrace{\mu^2_x}_{\sim g^2 T^2} + \underbrace{M^2(y)}_{\sim g \pi T^2} ) &= 
\frac{1}{2} \int_p \ln(p^2 + M^2(y)) 
+ \frac{1}{2} \mu^2_{x,3} \int_p \frac{1}{p^2 + M^2(y)} \nonumber \\
& - \frac{1}{4} \mu^4_{x,3} \int_p \frac{1}{(p^2 + M^2(y))^2} 
+ \mathcal{O}(\veps_\text{semi}^7 V_0)
\end{align}
where $M^2(y) \equiv \frac{1}{2} \lambda_{xy,3} y^2 $ is the semisoft contribution to the mass. Utilising such expansion in the soft mass, in Fig.~\ref{fig:semisoft-to-soft-veff} soft scalar propagators (dashed lines) are treated as massless, and solid lines have a semisoft-scale mass $M^2$. Soft mass insertions are depicted as two-point vertices (dots).

To compute the full N6LO piece would require performing similar expansions at two-loop level. For example, the pure scalar sunset integral is
\begin{align}
\label{eq:sunset}
S(m_x, m_x, m_y) \equiv \int_{p,k} \frac{1}{(p+m^2_x)(k^2+m^2_x)[(p+k)^2 + m^2_y]},
\end{align}
where $m^2_y$ is soft and $m^2_x = \mu^2_{x,3} + M^2(y)$ includes both soft and semisoft contributions. Again, for the matching one first expands the integrand in the soft-scale quantities $m^2_y$ and $\mu^2_{x,3}$, and then evaluates the integrals with semisoft momenta in the loops. For this, we can make use of the generic two-loop result given in Eq.~(C.81) of Ref.~\cite{Gorda:2018hvi} (c.f. also~\cite{Davydychev:2022dcw} and references therein) and define 
\begin{align}
S_{\alpha \beta \delta}(M) &\equiv \int_{p,k} \frac{1}{(p+M^2)^\alpha (k^2+M^2)^\beta [(\vec{p}+\vec{k})^2]^\delta} \nonumber \\
& =\Big(\frac{e^{\gamma}\Lambda_3^2}{4\pi}\Big)^{2\epsilon}\frac{(M^2)^{d-\alpha-\beta-\delta}}{(4\pi)^d} 
\frac{\Gamma\left(\frac d2-\delta\right)\Gamma\left(\alpha+\delta-\frac d2\right)\Gamma\left(\beta+\delta-\frac d2\right)\Gamma(\alpha+\beta+\delta-d)}{\Gamma\left(\frac d2\right)\Gamma(\alpha)\Gamma(\beta)\Gamma(\alpha+\beta+2\delta-d)}.
\end{align}
Expanding the integrand of Eq.~\eqref{eq:sunset} in soft masses $m^2_y$ and $\mu^2_{x,3}$, and expressing the result in terms of $S_{\alpha \beta \delta}$, we obtain
\begin{align}
\label{eq:sunset-expanded} 
S(m_y, m_x, m_x) &\approx
S(0, M, M) - m^2_y S_{112}(M) - \mu^2_{x,3} \Big( S_{211}(M) + S_{121}(M)  \Big)
 \nonumber \\
&= \frac{1}{(4\pi)^2} \bigg( \frac{1}{4\epsilon} + \frac{1}{2}  + \ln \Big( \frac{\Lambda_3}{M} \Big) \bigg) + \frac{m^2_y - 4 \mu^2_{x,3}}{(4\pi)^2 8 M^2} + \ldots.
\end{align}
Here the last term shown describes soft mass insertions, that contribute at N6LO. Other two-loop diagrams could be treated in an analogous manner, first expanding the integrand with respect to soft masses and only then computing the resulting integrals. However, a similar treatment with integrals involving gauge field propagators is somewhat more laborious, and we have decided to truncate our computation to N5LO in this work at hand, and leave higher orders for future. We give a more detailed description of the computation depicted in Fig.~\ref{fig:semisoft-to-soft-veff} in Appendix \ref{sec:semisoft-matching}, working through a concrete example for the SM augmented with a real triplet. 

Finally, in the computation of the one-loop correction to the effective potential in the soft scale EFT, the mass of the soft field needs to be resummed as
\begin{align}
\label{eq:soft-resummation}
\widetilde{m}^2_y &= \frac{d^2}{dy^2} V_\text{EFT}^\text{soft}
= \frac{d^2}{dy^2}\Big( V_0 + V_1 + \ldots \Big).
\end{align}
Resummation by the $V_1$ contribution (but not $V_2$ or higher) is needed at N5LO. The effect of matching the momentum-dependent field normalisation will contribute first at N7LO for observables depending only on homogeneous backgrounds fields.

\section{Strict perturbative expansions}
\label{sec:methods}

In the previous section, we have discussed the construction of EFTs to describe first-order phase transitions, and the calculation of their effective potentials. With this in hand, there are a number of different calculational approaches which one could adopt to analyse the phase structure and thermodynamics.

The equilibrium thermodynamics of a model can be derived from the pressure -- determined by the effective potential evaluated at its minima -- as a function of temperature, and its derivatives with respect to temperature. The pressure can be written as \cite{Tenkanen:2022tly}
\begin{align}
p(T) = p_{0}(T) - T \; V_{\text{eff}},
\end{align}
where $p_{0}$ is the coefficient of the unit operator in the construction of the 3d EFT \cite{Braaten:1995cm} and $V_{\text{eff}}$ is the effective potential within the 3d EFT \cite{Farakos:1994kx}. This coefficient of the unit operator is linked to the symmetric phase pressure as $ p_{\text{sym}}(T) = p_{0} - T \; V_{\text{eff}}(0)$, where the effective potential is evaluated at the origin \cite{Gynther:2005av, Tenkanen:2022tly}. The critical temperature $T_\text{c}$ is defined as the temperature where the pressure difference between two phases vanishes $\Delta p(T_\text{c}) = 0$ and this translates to a condition that the effective potentials at different minima are degenerate. Gauge invariant condensates were already discussed in Sec.~\ref{sec:intro}, and they can be computed as derivatives of $V_{\text{eff}}$ with respect to the parameters of the 3d EFT. The strength of the phase transition can be characterised in terms of released latent heat, which is related to the pressure as $L = T \Delta p'$, where prime denotes a derivative with respect to temperature. All these quantities depend on differences between phases, and hence we do not need to compute $p_{0}$. However, note that the speed of sound in each phase depends directly on $p_{0}$, and this is relevant for the determination of the gravitational wave power spectrum, see \cite{Giese:2020znk, Tenkanen:2022tly}. 

Next, we turn to different calculational approaches. The most direct approach would be to just numerically minimise the effective potential. There are however a number of issues with this approach. First, the effective potential is generically complex, with imaginary parts arising away from the minima of the leading-order potential where squared masses can become negative \cite{Weinberg:1987vp}. Typically this issue is simply ignored by working only with the real part, or by replacing squared masses by their absolute magnitude.%
\footnote{Note that these two options for avoiding imaginary parts differ numerically.}
Second, the effective potential is gauge dependent, and so are its minima, when computed directly.

An alternative approach is to adhere strictly to the confines of the perturbative expansion, and to perform a strict expansion in powers of $\veps_{\text{eff}}$. This approach is sometimes called the $\hbar$ expansion, though in general the expansion parameter need not have anything to do with Planck's constant. Rather than directly minimising the full effective potential, one first minimises the leading-order effective potential, and then includes the corrections from higher orders perturbatively. This approach has the benefit of being exactly gauge invariant order-by-order \cite{Nielsen:1975fs, Fukuda:1975di}. It is also manifestly real. The difference between this approach and direct minimisation is due to a subset of higher order terms in the expansion of the minima (the tadpole expansion), which are resummed by direct minimisation.

Further possibilities arise for quantities which require additional intermediate steps in their computation from the effective potential, such as the critical temperature. In these cases, one can choose to make an additional strict expansion, or to mix the strict expansion of the potential with a direct approach at solving $\Delta p (T) = 0$ for the critical temperature. Unlike for the minima of the effective potential, in this case both possibilities are real and gauge invariant. For the critical temperature, there is however an important difference between these two approaches: If the critical temperature at some higher order is not within the range in which there is metastability at leading order, then the direct approach fails, while the strict perturbative approach continues to work. This issue is discussed further below.

\begin{table}[t]
\centering
\begin{tabular}{|c|c|c|}
\hline
Method & Gauge invariant & Real \\
\hline
direct & \xmark & \xmark \\
\hline 
mixed & \cmark & for sufficiently small $\veps_{\text{eff}}$ \\ 
\hline
strict& \cmark & \cmark \\
\hline
\end{tabular}
\caption{
Basic theoretical properties of different perturbative methods described in the text. The column headings here refer to the properties of real physical quantities computed using these methods, such as the free energy or critical temperature. Note that, while one can always get a real result from a complex quantity simply by discarding the imaginary part by hand, we do not consider such a result to be a genuinely real prediction of a given method. As Goldstone squared masses go through zero at the LO broken phase, the direct method can yield spurious imaginary parts even when perturbative corrections are arbitrarily small. The mixed method yields real physical results when the expansion parameter $\veps_{\text{eff}}$ is sufficiently small, but as we argue around Eq.~\eqref{eq:DeltaT} below, it can yield spurious imaginary parts when higher-order corrections exceed some finite bound.
}
\label{tab:pros_cons}
\end{table} 

Table \ref{tab:pros_cons} summarises the theoretical properties of the different perturbative expansion schemes we have considered. In Section \ref{sec:results}, we will further describe and test all these different approaches for a numerical example, comparing them to lattice Monte-Carlo data. For the remainder of this section, we formulate strict expansions for the effective potentials of the previous sections, as well as expansions for thermodynamic quantities of interest. In addition, we describe mixed approaches that combine direct and strict methods. 

\paragraph{Strict expansions for a soft field}
\label{sec:soft-EFT-expansion} 

In the case of a soft-scale field undergoing a phase transition, the effective potential has an expansion in the effective expansion parameter $\veps_{\text{soft}}$,  
\begin{align}
\label{eq:Veff-soft}
V_{\text{eff}}(v) &= V_0(v) + \veps_{\text{soft}} V_1(v)  + \veps^2_{\text{soft}}  V_2(v)  + \mathcal{O}(\veps^3_{\text{soft}} V_0(v)).
\end{align} 
The minima of the potential can also be expanded as
\begin{align}
\label{eq:minimum-expansion-soft}
v^{\text{min}} = v_{0} + \veps_{\text{soft}} v_{1} + \veps^2_{\text{soft}} v_{2} + \mathcal{O}(\veps^3_{\text{soft}} v_0) ,
\end{align} 
where the coefficients $v_{i}$ are determined by solving for the minima of the potential as an expansion in $\veps_{\text{soft}}$. For a single field, this results in \cite{Fukuda:1975di}
\begin{align}
\label{eq:Veff-strict}
V_{\text{eff}}(v^{\text{min}}) &=
V_0
+ \veps_{\text{soft}} V_1
+ \veps^2_{\text{soft}} \left\{ V_2 - \tfrac{1}{2} v_1 \cdot \partial_v^2 V_0 \cdot v_1 \right\}
+ \mathcal{O}(\veps^3_{\text{soft}} V_0)
, \\
v_1 &= - \frac{1}{\partial_v^2 V_0} \cdot \partial_v V_1.
\label{eq:v-strict}
\end{align}
where we have introduced $\partial_v \equiv \frac{\partial}{\partial v}$, and all terms on the right-hand side of the equalities are evaluated at $v_0$. This last point is crucial for the desirable properties of this expansion, such as order-by-order reality and gauge invariance. In analogy, we can write the pressure as
\begin{align}
p = - T V_{\text{eff}}(v^{\text{min}}) = p_0 + \veps_{\text{soft}} p_1 + \veps^2_{\text{soft}} p_2 + \mathcal{O}(\veps^3_{\text{soft}} T^4),
\end{align} 
where the coefficients incorporate corrections from the expansion of minima $p_0 = -T V_0$, $p_1 = -T V_1$ and $p_2 =  -T (V_2 + \tfrac{1}{2} v_1 \cdot \partial_v^2 V_0 \cdot v_1) $.

The generalisation to multiple scalar fields follows by upgrading the multiplications in Eqs.~\eqref{eq:Veff-strict} and \eqref{eq:v-strict} to matrix multiplications. The background field becomes a vector $v^a$ with index $a$, and $\partial_a \equiv \frac{\partial}{\partial_{v^a}}$ is the corresponding gradient operator. The second derivative of the potential then becomes a matrix, and we find
\begin{align}
\label{eq:Veff-strict-nfields}
V_{\text{eff}}(v^{\text{min}}) &= V^{}_{0}
+ \veps_{\text{soft}} V^{}_{1}
+ \veps^2_{\text{soft}} \left\{ V^{}_{2} - \tfrac{1}{2} v_1^a\cdot \partial_{a}\partial_{b} V_0 \cdot v_1^b  \right\} + \mathcal{O}(\veps^3_{\text{soft}} V^{}_{0}), \\
v_1^a &= - (\partial_{a}\partial_{b} V_0)^{-1} \cdot \partial_b V_1,
\label{eq:v-strict-nfields}
\end{align}
where $(\partial_{a}\partial_{b} V_0)^{-1}$ is the matrix inverse of the Hessian matrix $\partial_{a}\partial_{b} V_0$, and we have used Einstein summation convention. As above, all terms on the right-hand sides of these equations are evaluated at the LO minimum $v^a_0$.

For a 2-field model, with $v=(x, y)$, one can invert the Hessian matrix explicitly, resulting in \cite{Niemi:2020hto}
\begin{align}
\label{eq:Veff-strict-2fields}
V_{\text{eff}}(v^{\text{min}}) =&
V^{}_{0}
+ \veps_{\text{soft}} V^{}_{1}
+ \veps^2_{\text{soft}} \bigg\{ V^{}_{2} \nonumber  \\ 
& + \frac{1}{2} \bigg( \Big( \frac{\partial^2 V^{}_{0}}{\partial x \partial y} \Big)^2 - \Big( \frac{\partial^2 V^{}_{0}}{\partial x^2} \Big) \Big( \frac{\partial^2 V^{}_{0}}{\partial y^2} \Big) \bigg)^{-1} \nonumber \\
& \times \bigg( \Big( \frac{\partial V^{}_{1}}{\partial x} \Big)^2 \Big( \frac{\partial^2 V^{}_{0}}{\partial y^2} \Big)
+ \Big( \frac{\partial V^{}_{1}}{\partial y} \Big)^2 \Big( \frac{\partial^2 V^{}_{0}}{\partial x} \Big) \nonumber \\
& - 2 \Big( \frac{\partial V^{}_{1}}{\partial x} \Big)\Big( \frac{\partial V^{}_{1}}{\partial y} \Big) \Big( \frac{\partial^2 V^{}_{0}}{\partial x \partial y} \Big)   \bigg) \bigg\} + \mathcal{O}(\veps^3_{\text{soft}} V^{}_{0}), 
\end{align}
where again the right-hand side is evaluated at the LO minima $v^a_0 = (x_0, y_0)$. 

The LO minima $v_0$ describe different phases of the system, and the most likely phase in thermal equilibrium corresponds to the global minimum.  Later on, we refer to Eqs.~\eqref{eq:Veff-strict}, \eqref{eq:Veff-strict-nfields} and \eqref{eq:Veff-strict-2fields} as the \textit{soft EFT expansion} for the effective potential. Technically, this is a \textit{strict} expansion of the effective potential around its LO minima, in a 3d EFT at the soft scale. Notably, this expansion is gauge invariant order by order \cite{Fukuda:1975di, Laine:1994zq}, since it satisfies the Nielsen-Fukuda-Kugo identities within the EFT \cite{Nielsen:1975fs, Fukuda:1975di}, and the construction of the EFT through dimensional reduction is gauge invariant \cite{Croon:2020cgk, Hirvonen:2021zej}.

The strict expansion strategy can be extended to determining the critical temperature \cite{Laine:1994zq}, by writing
\begin{align}
T_\text{c} = T_0 + \veps_{\text{soft}} T_1 + \veps^2_{\text{soft}} T_2 + \mathcal{O}(\veps^3_{\text{soft}} T_0), 
\end{align}
where $T_0$ is solved from
\begin{align}
\Delta p_0(T_0) = 0,
\end{align} 
or equivalently $-\Delta V_0(T_0) = 0$.
The next two orders in the expansion are%
\footnote{
We remind the reader that prime is used to denote the temperature derivative, $V_0' \equiv \frac{d V_0}{dT}$. 
}
\begin{align}
\label{eq:T1-soft}
T_1 &= - \Delta p_1(T_0)/\Delta p_0'(T_0) , \\
\label{eq:T2-soft}
T_2 &= \Big( -\Delta p_2(T_0) - T_1 \Delta p_1'(T_0) - \frac{1}{2} T_1^2 \Delta p''_0(T_0) \Big) /\Delta p_0'(T_0). 
\end{align} 
Similarly, expanding the latent heat 
\begin{align}
L = L_0 + \veps_{\text{soft}} L_1 + \veps^2_{\text{soft}} L_2 + \mathcal{O}(\veps^3_{\text{soft}} T^4),
\end{align} 
where 
\begin{align}
L_0 &= T_0 \Delta p_0'(T_0), \\
\label{eq:L1-soft}
L_1 &= T_1 \Delta p_0'(T_0) +  T_0 \Big( \Delta p_1'(T_0) + T_1 \Delta p_0''(T_0) \Big) , \\
\label{eq:L2-soft}
L_2 &= T_2 \Delta p_0'(T_0) + T_0 \Delta p_2'(T_0) + T_1 \Delta p_1'(T_0) \nonumber \\
& + \Big( T^2_1 + T_0 T_2 \Big) \Delta p_0''(T_0) +  T_0 T_1 \Big( \Delta p_1''(T_0) + \frac{1}{2}T_1 \Delta p_0'''(T_0)  \Big). 
\end{align}
We emphasize that in consistent EFT expansions we do not encounter the problem reported in \cite{Laine:1994zq}, whereby the condition $\Delta p_0(T_0) = 0$ leads to a vanishing mass parameter $\mu^2_3(T_0) = 0$ which then results in spurious IR divergences at two-loop order for $T_2$. This problem has its roots in the fact that the leading order potential used in \cite{Laine:1994zq} is the one at tree-level at the soft scale, and does not have a barrier, but describes instead a second order phase transition. In the approach we have advocated, the assumption of a barrier in the LO potential is build-in to the EFT construction. In the case of an apparently second-order transition at the soft scale, the correct description for the transition is the supersoft scale EFT, cf.\ Secs.~\ref{sec:supersoft-EFT} and the discussion below in this section. Furthermore, EFT expansions are manifestly real: since all expressions are evaluated at LO minima, all squared mass eigenvalues are non-negative and hence no imaginary parts arise from the computation of higher order corrections.

The combination of the soft EFT expansion for the effective potential and strict expansions for $T_\text{c}$ and $L$ is the strict method of Table~\ref{tab:pros_cons}. This method is computationally very efficient on an algorithm level, since once $T_0$ is solved numerically from the condition $\Delta p_0(T_0) = 0$, all corrections are simply evaluated at $T_0$ from expressions that are known analytically.%
\footnote{
Of course, expressions such as Eq.~\eqref{eq:T2-soft} for $T_2$ can have very long expressions in practice, but nevertheless this can be handled analytically by symbolic calculation tools. This is much more efficient than numerical minimisation of complicated potentials.
}  

Expansions in $\veps_{\text{soft}}$ for the condensates follow naturally from the EFT expansion for the effective potential. Concretely, for the scalar quadratic condensate 
\begin{align}
\langle \phi^\dagger \phi \rangle &= \langle \phi^\dagger \phi \rangle_{0} + \veps_{\text{soft}} \langle \phi^\dagger \phi \rangle_{1} + \veps^2_{\text{soft}} \langle \phi^\dagger \phi \rangle_{2} + \mathcal{O}(\veps^3_{\text{soft}} \langle \phi^\dagger \phi \rangle_{0}),
\end{align}
where $\langle \phi^\dagger \phi \rangle_{i} = \frac{\partial V_i(v_0)}{\partial \mu^2_3}$ and where for simplicity we assumed that the scalar is a Higgs doublet. Note that since $v_0$ often depends on the 3d mass parameter $\mu^2_3$, one first evaluates the effective potential in each phase, and only then differentiates.

Alternatively, following a strategy of \cite{Patel:2011th, Niemi:2020hto, Schicho:2022wty}, one could determine the critical temperature by computing $\Delta p(T)$ using the soft EFT expansion, and then numerically solving for its root $\Delta p(T_\text{c}) = 0$. Such a \textit{direct} determination of $T_\text{c}$ is indeed computationally efficient, as it does not require numerical minimisation of a complicated effective potential. This is the mixed method of Table~\ref{tab:pros_cons}: the effective potential is evaluated in a strict expansion around the leading order minima, while thermodynamic quantities are determined directly as functions of temperature. Indeed, the same method can be applied to the determination of the latent heat~\cite{Niemi:2020hto, Schicho:2022wty}.

With the mixed method, or more generally if we consider evaluating quantities at temperatures other than $T_0$, we come across the following problem: (here we denote the expansion parameter again by $\veps_{\text{eff}}$ for generality)
\begin{itemize}
\item[] 
The LO minima $v_0$ exist in some range of temperatures $T \in [T_0^L, T_0^R]$. The LO critical temperature lies in this range $T_0 \in [T_0^L, T_0^R]$. In a power counting sense the width of the range is $T_0^R - T_0^L = O(T_0)$, so for $\veps_{\text{eff}} \to 0$ the full critical temperature $T_\text{c}=T_0 + \veps_{\text{eff}} T_1 + \veps^2_{\text{eff}} T_2 + ...$ must also lie in this range. However, for a finite expansion parameter $\veps_{\text{eff}}$ it is possible that $T_\text{c}=T_0 + \veps_{\text{eff}} T_1 + \veps^2_{\text{eff}} T_2 + ...$ is outside the range. This leads to the problem that for any temperature dependent function $F(T_\text{c})$ an expansion $F(T_\text{c}) = F_0(T_\text{c}) + \veps_{\text{eff}} F_1(T_\text{c}) + ...$ does not exist. Furthermore, even if the LO minima exist at $T_\text{c}$, in the mixed method the range of existence of a given phase is fixed at LO, and does not change  at higher orders.
\end{itemize}

The proposed solution is to consider physical quantities as functions of
\begin{align} \label{eq:DeltaT}
\Delta T = T - T_\text{c},
\end{align}
to treat the difference as of leading order $\Delta T = O(T_0)$, and then to power expand everything in $\veps_{\text{eff}}$. The origin of the independent variable $\Delta T$ is fixed to the critical temperature order-by-order, so that the expansion cannot cause $T-T_\text{c}$ to change sign, or to grow too large in magnitude. This helps to extend the principles of the strict method to temperatures other than the critical temperature. Note that for a fixed $T$, the corresponding value of $\Delta T$ depends on the order to which we compute $T_\text{c}$. It is the difference from the critical temperature in a given approximation.

Now consider the expansion of some generic quantity $F$:
\begin{align}
F(T) &= F(\Delta T + T_\text{c}), \\
&= F_0(\Delta T + T_0) + \veps_{\text{eff}} (F_1(\Delta T + T_0) + F'_0(\Delta T + T_0) T_1) + ...
\label{eq:F_Delta_T}
\end{align}
Evaluating this at $\Delta T = 0$ gives
\begin{align}
F(T_\text{c}) &= F_0(T_0) + \veps_{\text{eff}} (F_1(T_0) + F'_0(T_0) T_1) + ... ,
\end{align}
which reproduces the strict expansion at $T_\text{c}$. Note also that everything on the RHS is evaluated at $T_0$ and hence within the range $[T_0^L, T_0^R]$, so it always exists. Next consider the range of existence of the phases beyond LO. The RHS of Eq.~\eqref{eq:F_Delta_T} exists for $\Delta T + T_0 \in [T_0^L, T_0^R]$ and hence for
\begin{align}
T \in T_\text{c} + [T_0^L - T_0, T_0^R - T_0]
\end{align}
or, equivalently,
\begin{align}
T \in [T_0^L, T_0^R] + \veps_{\text{eff}} T_1 + \veps^2_{\text{eff}} T_2 + ...
\end{align}
So, the range of existence of phases is shifted at each order, by the amount that $T_\text{c}$ changes at that order. This essentially solves the problem of the static range. Note however that the width of the range does not change from order to order; a feature which still seems undesirable.

\paragraph{Strict expansions for a supersoft field}
\label{sec:supersoft-EFT-expansion} 

In the case of a supersoft field, EFT expansions follow essentially the same logic. However, due to the different structure of the effective expansion, the resulting expressions are slightly simpler. The effective potential consists of two expansions 
\begin{align}
V^{\text{supersoft}}_{\text{eff}} \simeq V^{\text{supersoft}}_{\text{EFT,LO}} \Big( 1 + \veps_{\text{soft}} + \mathcal{O}(\veps^2_{\text{soft}}) \Big)
+ \veps_{\text{super}} V^{\text{supersoft}}_{\text{EFT,LO}} + \ldots 
\end{align}
Since $\veps_{\text{super}} \sim \veps^{\frac{3}{2}}_{\text{soft}}$, following Ref.~\cite{Ekstedt:2022zro} we can write formally 
\begin{align}
\label{eq:Veff-supersoft}
V^{\text{supersoft}}_{\text{eff}} &= V_0 + \veps V_1 + \veps^2 V_2 + \veps^3 V_3 + \mathcal{O}(\veps^4 V_0),
\end{align}
where $\veps \sim \sqrt{\veps_{\text{soft}}}$ and $V_1 = 0$ identically. In this expansion, the soft and supersoft expansions are mixed together. In principle one could do everything in a fully EFT way, essentially resumming $V^{\text{supersoft}}_{\text{eff}} = \veps^0_{\text{super}} (V_0 + V_2) + \veps_{\text{super}} V_3$, i.e.~both $V_0$ and $V_2$ are treated as LO within the supersoft EFT. This is in analogy to not mixing the hard and soft expansions in dimensional reduction. However, here we choose the former option and mix the expansions together, so the supersoft scale EFT should be understood in the sense of this mixed expansion. We note, that these two alternatives agree up to the order computed \cite{Lofgren:2023sep}, yet resum different sets of formally higher order corrections.  

We emphasize that the order $\veps^1$ is not present in the effective potential, and this leads to multiple simplifications in formulae below. We formally expand the minima as
\begin{align}
\label{eq:minimum-expansion}
v^{\text{min}} = v_{0} + \veps v_{1} + \veps^2 v_{2} + \veps^3 v_{3} + \mathcal{O}(\veps^4 v_0) ,
\end{align} 
where $v_{1} = 0$ since $V_1 = 0$. The expansion for the potential evaluated at the minimum reads
\begin{align}
\label{eq:supersoft-EFT-expansion}
V^{\text{supersoft}}_{\text{eff}}(v^{\text{min}}) &= V_0(v_{0}) + \veps^2 V_2(v_{0}) + \veps^3  V_3(v_{0}) + \mathcal{O}(\veps^4 V_0(v_{0})).
\end{align} 
This expression is particularly simple up to and including $\mathcal{O}(\veps^3)$, since subleading corrections to the minimum start at $v_2 = \mathcal{O}(\veps^2 v_0)$, and the condition $\partial_v V_0 = 0$ at $v_{0}$ ensures that this does not contribute to the potential until $\mathcal{O}(\veps^4 V_0(v_0))$.

Later on, we refer to Eq.~\eqref{eq:supersoft-EFT-expansion} as the \textit{supersoft EFT expansion} for the effective potential. Technically, this is a \textit{strict} expansion of the effective potential around its LO minima, in a 3d EFT at the supersoft scale. In analogy to case of the soft field in the previous section, this expansion is gauge invariant \cite{Lofgren:2021ogg, Hirvonen:2021zej, Ekstedt:2022zro}, and Eq.~\eqref{eq:supersoft-EFT-expansion} can be utilised in the mixed or strict methods of Table~\ref{tab:pros_cons}. 

For the strict method, we expand
\begin{align}
T_\text{c} = T_0 + \veps T_1 + \veps^2 T_2 + \veps^3 T_3 + \mathcal{O}(\veps^4 T_0).
\end{align}
In analogy to the expansion of the minimum in Eq.~\eqref{eq:minimum-expansion}, here $T_1=0$. The leading order $T_0$ is solved from $\Delta p_0(T_0) = 0$ or $-\Delta V_0(T_0) = 0$. Higher order corrections are then obtained iteratively
\begin{align}
T_1 &= 0, \\ 
T_2 &= - \Delta V_2(T_0)/\Delta V_0'(T_0) , \\
T_3 &= -\Delta V_3(T_0)/\Delta V_0'(T_0). 
\end{align}
Again, we emphasize, that these EFT expansions are free from spurious IR divergencies reported in \cite{Laine:1994zq, Niemi:2020hto}. In the EFT expansion, a radiative barrier provided by the soft fields is included to the LO effective potential, and the condition for $T_0$ does not lead to a vanishing mass parameter. Hence, the spurious singularities at higher orders are avoided, and concretely $T_2$ and $T_3$ are finite.

Similarly, the latent heat has an expansion
\begin{align}
L = L_0 + \veps L_1 + \veps^2 L_2 + \veps^3 L_3 + \mathcal{O}(\veps^4 L_0).
\end{align}  
By writing the pressure as
\begin{align}
p = p_0 + \veps p_1 + \veps^2 p_2 + \veps^3 p_3 + \mathcal{O}(\veps^4 p_0),
\end{align}   
where 
$p_0 \equiv -T V_0(v_{0})$,
$p_1 = 0$,
$p_2 \equiv -T V_2(v_{0})$
and
$p_3 \equiv -T V_3(v_{0})$, 
we obtain 
\begin{align}
L_0 &= T_0 \Delta p_0'(T_0), \\
L_1 &= 0, \\
L_2 &= T_2 \Delta p_0'(T_0) + T_0 \Big( \Delta p_2'(T_0) + T_2 \Delta p_0''(T_0) \Big) , \\
L_3 &= T_3 \Delta p_0'(T_0) + T_0 \Big( \Delta p_3'(T_0) + T_3 \Delta p_0''(T_0) \Big) . 
\end{align}

Finally, the quadratic condensate has an expansion
\begin{align}
\langle \phi^\dagger \phi \rangle &= \langle \phi^\dagger \phi \rangle_{0} + \veps^2 \langle \phi^\dagger \phi \rangle_{2} + \veps^3 \langle \phi^\dagger \phi \rangle_{3} + \mathcal{O}(\veps^4 \langle \phi^\dagger \phi \rangle_{0}),
\end{align}
where
$\langle \phi^\dagger \phi \rangle_{n} = \frac{\partial V_n(v_0)}{\partial \mu^2_3}$.

\paragraph{Strict expansions for a soft field, induced by semisoft scale}
\label{sec:strict-expansion-semisoft} 

At this point, after the previous discussions, the methodology of strict expansions should be clear. However, for the sake of completeness we repeat the corresponding discussion here. The only difference is the form of the expansion of the potential, which in the case of the semisoft-induced soft-scale EFT reads    
\begin{align}
\label{eq:Veff-soft-semisoft}
V^{\text{soft}}_{\text{eff}} &= V^{}_{0} + \veps_{\text{semi}} V^{}_{1}  + \veps^3_{\text{semi}}  V^{}_{3} + \veps^4_{\text{semi}}  V^{}_{4} + \veps^5_{\text{semi}}  V^{}_{5}   + \mathcal{O}(\veps^6_{\text{semi}} V^{}_{0} ).
\end{align}
Compared to the previous sections, we have many more orders available, and this time it is the N2LO term that is missing (i.e.~$V_2 = 0$) due to the nature of the matching between semisoft and soft scales. From this expansion all else follows by Taylor expansion. The minima, pressure, latent heat, field condensates and other thermodynamic quantities can all be expanded as
\begin{align}
F = F_0 + \veps_{\text{semi}} F_1 + \veps_{\text{semi}}^2 F_2 + \veps_{\text{semi}}^3 F_3 + \veps_{\text{semi}}^4 F_4 + \veps_{\text{semi}}^5 F_5 + \mathcal{O}(\veps^6_{\text{semi}} F_0).
\end{align}
First, one solves $\partial_v V=0$ for the minima, with each successive order determined by a linear algebraic equation, $a v_n + b = 0$ where $a$ and $b$ are given in terms of lower orders. From this one can construct the pressure, the first few orders of which are
\begin{align}
p_{0} &= -T \left( V_0 \right), \qquad
p_{1} = -T \left( V_1 \right) , \qquad
p_{2} = -T \left( -\frac{1}{2} \frac{(\partial_vV_1)^2}{\partial_v^2V_0} \right) , \\
p_{3} &= -T \left( V_3 - \frac{1}{6} \frac{(\partial_v V_1)^2}{(\partial_v^2V_0)^3} \Big( -3 \partial_v^2 V_0 \partial_v^2 V_1 + \partial_v V_1 \partial_v^3 V_0 \Big)   \right)  , 
\end{align} 
where as always the expressions on the right hand sides are evaluated at $v_0$ and higher orders can be generated iteratively. Notably, unlike in the supersoft EFT case, where $p_1$ was zero due to $V_1$ being zero, in this case all orders for the pressure are nonzero despite $V_2$ being zero. 
Hence, solving $\Delta p(T_\text{c}) = 0$, 
and computing $L = T_c \Delta p'(T_\text{c})$, the order-by-order results reproduce the same expressions at the first few orders that were already encountered in Eqs.~\eqref{eq:T1-soft} and \eqref{eq:T2-soft}, as well as Eqs.~\eqref{eq:L1-soft} and \eqref{eq:L2-soft}.
Finally, the expansion coefficients for quadratic scalar condensates follow from that of the pressure by their defining relations, $\langle \phi^\dagger \phi \rangle_{n} = \frac{\partial}{\partial \mu^2_3} \left( -T^{-1} p_n \right)$; see Eqs.~\eqref{eq:cond-phi} and \eqref{eq:cond-sigma}.

This completes the formal outline of our setup, and next we turn to applications.

\section{Cosmological phase transitions}
\label{sec:results}

In this section, we will test the EFTs and perturbative expansions presented above, as applied to possible cosmological thermal histories. As introduced in Sec.\ref{sec:intro} we use the real-triplet extended SM as our concrete playground, in order to compare our perturbative EFT methods to the lattice results of Ref.~\cite{Niemi:2020hto}. We will also use the renormalisation scale dependence of our perturbative results to provide an intrinsic measure of their uncertainty. However, for the thermodynamics of this model, the lattice results are expected to be correct up to very small statistical uncertainties, so they are the ultimate arbiter.

We would like to comment on a subtlety in comparing to Ref.~\cite{Niemi:2020hto}. The lattice simulations of \cite{Niemi:2020hto} were performed for a 3d EFT without the temporal components of gauge fields, their effects being captured by the parameters of the EFT. Such EFTs are commonly referred as ``ultrasoft'' scale EFTs, and have been studied using non-perturbative lattice simulations e.g.~in \cite{Farakos:1994xh, Kajantie:1995kf, Kajantie:1997hn, Laine:2000rm, Kainulainen:2019kyp}. However, the derivation of such theories does not require scalar masses to be ultrasoft. Indeed, the only necessary assumptions therein are that (i) $m_3^2\ll m_D^2$, i.e.~scalar masses are much lighter than Debye masses, and (ii) $h_3 v^2 \ll m_D^2$, where $h_3$ is a generic portal coupling between scalars and temporal gauge field components, and $v$ is the Higgs background field. Based on the power counting arguments of the previous sections, the Higgs and triplet scalar fields are not expected to become ultrasoft, except in the near vicinity of a second-order phase transition. For a reasonably strong first-order phase transition, they are either of the soft or supersoft scales. As a consequence, the temporal components of the gauge fields should be treated as described in Sec.~\ref{sec:setup} in the construction of the EFT for the transitioning fields. In particular, assumption (ii) can break down when the background field $v^2$ becomes large. In this case, one should not integrate out temporal gauge field components as described in \cite{Kajantie:1995dw}.

Nevertheless, in order to provide an apples-to-apples comparison with the lattice results of \cite{Niemi:2020hto}, in all the numerical computations of this work, we incorporate contributions from the temporal components of gauge fields as in \cite{Niemi:2018asa, Niemi:2020hto}. Furthermore, in the lattice simulations of \cite{Niemi:2020hto}, dynamical effects of the U(1) subgroup in the 3d EFT were not included, hence we set $g_3'=0$ in perturbation theory as well, but note that we still keep effects of $g'$ in the dimensional reduction matching relations, as in \cite{Niemi:2020hto}.

The effective potential for the soft-scale EFT of the real-triplet extended SM can be found in \cite{Niemi:2020hto}, up to N2LO, or two-loops. Explicit results for the corresponding supersoft-scale effective potential are given in Appendix~\ref{sec:supersoft-matching}, and results for soft scale effective potentials including effects from the semisoft scale are collected in Appendix~\ref{sec:semisoft-matching}.

\subsection{One-step symmetry-breaking transition}

The two benchmark points studied non-perturbatively in Ref.~\cite{Niemi:2020hto} both showed a succession of two phase transitions, with the pattern: symmetric to triplet to Higgs phase.%
\footnote{
Here we use the phase \textit{phase transition} somewhat loosely. Depending on the mass and couplings of the triplet scalar, the symmetry-breaking transition may be a smooth crossover, like the liquid-gas transition of water above a pressure of 22 MPa.
}
For now, we will focus on the transition from the symmetric phase to the triplet phase.

As with other symmetry-breaking transitions, the symmetric to triplet transition appears to be of second order at tree level in the soft-scale EFT. Following the arguments of Sec.~\ref{sec:setup}, this implies that in fact the transition takes place at lower energies, so the soft scale should be integrated out. Barring further cancellations, we thus expect the transition to take place at the supersoft scale. However, for completeness, and for comparison to the previous literature, we also consider the possibility that the transition takes place at the soft scale (though this will lead to IR divergences). For each EFT we consider each of the perturbative methods introduced in the previous section. The matrix of possibilities are summarised in Table \ref{tab:methods1}.
\begin{table}[t]
\centering
\begin{tabular}{|c|c|c|c|}
\hline
inducing$\to$transitioning & direct & mixed & strict \\
\hline
hard$\to$soft & e.g.~\cite{Niemi:2021qvp} & \cite{Croon:2020cgk, Gould:2021oba, Schicho:2022wty} & \cite{Laine:1994zq}  \\
\hline 
soft$\to$supersoft & -- & \cite{Schicho:2022wty} & \cite{Gould:2022ran} \\ 
\hline 
\end{tabular}
\caption{
Table of perturbative approaches to the study of a symmetry-breaking phase transition. The column labels, direct, mixed and strict, refer to different approaches to carrying out the perturbative computation. The row labels, hard$\to$soft and soft$\to$supersoft refer to different EFTs. Here \textit{inducing} refers to the lowest energy scale of fields which are integrated out, and which induce the temperature-dependent barrier between phases; \textit{transitioning} refers to the energy scale of the transitioning fields. Each element in the table lists references where these approached have been used in the literature.
}
\label{tab:methods1}
\end{table}

\begin{figure}[t]
\begin{subfigure}{0.5\textwidth}
    \centering
    \includegraphics[width=\textwidth]{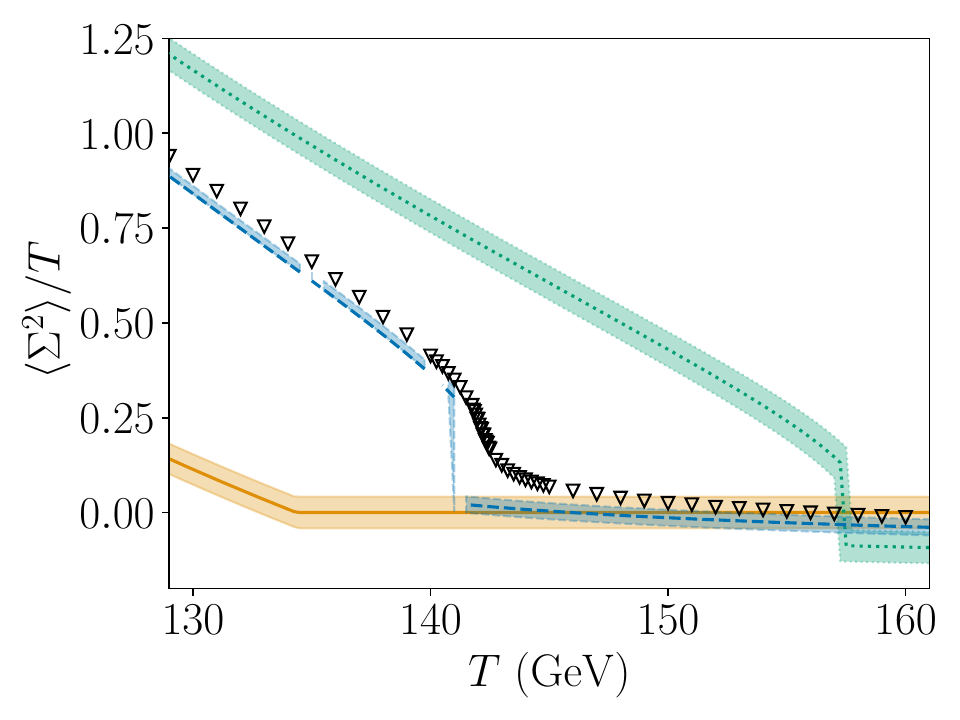}
    \caption{Hard$\to$soft EFT, direct method}
    \label{fig:BM1_onestep_a}
\end{subfigure}
\begin{subfigure}{0.5\textwidth}
    \centering
    \includegraphics[width=\textwidth]{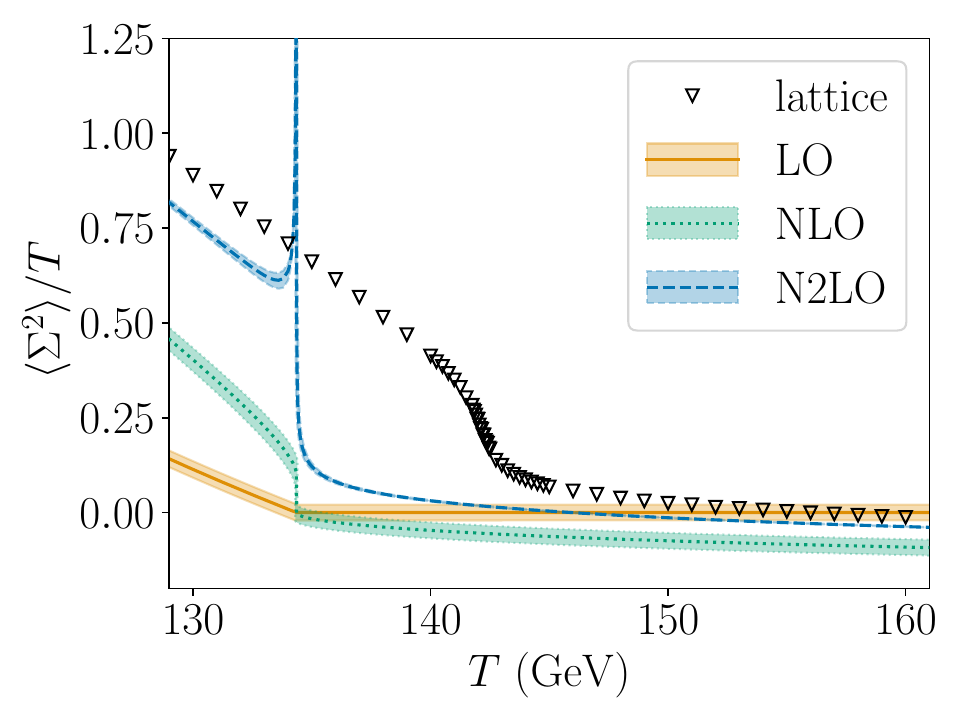} 
    \caption{Hard$\to$soft EFT, mixed method}
    \label{fig:BM1_onestep_b}
\end{subfigure} 
\begin{subfigure}{0.5\textwidth}
    \centering
    \includegraphics[width=\textwidth]{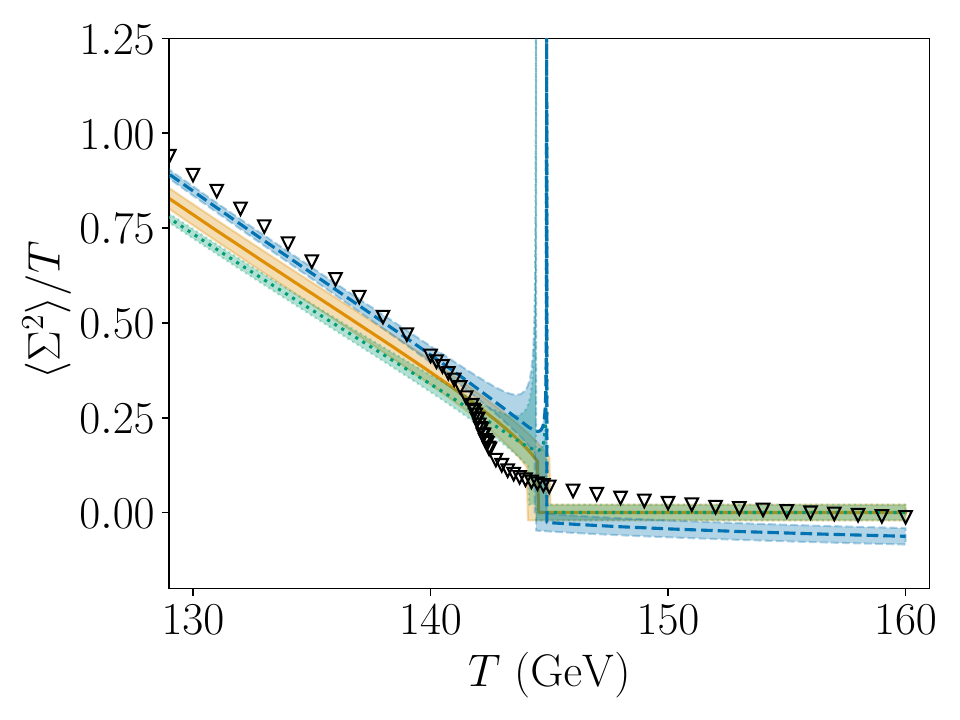}
    \caption{Soft$\to$supersoft EFT, mixed method}
    \label{fig:BM1_onestep_c}
\end{subfigure}
\begin{subfigure}{0.5\textwidth}
    \centering
    \includegraphics[width=\textwidth]{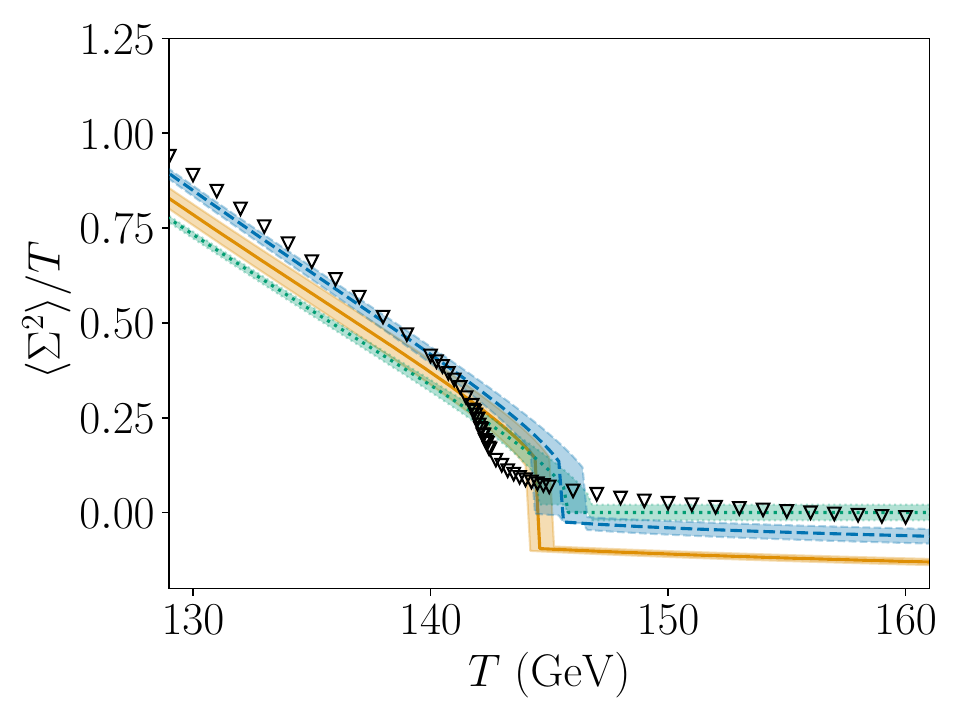}
    \caption{Soft$\to$supersoft EFT, strict method}
    \label{fig:BM1_onestep_d}
\end{subfigure}
\caption{
The triplet scalar condensate as a function of temperature, in different perturbative approaches for the triplet transition in BM1, in the presence of a soft Higgs field that is not dynamical for this transition. Bands depict variation due to RG scale, as explained in the main text. Note that in panel (a) there are a number of data points missing in the N2LO result, due to numerical difficulties.
}
\label{fig:BM1-1step-methods}
\end{figure}

\begin{figure}[t]
\begin{subfigure}{0.5\textwidth}
    \centering
    \includegraphics[width=\textwidth]{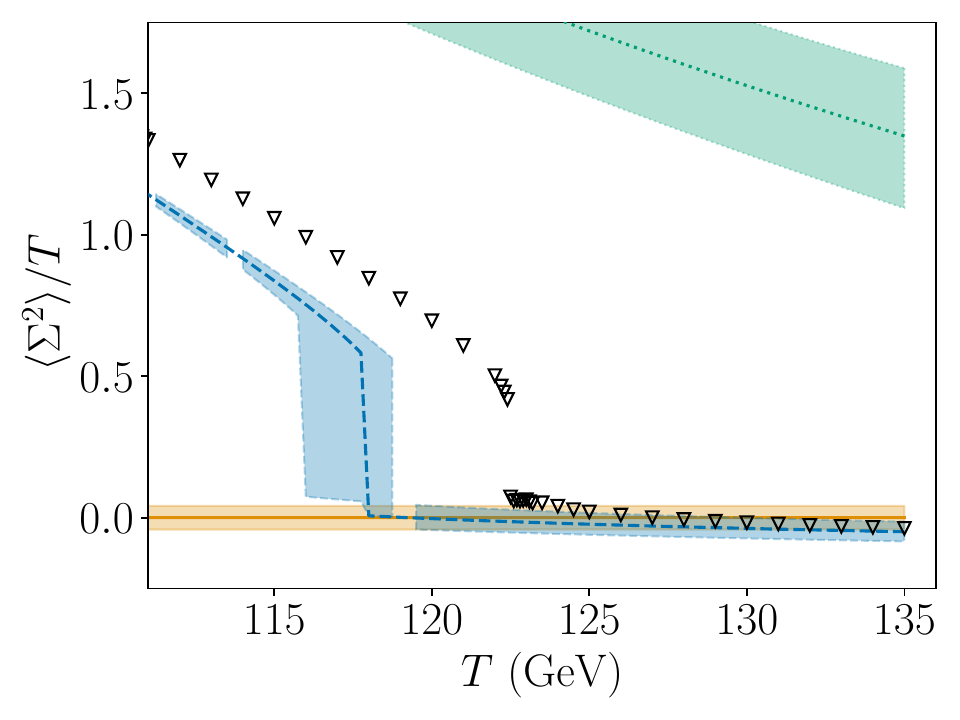}
    \caption{Hard$\to$soft EFT, direct method}
    \label{fig:BM2_onestep_a}
\end{subfigure}
\begin{subfigure}{0.5\textwidth}
    \centering
    \includegraphics[width=\textwidth]{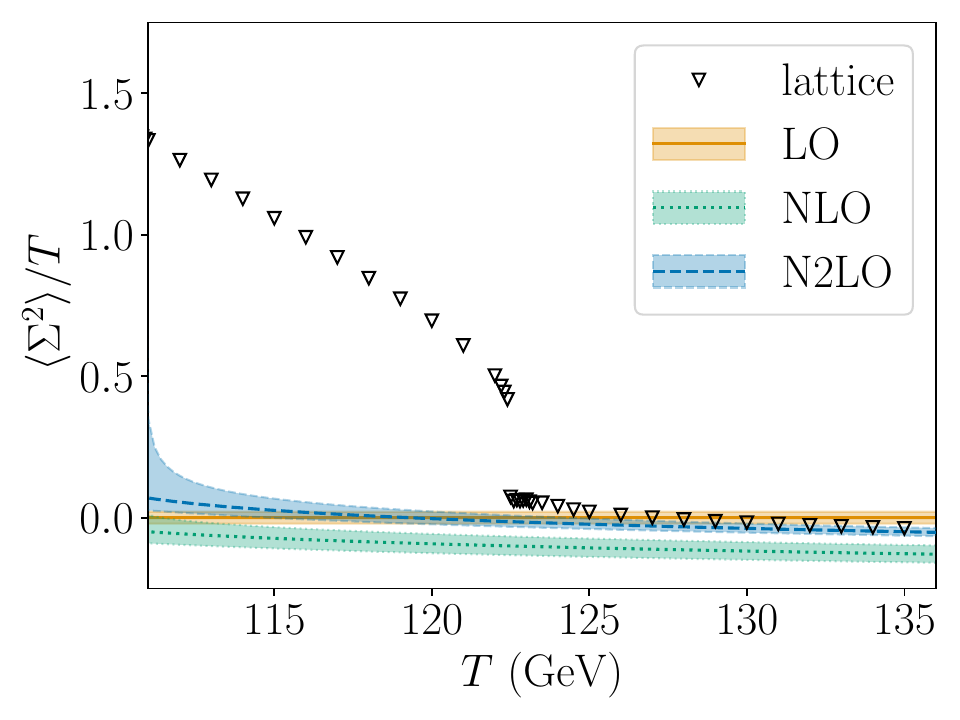} 
    \caption{Hard$\to$soft EFT, mixed method}
    \label{fig:BM2_onestep_b}
\end{subfigure} 
\begin{subfigure}{0.5\textwidth}
    \centering
    \includegraphics[width=\textwidth]{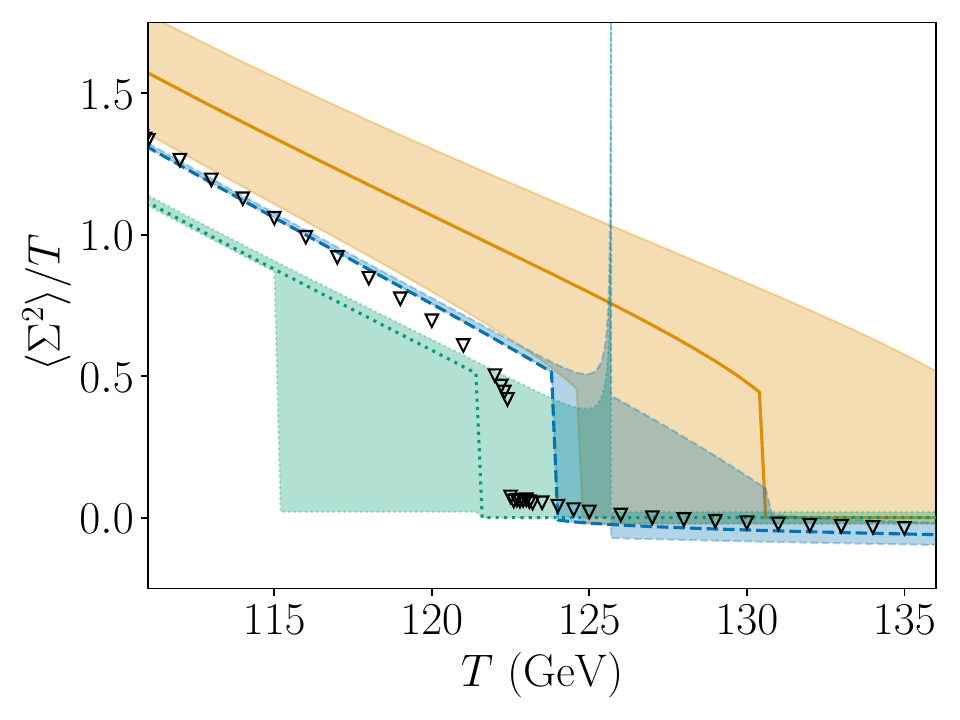}
    \caption{Soft$\to$supersoft EFT, mixed method}
    \label{fig:BM2_onestep_c}
\end{subfigure}
\begin{subfigure}{0.5\textwidth}
    \centering
    \includegraphics[width=\textwidth]{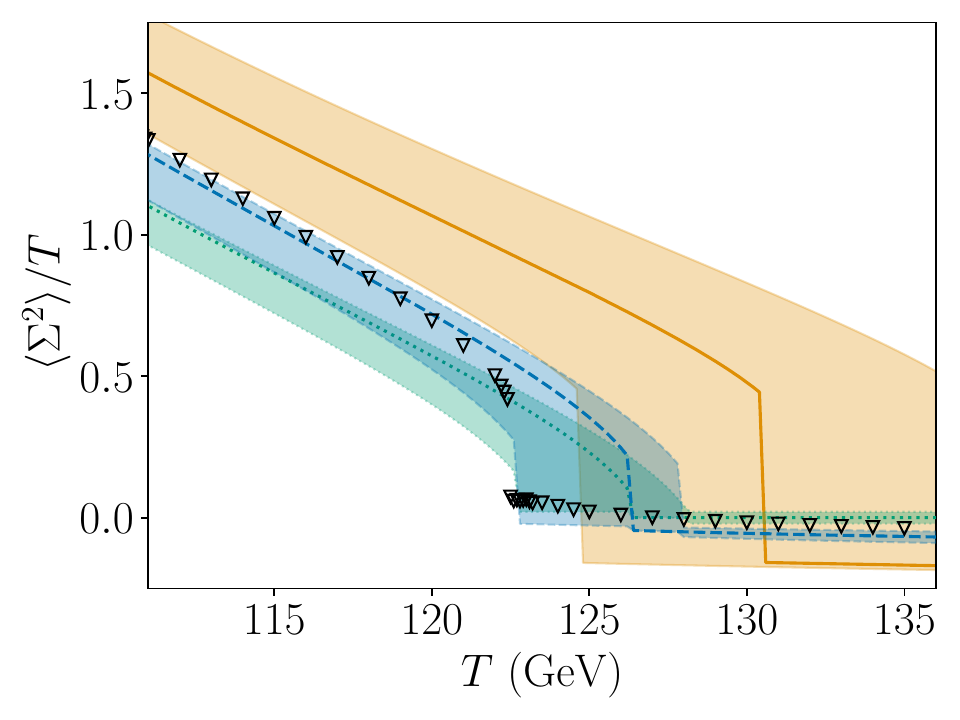}
    \caption{Soft$\to$supersoft EFT, strict method}
    \label{fig:BM2_onestep_d}
\end{subfigure}
\caption{
As Fig.~\ref{fig:BM1-1step-methods}, but for BM2. Note that the substantially larger portal coupling in this case leads to larger uncertainties at lower orders, and slower convergence. Nevertheless, the soft$\to$supersoft strict method shows good agreement with the lattice at N2LO. 
}
\label{fig:BM2-1step-methods}
\end{figure}

We have computed the triplet condensates according to each of the matrix of possibilities shown in Table \ref{tab:methods1}. Our results, together with the lattice results of Ref.~\cite{Niemi:2020hto} are shown in Fig.~\ref{fig:BM1-1step-methods} for BM1, and in Fig.~\ref{fig:BM2-1step-methods} for BM2. Exceptions are made for the strict method in the hard$\to$soft EFT as well as the direct method in the soft$\to$supersoft EFT, for which we do not show the results in Figs.~\ref{fig:BM1-1step-methods} and \ref{fig:BM2-1step-methods}. For the former approach, strict expansions fail as originally realised in \cite{Laine:1994zq}. At leading order in the soft theory the transition is of second order, and strict loop corrections do not modify the position of this transition, so that the mixed and strict methods in the hard$\to$soft EFT are equivalent. While the use of direct minimisation in the soft$\to$supersoft EFT has not appeared before in the literature, due to its gauge dependence there is no clear reason to prefer it to the other methods applied to the supersoft EFT, and we relegate the results to Appendix~\ref{sec:comparing-2steppers}.

In the following, we discuss in turn the panels of Figs.~\ref{fig:BM1-1step-methods} and \ref{fig:BM2-1step-methods}, focusing on the successes and failures of the different perturbative approaches. In all figures shaded bands correspond to the range of predictions from varying the 3d RG scale over the set $\scaleft \in \{0.5 T, T, 2T\}$, and therefore give an intrinsic measure of the theoretical uncertainty.

\begin{itemize}

\item[(a)] 
\textbf{Hard$\to$soft EFT, direct method:}
At LO there is a 2nd order transition at $T_\text{c} \approx 134.5$ GeV, which is the temperature where the triplet 3d mass vanishes, and is below the lattice transition temperature. The NLO result shifts the critical temperature significantly above the lattice result, and the transition is of first order. Extending to N2LO yields much closer agreement with the lattice results for both $T_\text{c}$ and the values of the condensates, yet note that there are a number of data points missing, appearing as cuts in the otherwise continuous result. This is due to the possibility of our direct minimisation algorithm%
\footnote{
We used Mathematica's \texttt{NMinimize} function, adopting the differential evolution method and choosing tolerance parameters so that producing the N2LO data for one benchmark point took around one hour on a laptop, with temperature steps $\Delta T = 0.25$ GeV.
} 
failing. While this feature could be ameliorated by improving the minimisation algorithm, we have stuck to the aforementioned algorithm for the following reasons: the failure of direct minimisation at N2LO is a fairly common occurrence compared to using the same algorithm for the potential at lower orders. This is due to IR-sensitive logarithmic terms at two loops, that can result in spuriously large contributions to the potential at field values where the corresponding mass eigenvalues in the logarithm vanish, possibly preventing convergence to the actual minimum. Such an issue may be mitigated with a cost in performance time, thought this would be a limiting issue for model parameter space scans. To emphasize this aspect, we have used the same minimisation algorithm at all orders, despite the algorithm occasionally failing at N2LO, causing gaps in the corresponding plots in Figs.~\ref{fig:BM1-1step-methods} and \ref{fig:BM2-1step-methods}. Note that RG improvement kicks in for the first time at N2LO, due to the structure of running in these 3d EFTs \cite{Farakos:1994kx, Gould:2021oba}. The width of the error bands at LO and NLO are comparable, and in both cases significantly underestimates the theoretical error.

\item[(b)]
\textbf{Hard$\to$soft EFT, mixed method:}
This method is known to fail when the LO potential does not have a barrier between the minima \cite{Laine:1994zq, Niemi:2020hto}. From Figs.~\ref{fig:BM1_onestep_b} and \ref{fig:BM2_onestep_b} one can see two clear failures of this approach: $T_\text{c}$ is unchanged by higher orders, and at N2LO there is a spurious divergence at $T_\text{c}$. The broken minimum exists only after the triplet 3d mass parameter ($\mu^2_{\Sigma,3}$) becomes negative (when it is positive the value of the effective potential in the triplet phase is imaginary), at which point the triplet minimum immediately becomes the global one: the critical temperature is therefore erroneously identified -- at all orders -- with the condition that the triplet 3d mass parameter vanishes. The divergence at the critical temperature arises from a logarithm of the triplet mass parameter as it goes through zero. At higher orders in this expansion it is expected that stronger IR diverges will occur.

\item[(c)]
\textbf{Soft$\to$supersoft EFT, mixed method:}
While this method is gauge-independent, one can see from Figs.~\ref{fig:BM1_onestep_c} and \ref{fig:BM2_onestep_c} that it yields spurious divergences at NLO and N2LO for some RG scales. This problem arises at the edge of the range of temperatures where the LO result ceases to exist, resulting in ill-defined behaviour for the condensate. This problem was further discussed around Eq.~\eqref{eq:DeltaT} in Sec.~\ref{sec:methods}, where the strict method was proposed as a general solution.

\item[(d)] 
\textbf{Soft$\to$supersoft EFT, strict method:}
Finally, this approach resolves all the theoretical problems encountered by the other approaches, and seemingly converges towards the lattice results with impressive accuracy. For the more weakly coupled BM1, the LO result in the supersoft EFT already agrees well with the lattice, and higher orders lead to small improvements, especially noticeable in the broken phase. However, it is for the more strongly coupled BM2 where this method clearly outshines the others, especially in the vicinity of the phase transition, where this method converges towards the lattice without spurious artefacts.

\end{itemize}

We highlight that despite the respective successes of different EFT expansions in predicting $T_\text{c}$ and the value of the triplet condensate as a function of temperature, in all approximations perturbation theory incorrectly predicts the character of the transition for BM1. The transition is a smooth crossover, as measured on the lattice, whereas perturbation theory predicts a weak first-order transition. In BM2 the transition is actually first order, and perturbation theory predicts it correctly.

\subsection{Two-step transition}

Finally, we turn to a two-step phase transition, for which we suggest a novel prescription in terms of two separate EFTs for consequent transitions. From the previous section, we know that for the first transition to the triplet phase we need to use the supersoft EFT. For the second transition, we have multiple options.

First, there is a possibility that the second transition happens at the soft scale. As discussed in Sec.~\ref{sec:semisoft-to-soft-EFT}, such a transition can be induced by the semisoft scale in addition to the hard scale. We depict the results based on this approach in Fig.~\ref{fig:semisoft}.%
\footnote{
For comparison, in Appendix \ref{sec:comparing-2steppers} we discuss results assuming that there is no enhancement from the semisoft scale, and the transition at the soft scale is solely induced by the hard scale.
}
\begin{figure}[t]
\begin{subfigure}[t]{0.69\textwidth}
    \centering
    \includegraphics[width=\textwidth]{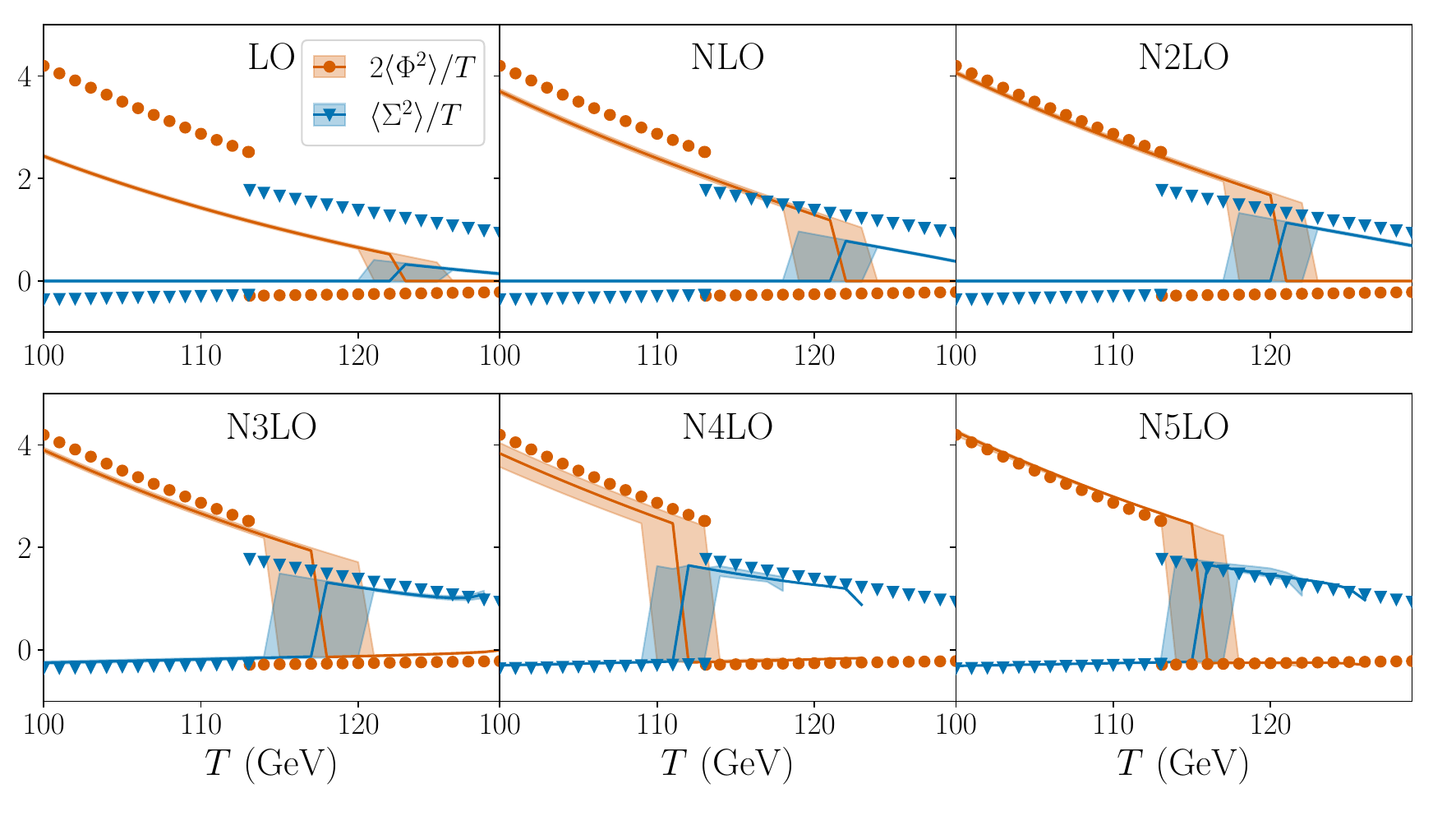}
    \caption{BM1}
\end{subfigure}
\begin{subfigure}[t]{0.29\textwidth}
    \centering
    \raisebox{17mm}{
    \includegraphics[width=\textwidth]{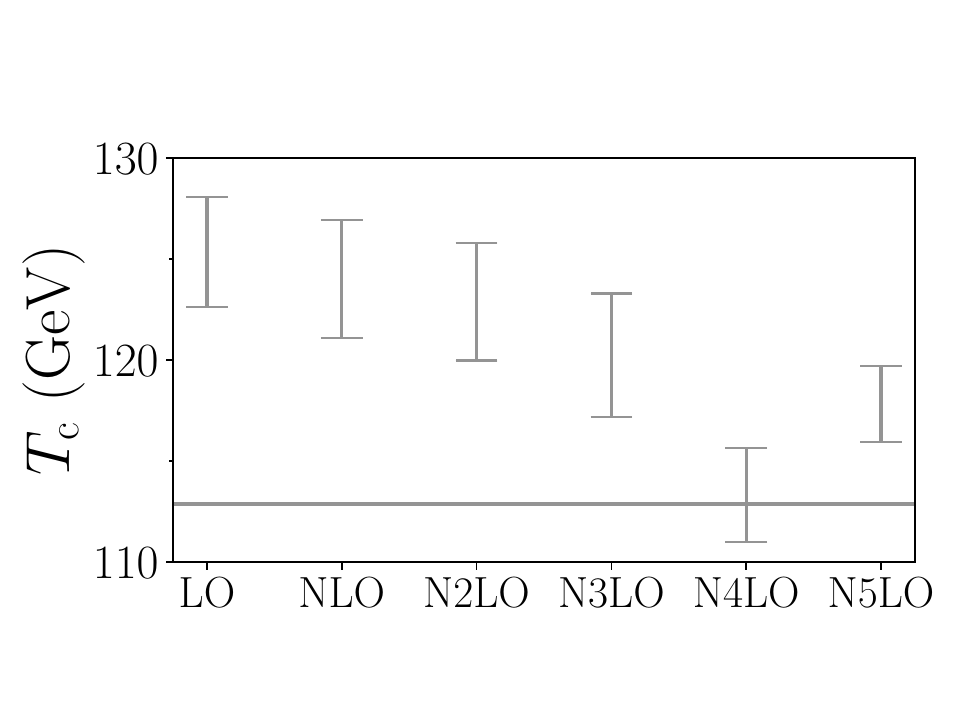}
    }%
\end{subfigure}
\begin{subfigure}[t]{0.69\textwidth}
    \centering
    \includegraphics[width=\textwidth]{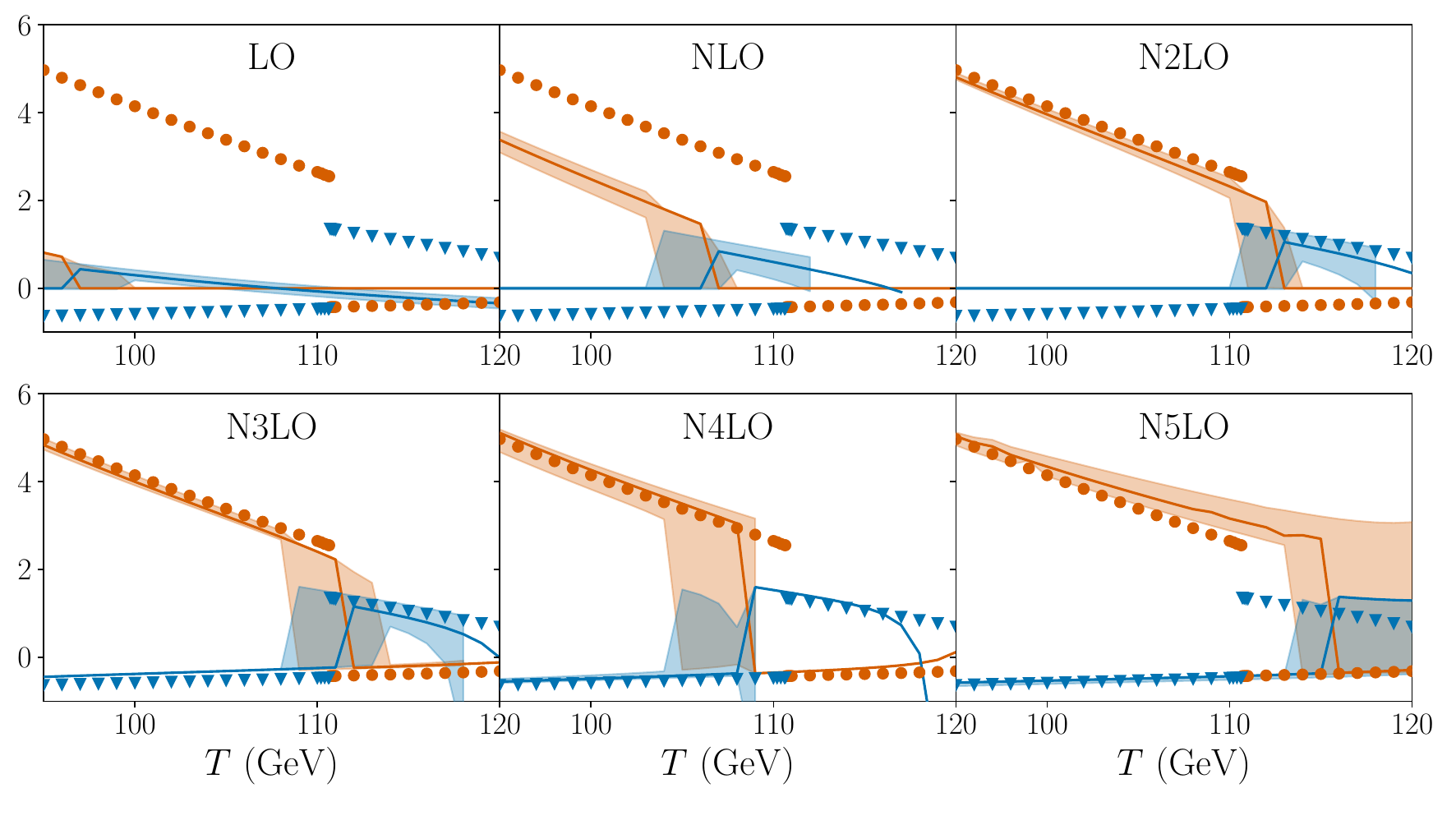}
    \caption{BM2}
\end{subfigure}
\begin{subfigure}[t]{0.29\textwidth}
    \centering
    \raisebox{17mm}{
    \includegraphics[width=\textwidth]{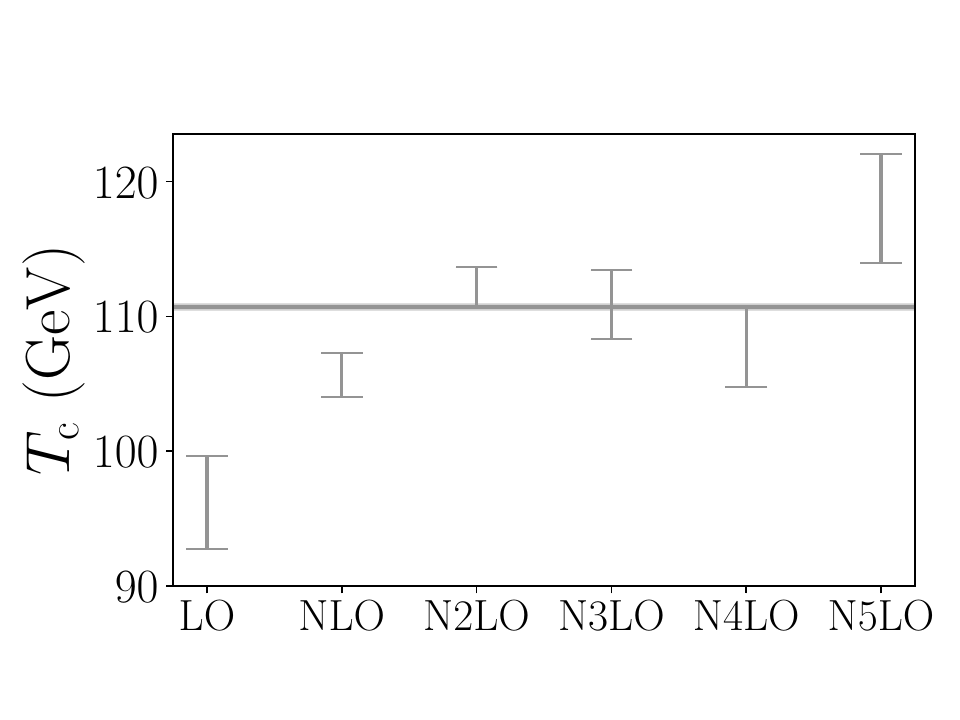}
    }%
\end{subfigure}
\caption{
Behaviour of the semisoft to soft expansion for the triplet to Higgs transition. In the plots of the critical temperature, the horizontal band is the lattice result together with its statistical uncertainty, and the bars show perturbative results at each order in the EFT expansion with uncertainty due to varying the RG scale.
}
\label{fig:semisoft}
\end{figure}
This figure depicts both condensates as well as the critical temperature for the triplet to Higgs transition, at each order in the expansion. The plots indicate some degree of convergence, yet even at the highest orders we have computed they do not provide striking agreement with the lattice results. This signals either that still higher order contributions should be included (especially since several RG improvements kick in only at N6LO) or that the assumption that the transitioning fields live at the soft scale is not correct, for the benchmark points in question.  
 
\begin{figure}[t]
\centering
\begin{subfigure}[t]{0.71\textwidth}
    \centering
    \includegraphics[width=\textwidth]{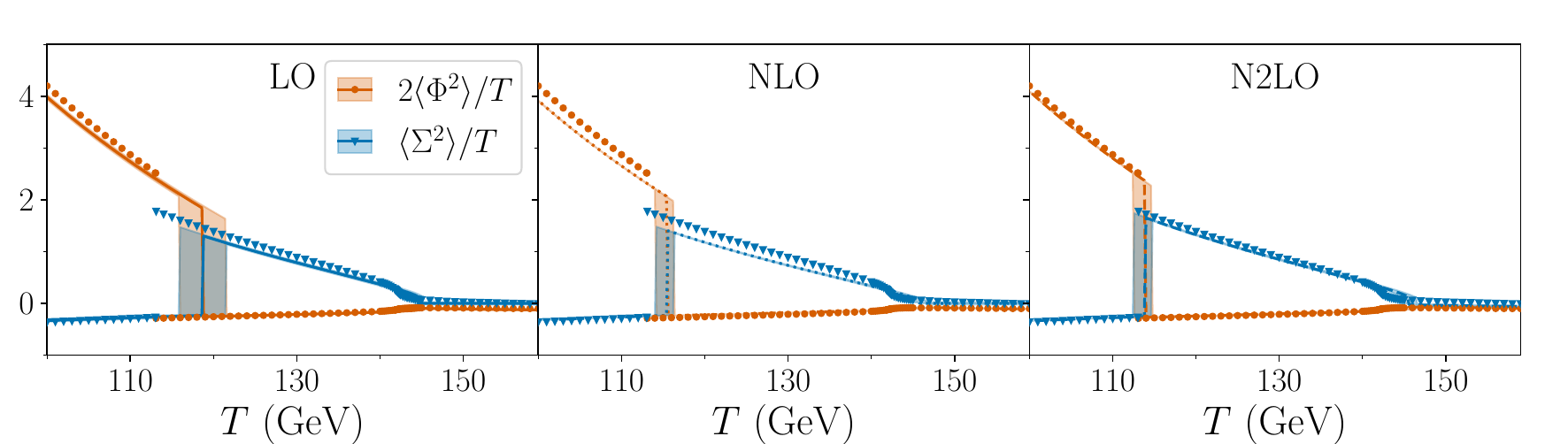}
    \caption{BM1}
    \label{fig:BM1_supersoft_strict_strict}
\end{subfigure}
\begin{subfigure}[t]{0.27\textwidth}
    \centering
    \raisebox{0mm}{
    \includegraphics[width=\textwidth]{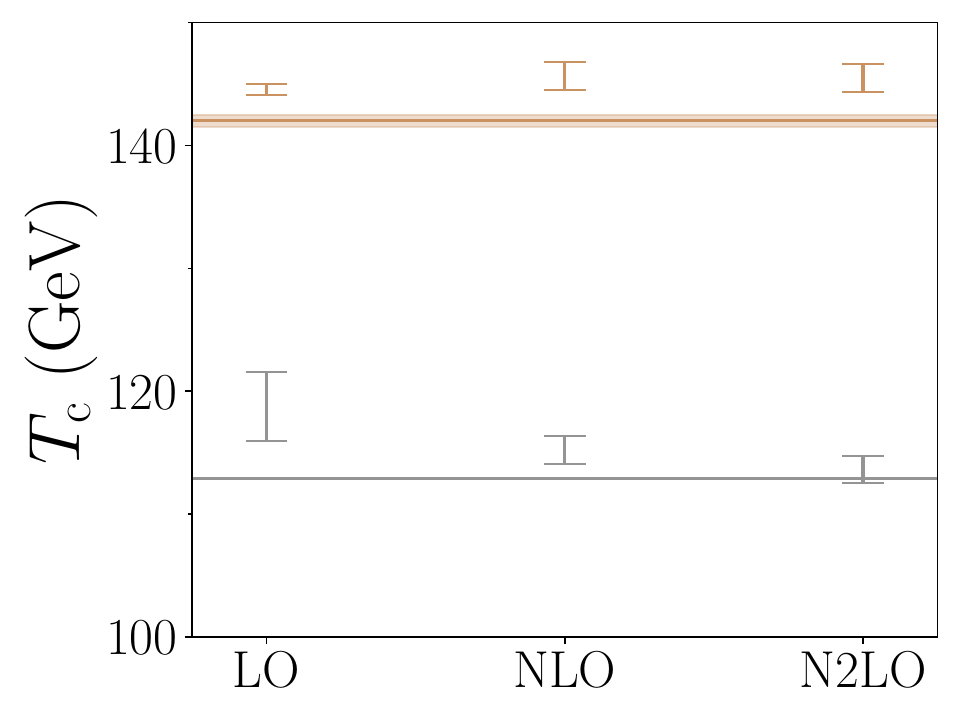}
    }%
\end{subfigure}
\hfill
\begin{subfigure}[t]{0.71\textwidth}
    \centering
    \includegraphics[width=\textwidth]{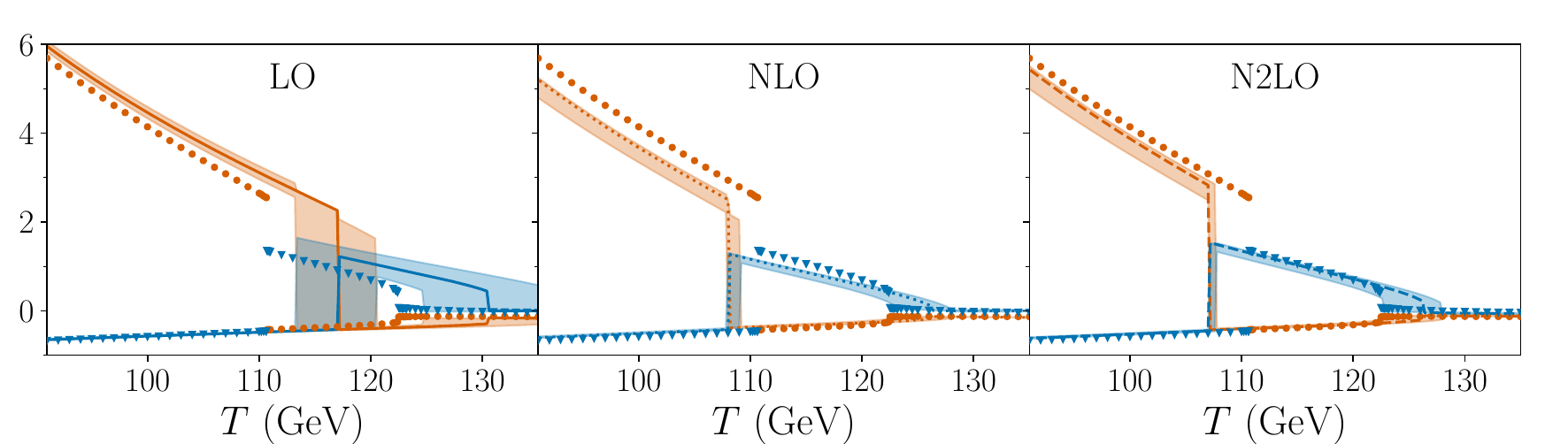} 
    \caption{BM2}
    \label{fig:BM2_supersoft_strict_strict}
\end{subfigure}
\begin{subfigure}[t]{0.27\textwidth}
    \centering
    \raisebox{0mm}{
    \includegraphics[width=\textwidth]{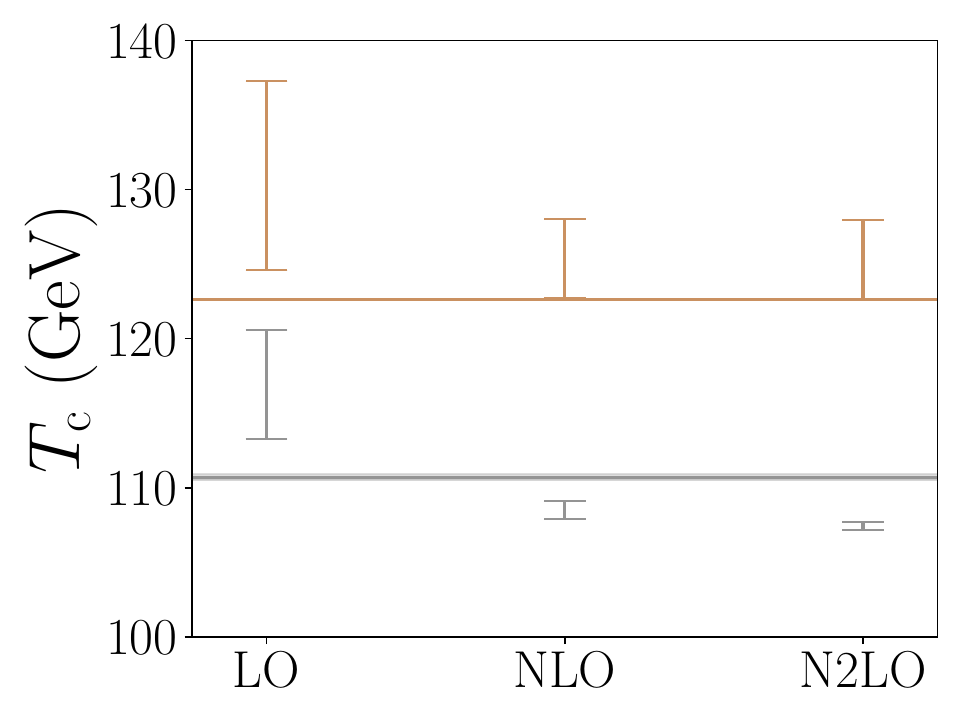}
    }%
\end{subfigure}
\caption{
Convergence of supersoft strict approximations, where in both cases $\Delta T$ has been defined relative to the higher temperature transition. Note that the higher temperature transition for BM1 is a crossover, so $T_\text{c}$ here corresponds to a pseudo-critical temperature, defined as the peak in the susceptibility for the scalar condensates.
}
\label{fig:supersoft_strict_strict}
\end{figure}

Hence, we next test the possibility that the transitioning fields of the second transition live at the supersoft scale. This requires two separate EFTs for each of the triplet and Higgs phases. Comparison of the N2LO result to lower orders and convergence of the expansion is depicted side-by-side in the triptychs of Fig.~\ref{fig:supersoft_strict_strict}.
These triptychs of plots indicate reasonably good convergence, from LO results which only rough accord with the lattice data, each additional order yields closer agreement.
We observe that at N2LO our novel perturbative computation of the scalar condensates using EFT expansions at the supersoft scale provides a striking correspondence with lattice results.
While the convergence is clearer in BM1, it is also the case in BM2 which has a portal coupling more than twice as large, albeit the critical temperature for the second transition is further away from the lattice result.
In analogy to Fig.~\ref{fig:soft_strict_direct} in Sec.~\ref{sec:motivation} we summarise our analysis in Fig.~\ref{fig:condensates} which recollects the N2LO result of Fig.~\ref{fig:supersoft_strict_strict}.
This plot indeed demonstrates the key result of this article: EFT expansions resolve all theoretical blemishes that have haunted perturbative predictions of the past, and while doing so, provide results that are not far away from those obtained on the lattice.

\section{Discussion}
\label{sec:discussion}

In the past few decades, a rich patchwork of perspectives and insights have been developed regarding the reliability of perturbation theory to describe cosmological phase transitions. In one thread of the inquiry, a range of thermal hierarchies of scale were identified, and corresponding resummation schemes to correctly account for them. The early development of high-temperature dimensional reduction was based around the hard, soft and ultrasoft scales \cite{Kajantie:1995dw, Braaten:1995jr}, yet another scale, the supersoft scale, was identified as central to first-order phase transitions \cite{Arnold:1992rz}. In a separate thread of inquiry, a range of different perturbative methods were developed for the study of equilibrium thermodynamics, from direct minimisation of the thermal effective potential to strict $\hbar$-expansions. Concerns were raised that the direct minimisation method led to gauge-dependent results \cite{Laine:1994zq}, while strict $\hbar$-expansions appeared to lead to IR divergences \cite{Laine:1994zq}. Concern about gauge dependence was then later revived in \cite{Patel:2011th}, further inspiring \cite{Ekstedt:2020abj}. In recent years, significant progress towards resolving the aforementioned problems was made in Refs.~\cite{Croon:2020cgk, Gould:2021dzl, Gould:2021oba, Lofgren:2021ogg,Hirvonen:2021zej, Schicho:2022wty, Gould:2021ccf, Ekstedt:2022zro,Ekstedt:2021kyx,Ekstedt:2022tqk,Ekstedt:2022ceo, Hirvonen:2022jba}. In this work at hand, we have unified, generalised and expanded most of this progress to a revised EFT framework for equilibrium thermodynamics that builds from the dimensionally reduced 3d EFTs \cite{Farakos:1994kx,Farakos:1994xh,Kajantie:1995dw,Braaten:1995cm,Kajantie:1995kf,Kajantie:1996qd}, but also consistently applies strict power-counting expansions in perturbation theory \cite{Ekstedt:2022zro}.

Concretely, we have simultaneously tested both these threads of inquiry, and have found a consistent resolution to all the concerns in terms of self-consistent perturbative EFT expansions. The results of recent lattice Monte-Carlo simulations at two benchmark points in the real-triplet extended SM \cite{Niemi:2020hto} have formed the bedrock of these tests. This has allowed us to obtain an unambiguous measure of the error in different perturbative approaches.

\begin{figure}[t]
\begin{subfigure}{0.5\textwidth}
    \centering
    \includegraphics[width=\textwidth]{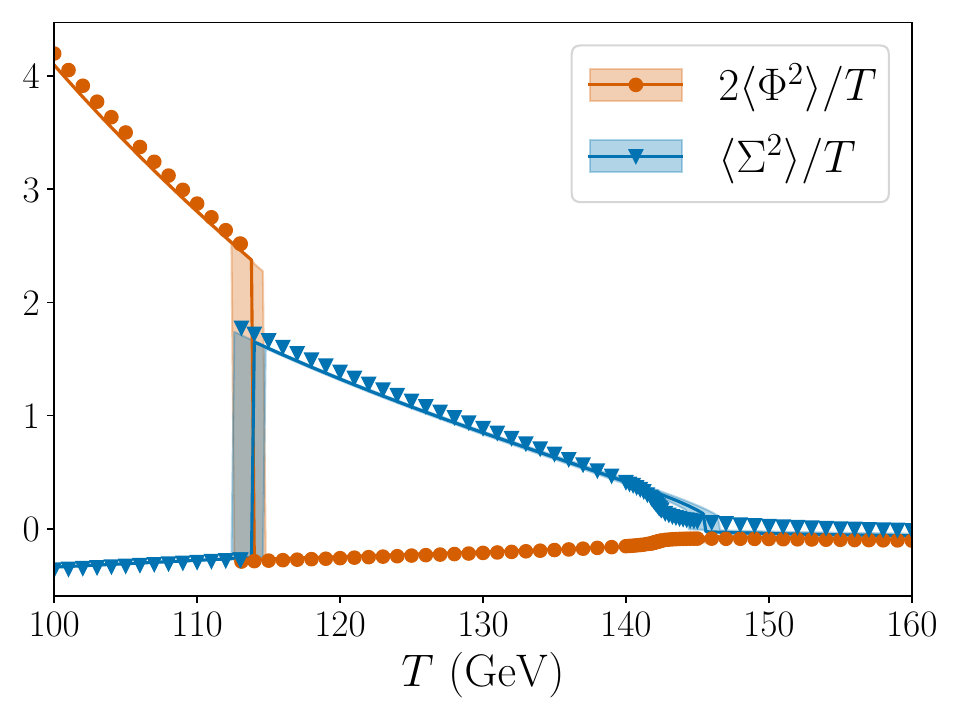}
    \caption{BM1}
    \label{fig:beta_BM1}
\end{subfigure}
\begin{subfigure}{0.5\textwidth}
    \centering
    \includegraphics[width=\textwidth]{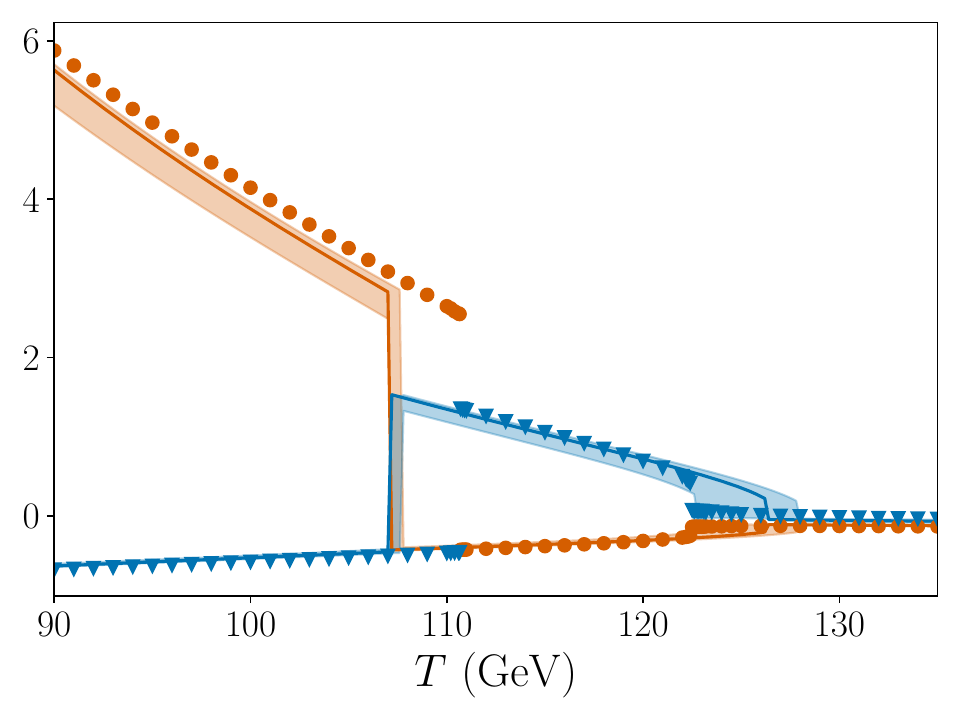}
    \caption{BM2}
    \label{fig:beta_BM2}
\end{subfigure}
\caption{
Comparison of the thermal evolution of scalar condensates in the real-triplet extended Standard Model. Solid lines show the N2LO results of strict perturbation theory within supersoft-scale EFTs, with corresponding bands giving the renormalisation scale dependence. Scatter points show the lattice results of Ref.~\cite{Niemi:2020hto}. The supersoft EFTs are constructed separately in each broken phase, with the triplet field treated as supersoft within the triplet phase and the Higgs field treated as supersoft within the Higgs phase. This approach yields gauge-invariant results, in good agreement with the lattice, and a significant improvement over previous perturbative approaches (see Fig.~\ref{fig:soft_strict_direct}).
}
\label{fig:condensates}
\end{figure}

Our results, summarised in Fig.~\ref{fig:condensates}, attest to the correctness of a particular perturbative approach, which is both theoretically consistent and numerically reliable. This approach is rather simple, and in hindsight obvious. It is the following:
\begin{enumerate}
\item Successively integrate out UV modes, starting from the hard scale, and working towards the IR, and stopping when one meets the mass scale of the transitioning fields.
\item The mass scale for the transitioning fields can be identified by power counting, applied to the tree-level potential of the EFT. If there is an apparent second-order phase transition at tree-level within this EFT, then more modes must be integrated out. 
\item Carry out strict perturbative expansions within the EFT for the transitioning field, ensuring to remain within the region of validity of the EFT.
\end{enumerate}
Point 1 ensures that all the necessary resummations are carried out. Consideration of point 2 has revealed that the transitioning fields often live at the supersoft scale. Finally, point 3 ensures that the final results are order-by-order gauge invariant, renormalisation scale invariant and real. Together this perturbative approach has demonstrated quantitative agreement with lattice Monte-Carlo simulations. We emphasize that in our strict EFT expansions, the underlying expressions for effective potentials and for thermodynamic quantities are astonishingly simple -- excepting complicated, yet closed form, expressions for LO broken mimima in cases of radiatively generated barriers -- and the striking agreement with lattice results highlights that the underlying physics is well captured in perturbation theory.%
\footnote{
Yet we emphasize again that perturbation theory cannot separate weak first-order transitions from crossovers, or describe purely non-perturbative phenomena related to phase transitions, such as condensation of monopoles \cite{Niemi:2022bjg}.  
}  

The good agreement of perturbation theory with the lattice shown in Fig.~\ref{fig:condensates} should be contrasted with that of Fig.~\ref{fig:soft_strict_direct}. The crucial difference is that in Fig.~\ref{fig:condensates} the supersoft scale has been correctly identified as the energy scale of the transitioning fields. Our results align with \cite{Gould:2021dzl, Ekstedt:2022zro}, which also compared perturbation theory to the lattice, and came to similar conclusions. Yet, in this work at hand we have for the first time applied these developments to a BSM theory, where the studied phase transition pattern is more complicated and leads to a rich chain of EFT setups. Indeed, in the course of this study, a number of further technical manoeuvres have been identified. We have shown how, when one is interested in thermodynamic observables away from the critical temperature, it is advantageous to re-express quantities in terms of deviations from the critical temperature $\Delta T = T - T_\text{c}$. This resolves a problem of the existence of the LO result required by a strict expansion, and underlies the difference between our mixed and strict approaches. We have also discovered that a new scale between the hard and soft scales, here dubbed the semisoft scale, arises naturally in strong two-step first-order phase transitions.

The strict perturbative EFT expansions presented in this work can still be extended by computing the final perturbative orders that are available before crashing against the Linde problem of non-Abelian gauge theories at four-loop order. Computing these final orders requires three-loop vacuum diagrams, and provides an intriguing future challenge analogous to that achieved in hot QCD \cite{Braaten:1995jr, Kajantie:2002wa,Kajantie:2003ax}. Yet another, different kind of challenge will be to incorporate the presented EFT expansions to parameter space scans of phenomenologically interesting models. Herein the challenge lies in the implementation: different EFTs may be required to study different parameter points, and even in a single parameter point there can be several EFTs in different temperature regimes. This issue has also been raised in the recent Ref.~\cite{Lofgren:2023sep} and is further discussed in Appendix \ref{sec:comparing-2steppers}. This reference indeed discusses many of the same ideas as detailed in our work at hand, yet our computation includes concrete applications to a BSM model, as well as comparison to lattice data. 

Finally, while we have limited ourselves to the study of equilibrium thermodynamics, EFT expansions are expected to carry over to the perturbative study of other properties of first-order phase transitions, such as the bubble nucleation rate and the bubble wall speed, as well as the sphaleron rate. For the bubble nucleation rate, Refs.~\cite{Gould:2021ccf,Hirvonen:2021zej, Ekstedt:2022ceo, Gould:2022ran} have in fact already used the approach proposed here. On the other hand, transferring what we have learnt here in this article to tunnelling will be challenging. While it is possible to utilise different EFTs in different phases for computing the free-energy of homogeneous phases, for bubble nucleation this must be generalised to non-trivial paths in field space. The formulation of the EFT description of bubble nucleation for such a two-step transition, warrants dedicated future studies.  

\section*{Acknowledgements}
The authors wish to thank Andreas Ekstedt, Joonas Hirvonen, Lauri Niemi, Johan L{\"o}fgren, Michael J. Ramsey-Musolf, Philipp Schicho, Bogumiła Świeżewska and Jorinde van de Vis for enlightening discussions. In addition, we would like to thank Lauri Niemi for correspondence related to the lattice data of Ref.~\cite{Niemi:2020hto}. O.G.\ was supported by U.K.~Science and Technology Facilities Council (STFC) Consolidated grant ST/T000732/1, a Research Leadership Award from the Leverhulme Trust and a Dorothy Hodgkin Fellowship from the Royal Society. The work of T.T.\ has been supported in part by grants from National Natural Science Foundation of China (grant nos. 11975150) from the Ministry of Science and Technology of China (grant no. WQ20183100522).

%
\appendix

\numberwithin{equation}{section}

\renewcommand{\thesection}{\Alph{section}}
\renewcommand{\thesubsection}{\Alph{section}.\arabic{subsection}}
\renewcommand{\theequation}{\Alph{section}.\arabic{equation}}

\section{EFT expansions with doublet and triplet fields}

In this appendix we present explicit expressions for EFT expansions of the effective potential for the real-triplet extended Standard Model. We can immediately read the effective potential for the soft scale EFT from \cite{Niemi:2020hto}. Therein, the presented ``$\hbar$-expansion'' matches the soft-scale strict expansion of Sec.~\ref{sec:methods}. Our task is then to compute the effective potentials at the supersoft scale EFT, as well as the soft scale EFT in the presence of fields at the semi-soft scale.

For this computation, we can read from \cite{Niemi:2020hto} all the two-loop diagrams we need, which we reorganise into EFT expansions. In this section, all masses $m_i$ are mass eigenvalues in the 3d EFT, and are not to be confused with physical pole masses. We use Landau gauge throughout.

\subsection{Supersoft EFT}
\label{sec:supersoft-matching}

In this section, we use the following notation for the supersoft scale effective potentials in the EFT expansion
\begin{align}
V^{\text{supersoft}}_{\text{eff}} = V_0 + V_1 + V_2 + V_3 + \mathcal{O}(\veps^4 V_0), 
\end{align}
where each order beyond $V_0$ at LO is suppressed by $\veps \sim \sqrt{\veps_\text{soft}}$ relative to the previous order; see the discussion around Eq.\eqref{eq:Veff-supersoft}. Notably, $V_1$ is identically zero.

\paragraph{Supersoft Higgs doublet}
For simplicity, we start with the 3d EFT of the Standard Model with SU(2) and U(1) gauge fields. We parametrise the Higgs doublet in the 3d EFT as
\begin{align}
\label{eq:phi}
\phi &=
\begin{pmatrix}
  G^+ \\ 
  \frac{1}{\sqrt{2}} (v + h + i z) 
\end{pmatrix} ,
\end{align}
i.e.~we compute the effective potential in terms of real background field $v > 0$.
The leading order contribution reads
\begin{align}
V^{\text{doublet}}_{0} = \frac{1}{2} \mu^2_3 v^2 + \frac{1}{4} \lambda_3 v^4 - \frac{1}{6\pi} \Big( 2 m^{3}_W + m^{3}_Z \Big), 
\end{align}
where $m_W = \frac{1}{2} g_3 v$ and $m_Z = \frac{1}{2} \sqrt{g^2_3 + {g_3'}^2} v$. The two-loop contribution from gauge fields and ghosts reads
\begin{align}
V^{\text{doublet}}_{2} &= \frac{1}{(4\pi)^2} \frac{1}{8 (g^2_3+{g_3'}^2)} \bigg( -\frac{1}{8} v^2 \Big( 9 g^6_3 + 21 g^4_3 {g_3'}^2 + 13 g^2_3 {g_3'}^4 + 3 {g_3'}^6 + 4 g^2_3 {g_3'}^4 \frac{m^2_W + m^2_Z}{m_W m_Z} \Big)  \nonumber \\ 
&+ g^2_3 {g_3'}^2 m^2_W - {g_3'}^4 m^2_Z  + g^4_3 \Big( 2 \frac{m^3_W}{m_Z} + 14 m_W m_Z + 14 m^2_Z  - 10 \frac{m^3_Z}{m_W} + \frac{m^4_Z}{m^2_W} \Big)  \nonumber \\ 
& + \Big( -32 g^2_3 {g_3'}^2 m^2_W + ( 3 g^6_3 + 5 g^4_3 {g_3'}^2 + 3 g^2_3 {g_3'}^4 + {g_3'}^6 ) v^2 \Big) \ln(2) \nonumber \\
& +  \frac{1}{2} g^2_3 {g_3'}^4 v^2 \frac{m^2_Z}{m^2_W} \ln\Big( \frac{m_W + m_Z}{m_Z} \Big) 
 + \frac{3}{2} g^4_3 ( g^2_3 + 3 {g_3'}^2 ) v^2 \ln\Big( \frac{m_W m^{\frac{1}{2}}_Z}{\Lambda^{\frac{3}{2}}_3} \Big) \nonumber  \\
& + \frac{21}{4} g^2_3 {g_3'}^4 v^2 \ln\Big( \frac{m^{\frac{3}{7}}_Z (m_W + m_Z)^{\frac{4}{7}}}{\Lambda_3} \Big) 
  + 40 g^2_3 {g_3'}^2 m^2_W \ln\Big( \frac{\Lambda_3}{m_W} \Big) \nonumber  \\
& + \frac{1}{4} {g_3'}^4 (8 m^2_Z + 3 {g_3'}^2 v^2 ) \ln\Big( \frac{m_Z}{\Lambda_3} \Big) 
 + g^4_3 \frac{m^6_Z}{m^4_W} \ln\Big( \frac{m_Z (2 m_W + m_Z)}{(m_W + m_Z)^2} \Big) \nonumber \\
& + 8 g^4_3 \frac{m^4_Z}{m^2_W} \ln\Big( \frac{2 m_W + m_Z}{m_W + m_Z} \Big) 
 + \frac{4 g^4_3 m^4_W - g^2_3 {g_3'}^4 m^2_W v^2 }{2 m^2_Z} \ln\Big( \frac{m_W}{m_W + m_Z} \Big)  \nonumber \\
& + 20 g^4_3 m^2_Z  \ln\Big( \frac{(m_W + m_Z)\Lambda_3}{(2 m_W + m_Z)^2} \Big) 
 + 8 g^4_3 m^2_W  \ln\Big( \frac{\Lambda^5_3}{(m_W + m_Z)(2 m_W + m_Z)^4} \Big)
\bigg).
\end{align}
This can be obtained from the result of Ref.~\cite{Niemi:2020hto} by dropping triplet contributions and by setting scalar masses to zero inside two-loop diagrams.
The resummed mass eigenvalues for the scalar fields -- Higgs ($h$) and Goldstone bosons ($G^\pm, z$) -- read 
\begin{align}
\widetilde{m}^2_h &= \frac{d^2}{dv^2} V^{\text{doublet}}_{0} = \mu^2_{h,3} + 3 \lambda_3 v^2 - \frac{v}{8\pi} \Big(2 g^3_3 + (g^2_3 + {g_3'}^2)^{\frac{3}{2}} \Big), \\
\widetilde{m}^2_G &= \frac{1}{v} \frac{d}{dv} V^{\text{doublet}}_{0} = \mu^2_{h,3} + \lambda_3 v^2 - \frac{v}{16\pi} \Big(2 g^3_3 + (g^2_3 + {g_3'}^2)^{\frac{3}{2}} \Big), 
\end{align}
where the Goldstone square mass eigenvalue is triply degenerate.
These expressions can be derived starting from a general potential written in terms of the gauge-invariant bilinear $\phi^\dagger \phi$, and expanding to quadratic order in fluctuations,
\begin{align}
V(\sqrt{2\phi^\dagger \phi}) = V(v) + \frac{1}{2} h^2 V''(v) + \frac{1}{2} z^2 \frac{1}{v}V'(v) + \ldots
\end{align}
where prime denotes a derivative with respect to $v$, and we have shown only the relevant bilinear terms.
In terms of resummed masses, the one-loop supersoft contribution reads
\begin{align}
\label{eq:NNLO-doublet}
V^{\text{doublet}}_{3} = -\frac{1}{12\pi} \Big( (\widetilde{m}^2_h)^{\frac{3}{2}} + 3 (\widetilde{m}^2_G)^{\frac{3}{2}} \Big). 
\end{align}

\paragraph{SU(2)+Higgs}
The above expressions for the supersoft effective potential becomes much more compact when the U(1) gauge sector is decoupled, i.e.\ in the limit ${g_3'} \rightarrow 0$, $m_Z \rightarrow m_W$,
\begin{align}
V^{\text{SU(2)+Higgs}}_{\text{eff}} &= \bigg( \frac{1}{2} \mu^2_3 v^2 + \frac{1}{4} \lambda_3 v^4 - \frac{1}{16\pi}g^3_3 v^3 \bigg)_{V_0} \nonumber \\
& + \frac{1}{(4\pi)^2} \bigg( -\frac{3}{64} g^4_3 v^2 \Big( -11 + 42 \ln\Big(\frac{3}{2}\Big) + 34 \ln\Big(\frac{g_3 v}{\Lambda_3}\Big) \Big) \bigg)_{V_2} \nonumber \\
& -\frac{1}{12\pi} \bigg( 3 \Big(  \mu^2_3 + \lambda_3 v^2 -\frac{3}{16\pi} g^3_3 v  \Big)^{\frac{3}{2}} + \Big(  \mu^2_3 + 3 \lambda_3 v^2 -\frac{3}{8\pi} g^3_3 v  \Big)^{\frac{3}{2}}  \bigg)_{V_3},
\end{align}
where we have substituted $m_W = \frac{1}{2} g_3 v$. This result has been previously obtained in Ref.~\cite{Ekstedt:2022zro}.

\paragraph{Supersoft doublet with a soft triplet}
Next, we include the effect of a soft triplet, i.e.\ the triplet remains at zero background field, and is integrated out together with the gauge fields. One-loop triplet contributions arise at leading order, so that
\begin{align}
\label{eq:veff-LO-doublet-triplet}
V^{\text{}}_{0} 
&= V^{\text{douplet}}_{0} + V^{\text{soft triplet}}_{0},
\end{align}
where
\begin{align}
V^{\text{soft triplet}}_{0}
&= - \frac{1}{12\pi} 3 \Big( m^2_\Sigma \Big)^{\frac{3}{2}},
\end{align}
and $m^2_\Sigma = \mu^2_{\Sigma,3} + \frac{1}{2} a_{2,3} v^2$. Lagrangian parameters are defined in Eq.~\eqref{eq:SigmaSM_Lagrangian}.
The triplet squared mass is triply degenerate, since the neutral and charged triplet have equal masses for vanishing triplet background field. Note, that here we have assumed that also $\mu^2_{\Sigma,3}$ is soft: this is relevant for one-step transitions directly to the Higgs phase in the presence of a soft triplet that can enhance the transition strength, but also for the second step of a two-step phase transition. In practice this requires that there is enough supercooling between the two transitions, and the magnitude of $\mu^2_{\Sigma,3}$ (which can and often will be negative) can increase parametrically from the supersoft to the soft scale after the first transition. In this case, the expression for the minimum of the LO potential of Eq.~\eqref{eq:veff-LO-doublet-triplet} becomes utterly complicated, yet it can still be found analytically. Below we comment on the case where $\mu^2_{\Sigma,3}$ is still at the supersoft scale, albeit $m^2_\Sigma$ is soft.

Resummed scalar masses related to the supersoft doublet read 
\begin{align}
\widetilde{m}^2_h &= 
\frac{d^2}{dv^2} V^{\text{}}_{0} = \mu^2_{h,3} + 3 \lambda_3 v^2 - \frac{v}{8\pi} \Big(2 g^3_3 + (g^2_3 + {g_3'}^2)^{\frac{3}{2}} \Big) - \frac{3}{8\pi} a_{2,3} \frac{\mu^2_{\Sigma,3} + a_{2,3} v^2}{m_\Sigma}, \\
\widetilde{m}^2_G &= 
\frac{1}{v} \frac{d}{dv} V^{\text{}}_{0} = \mu^2_{h,3} + \lambda_3 v^2 - \frac{v}{16\pi} \Big(2 g^3_3 + (g^2_3 + {g_3'}^2)^{\frac{3}{2}} \Big) - \frac{3}{8\pi} a_{2,3} m_\Sigma. 
\end{align}
$V_3$ is of same form as before in Eq.~\eqref{eq:NNLO-doublet}, but with the above modified supersoft masses. Next we have
\begin{align}
V^{\text{}}_{2} 
= V^{\text{doublet}}_{2} + V^{\text{soft triplet}}_{2},
\end{align}
where the triplet contributions read
\begin{align}
\label{eq:soft-triplet-2loop}
V^{\text{soft triplet}}_{2} &=
\frac{1}{ (4\pi)^2} \bigg( 
\frac{15}{4} b_{4,3} m^2_\Sigma  - \frac{3}{8} a^2_{2,3} v^2 - \frac{1}{2} g_3^2 (m^2_W - 4 m_W m_\Sigma  - 6 m^2_\Sigma) \nonumber  \\
&+ \frac{6 g^2_3 {g_3'}^2 m^2_\Sigma + g^4_3 (-m^2_Z + 4 m_Z m_\Sigma  + 6 m^2_\Sigma) }{4(g^2_3 + {g_3'}^2)} 
+  \Big(2 \frac{g^2_3 {g_3'}^2 m^2_\Sigma }{g^2_3 + {g_3'}^2} - \frac{3}{4} a^2_{2,3} v^2 \Big) \ln\Big( \frac{\Lambda_3}{2 m_\Sigma} \Big) \nonumber \\
&- g^2_3 (m^2_W - 4 m^2_\Sigma) \ln\Big( \frac{\Lambda_3}{m_W + 2 m_\Sigma} \Big)
-  \frac{g^4_3 (m^2_Z - 4 m^2_\Sigma)}{2(g^2_3 + {g_3'}^2)} \ln\Big( \frac{\Lambda_3}{m_Z + 2 m_\Sigma} \Big)
\bigg).
\end{align}

If $\mu^2_{\Sigma,3}$ is supersoft despite $m^2_\Sigma$ being soft, we can account for the effect of the triplet 3d mass parameter as a perturbative mass insertion in the matching. In this case, the LO potential has a simple expression
\begin{align}
V^{}_{0} = \frac{1}{2} \mu^2_3 v^2 + \frac{1}{4} \lambda_3 v^4 + C |v|^3,
\end{align}
where $C \equiv -\frac{1}{48\pi}\Big( 2g^3_3 + (g^2_3 + {g_3'}^2)^{\frac{3}{2}} + 3 \sqrt{2} a^{\frac{3}{2}}_{2,3}  \Big)$.
The broken minimum of this potential has the simple expression
\begin{align}
v^{\text{broken}} = \frac{-3 C + \sqrt{9C^2 - 4 \lambda_{h,3} \mu^2_{h,3}}}{2 \lambda_{h,3}}. 
\end{align}
Evaluating expressions in this LO minimum in strict expansions, results in relatively simple, closed-form expressions for many quantities. The resummed Higgs and Goldstone masses read
\begin{align}
\widetilde{m}^2_h &= \frac{d^2}{dv^2} V^{\text{}}_{0} = \mu^2_{h,3} + 3 \lambda_3 v^2 - \frac{v}{8\pi} \Big(2 g^3_3 + (g^2_3 + {g_3'}^2)^{\frac{3}{2}} + 3 \sqrt{2} a^{\frac{3}{2}}_{2,3} \Big), \\ 
\widetilde{m}^2_G &= \frac{1}{v} \frac{d}{dv} V^{\text{}}_{0} = \mu^2_{h,3} + \lambda_3 v^2 - \frac{v}{16\pi} \Big(2 g^3_3 + (g^2_3 + {g_3'}^2)^{\frac{3}{2}} + 3 \sqrt{2} a^{\frac{3}{2}}_{2,3} \Big).
\end{align}
Triplet NLO contributions are obtained by replacing Eq.~\eqref{eq:soft-triplet-2loop} as
\begin{align}
\label{eq:soft-triplet-2loop-IR-mass-insertion}
V^{\text{soft triplet}}_{2} &\rightarrow V^{\text{soft triplet}}_{2}|_{m^2_\Sigma \rightarrow M^2_\Sigma} - \frac{3 \mu^2_{\Sigma,3} M_\Sigma}{8\pi},
\end{align}
where $M^2_\Sigma \equiv \frac{1}{2} a_{2,3} v^2$, and we have added the triplet one-loop diagram with a single IR mass insertion, which contributes at $\mathcal{O}(\veps_{\text{soft}} V_0 )$ to the soft-to-supersoft matching.

\paragraph{Supersoft triplet}
In this section, we assume that the triplet is supersoft, whereas the Higgs is soft. For simplicity, we start with the case of a sole triplet, and only after add contributions from the Higgs. Note that the triplet is not charged under the U(1) gauge group. We parametrise the triplet field as 
\begin{align}
\vec \Sigma &=
\begin{pmatrix}
  \Sigma^1 \\ 
  \Sigma^2 \\ 
  x+\Sigma^3 \\ 
\end{pmatrix},
\end{align}
where the triplet background field is denoted by $x > 0$. The leading order effective potential reads 
\begin{align}
V^{\text{triplet}}_{0} = \frac{1}{2} \mu^2_{\Sigma,3} x^2 + \frac{1}{4} b_{4,3} x^4 - \frac{1}{6\pi} \Big( 2 m^{3}_W \Big), 
\end{align}
where $m_W = g_3 x$, i.e.~the W-boson contribution is resummed together with a tree-level potential. The Z boson is massless in the triplet phase. Two-loop diagrams with W bosons and ghosts yield an extremely simple result 
\begin{align}
V^{\text{triplet}}_{2} = -\frac{1}{(4\pi)^2} 2 g^4_3 x^2,
\end{align}
in which we have used $m_W = g_3 x$. Note that no logarithmic terms arise: in the sole triplet case the triplet mass parameter does not run at this order.%
\footnote{
In the 3d counterterm $\delta \mu^2_{\Sigma,3}$ presented in \cite{Niemi:2020hto}, the $g^4_3$ contribution comes solely from the Higgs doublet loops.
}  
In the supersoft scale EFT triplet masses are those resummed by the soft gauge-field contributions. For neutral and charged triplets in the broken triplet phase we have
\begin{align}
\widetilde{m}^2_{\Sigma_0} &= \frac{d^2}{dx^2} V^{\text{triplet}}_{0} = \mu^2_{\Sigma,3} + 3 b_{4,3} x^2 - \frac{2 x}{\pi} g^3_3, \\
\widetilde{m}^2_{\Sigma^\pm} &= \frac{1}{x} \frac{d}{dx} V^{\text{triplet}}_{0} = \mu^2_{\Sigma,3} + b_{4,3} x^2 - \frac{x}{\pi} g^3_3.
\end{align}
Consequently, the one-loop triplet diagrams yield
\begin{align}
\label{eq:NNLO-triplet}
V^{\text{triplet}}_{3} = -\frac{1}{12\pi} \Big( (\widetilde{m}^2_{\Sigma_0})^{\frac{3}{2}} + 2 (\widetilde{m}^2_{\Sigma^\pm})^{\frac{3}{2}} \Big). 
\end{align}

\paragraph{Supersoft triplet with soft Higgs doublet}
Finally, we include the soft Higgs doublet contributions. At leading order, the resummed supersoft scale effective potential reads
\begin{align}
V^{}_{0} &= V^{\text{triplet}}_{0} + V^{\text{soft doublet}}_{0},
\end{align}
where
\begin{align}
\label{eq:V0-triplet-soft-doublet}
V^{\text{soft doublet}}_{0} &= - \frac{1}{12\pi} 4 \Big( m^2_\phi \Big)^{\frac{3}{2}},
\end{align}
where $m^2_\phi = \mu^2_{h,3} + \frac{1}{2} a_{2,3} x^2 $ is the quadruply degenerate mass squared eigenvalue for the Higgs field. Again, we have first assumed here that $\mu^2_{h,3}$ is soft. Resummed masses for neutral and charged triplets get new contributions accordingly, 
\begin{align}
\label{eq:neutral-triplet-mass}
\widetilde{m}^2_{\Sigma_0} &= \frac{d^2}{dx^2} V^{\text{}}_{0} = \mu^2_{\Sigma,3} + 3 b_{4,3} x^2 - \frac{2 x}{\pi} g^3_3  - \frac{1}{2\pi} a_{2,3} \frac{\mu^2_{h,3} + a_{2,3} x^2}{m_\phi}, \\
\label{eq:charged-triplet-mass}
\widetilde{m}^2_{\Sigma^\pm} &= \frac{1}{x} \frac{d}{dx} V^{\text{}}_{0} = \mu^2_{\Sigma,3} + b_{4,3} x^2 - \frac{x}{\pi} g^3_3  - \frac{1}{2\pi} a_{2,3} m_\phi.
\end{align}
The form of $V_3$ is the same as before (Eq.~\eqref{eq:NNLO-triplet}), with the above mass squared eigenvalues. At two-loop, contributions involving doublet scalar diagrams result in
\begin{align}
V^{\text{}}_{2} &= V^{\text{triplet}}_{2} + V^{\text{soft doublet}}_{2},
\end{align}
where
\begin{align}
\label{eq:soft-doublet}
V^{\text{soft doublet}}_{2} &= \frac{1}{(4\pi)^2} \bigg( 
\frac{1}{8} g^2_3 ( -3 m^2_W + 8 m_W m_\phi + 18 m^2_\phi) \nonumber \\ 
& - \frac{1}{2} a^2_{2,3} x^2 + \frac{1}{8} g^4_3 x^2 + \frac{3}{4} m^2_\phi ({g_3'}^2 + 8 \lambda_{h,3}) \nonumber \\
& + \Big( (g^2_3 + {g_3'}^2) m^2_\phi - a^2_{2,3} x^2   \Big) \ln\Big( \frac{\Lambda_3}{2 m_\phi} \Big) \nonumber \\
& - \frac{1}{2} g^2_3 (m^2_W - 4 m^2_\phi) \ln\Big( \frac{\Lambda_3}{m_W + 2 m_\phi} \Big) \nonumber \\
& + \frac{1}{2} g^2_3 (g^2_3 x^2 - m^2_W) \Big[ 8 \ln(2) - \ln\Big( \frac{\Lambda_3}{m_W} \Big) \Big]
\bigg),
\end{align}
in which the last term in fact vanishes since $m_W = g_3 x$. 

Finally, if we assume $\mu^2_{h,3}$ to be supersoft, so that in Eqs.~\eqref{eq:V0-triplet-soft-doublet}, \eqref{eq:neutral-triplet-mass} and \eqref{eq:charged-triplet-mass} one can set $\mu^2_{h,3}$ to zero, and replace Eq.~\eqref{eq:soft-doublet} by
\begin{align}
\label{eq:soft-doublet-2loop-IR-mass-insertion}
V^{\text{soft doublet}}_{2} &\rightarrow V^{\text{soft doublet}}_{2}|_{m^2_\phi \rightarrow M^2_\phi} - \frac{\mu^2_{h,3} M_\phi}{2\pi},
\end{align}
where $M^2_\phi \equiv \frac{1}{2} a_{2,3} x^2$, and the last term is the one-loop Higgs diagram with one mass insertion, i.e.~the first correction in an expansion in the supersoft mass $\mu^2_{h,3}$. As above in the case of the supersoft doublet, in this case the LO potential leads to a simple analytic formula for the minimum at LO, and corresponding strict expansions have relatively simple analytical expressions.  

These expressions complete our derivation of effective potentials at the supersoft scale.

\subsection{Semisoft scale induced soft EFT}
\label{sec:semisoft-matching}

In this section, we set ${g_3'}=0$ from the get go. Then $m_Z = m_W$ in the Higgs phase. The diagrammatic power counting for this section is outlined in Fig.~\ref{fig:semisoft-to-soft-veff}.

\paragraph{Higgs phase}
The effective potential in the Higgs phase has an expansion
\begin{align}
V^{\text{soft}}_{\text{eff}}(v) &= V_0 + V_1 + V_2 + V_3 + V_4 +  V_5  + \mathcal{O}(\veps^6 V_0),
\end{align}
where $\veps \sim \sqrt{g/\pi}$. Mass eigenvalues read $m^2_W = \frac{1}{4} g^2_3 v^2 \sim (\sqrt{g \pi})T^2$, and a triple-degenerate $m^2_\Sigma = \mu^2_{\Sigma,3} + M^2_\Sigma$. Here the semisoft piece is $M^2_\Sigma \equiv  \frac{1}{2} a_{2,3} v^2 \sim (\sqrt{g \pi})T^2$, while $\mu^2_{\Sigma,3} \sim g^2 T^2$ is soft.
The soft masses read $m^2_h = \mu^2_{h,3} + 3 \lambda_3 v^2$ and $m^2_G = \mu^2_{h,3} + \lambda_3 v^2$.
The LO contribution is just the tree-level potential (for vanishing triplet background field)
\begin{align}
V_0(v) &= \frac{1}{2} \mu^2_{h,3} v^2 + \frac{1}{4} \lambda_{3} v^4.
\end{align}
The NLO potential is the one-loop contribution with semisoft masses
\begin{align}
V_1(v) &= -\frac{1}{12\pi}\Big( 6 m^3_W + 3 M^2_\Sigma \Big).
\end{align}
N2LO vanishes, $V_2 = 0$, and the result at N3LO is given by one soft-mass insertion of the triplet to the one-loop bubble diagram
\begin{align}
V_3(v)  &= -\frac{3 M_\Sigma \mu^2_{\Sigma,3}}{8\pi}.
\end{align}

All higher order terms include contributions from the soft EFT, and we highlight these contributions separately below, in addition to matching contributions from the semisoft scale. At N4LO we get
\begin{align}
V_4(v)  &=  \frac{1}{(4\pi)^2} \frac{3}{64} \bigg( 64 g^2_3 m_W M_\Sigma + g^4_3 v^2 \Big( -3 + 8\ln(2) - 6 \ln\Big(\frac{\Lambda_3}{m_W}\Big) \Big) \nonumber \\ 
& + 8 g^2_3 \Big( 12 M^2_\Sigma + m^2_W (5-21 \ln(3)) + 20 m^2_W \ln\Big(\frac{\Lambda_3}{m_W}\Big) -4 (m^2_W - 4 M^2_\Sigma) \ln\Big(\frac{\Lambda_3}{m_W + 2 M_\Sigma}\Big)  \Big) \nonumber \\
& + 8 a^2_{2,3} v^2 \Big( -1 - 2 \ln\Big(\frac{\Lambda_3}{2 M_\Sigma}\Big) \Big) \bigg)  + \bigg( -\frac{1}{12\pi} \Big( m^3_h + 3 m^3_G \Big) \bigg)_{\rmii{soft EFT}},
\end{align}
Here the soft EFT contribution results simply from one-loop bubble diagrams with unresummed masses.
The matching contribution comprises of two-loop diagrams without soft mass insertions.
At N5LO there are contributions from the triplet one-loop bubble with two soft mass insertions, as well as a contribution from within the soft EFT,
\begin{align}
V_5(v)  &= \frac{1}{(4\pi)^2} \Big( -\frac{3 \pi \mu^2_{\Sigma,3}}{2 M_\Sigma}  \Big) 
 + \frac{1}{(4\pi)^2} \bigg( \frac{3}{8} (\sqrt{2} a^{\frac{3}{2}}_{2,3} + g^{\frac{3}{2}}_3) (3 m_G + 2 m_h) v \bigg)_{\rmii{soft EFT}}.
\end{align}
The soft EFT pieces at N5LO result from resummations of $V_1$. That is using these resummed masses,
\begin{align}
\widetilde{m}^2_h &= \frac{d^2}{dv^2} \Big( V_0 + V_1 \Big), \\
\widetilde{m}^2_G &= \frac{1}{v} \frac{d}{dv} \Big( V_0 + V_1 \Big), 
\end{align}
inside one-loop bubble diagrams, and then re-expanding in $\veps$.  We higlight that our result for the soft EFT expansion of the effective potential is RG invariant at the order we truncate our computation.

\paragraph{Triplet phase}
The effective potential in the triplet phase has an expansion
\begin{align}
V_\text{eff}^\text{soft}(x) &= V_0 + V_1 + V_2 + V_3 + V_4 +  V_5 + \mathcal{O}(\veps^6 V_0),
\end{align}
where $\epsilon \sim \sqrt{g/\pi}$.
Mass eigenvalues read $m^2_W = g^2_3 x^2 \sim (\sqrt{g \pi})T^2$, and a quadruply degenerate $m^2_\phi = \mu^2_{h,3} + M^2_\phi$, where the semisoft piece is $M^2_\phi \equiv  \frac{1}{2} a_{2,3} x^2 \sim (\sqrt{g \pi})T^2$ and $\mu^2_{h,3} \sim g^2 T^2$ is soft.
Soft masses are $m^2_{\Sigma_0} = \mu^2_{\Sigma,3} + 3 b_{4,3} x^2$ and $m^2_{\Sigma^\pm} = \mu^2_{\Sigma,3} + b_{4,3} x^2$.
In analogy to the computation for the Higgs phase result, we get the following expressions,
\begin{align}
V_0(x) &= \frac{1}{2} \mu^2_{\Sigma,3} x^2 + \frac{1}{4} b_{4,3} x^4,
\end{align}
and
\begin{align}
V_1(x) &= -\frac{1}{12\pi}\Big( 4 m^3_W + 4 M^2_\phi \Big),
\end{align}
at LO and NLO, respectively. Again, the N2LO contribution vanishes $V_2(x) = 0$, and at N3LO we have
\begin{align}
V_3(x)  &= -\frac{M_\phi \mu^2_{h,3}}{2\pi}.
\end{align}
At higher orders, in analogy to the computation in the Higgs phase, we get 
\begin{align}
V_4(x)  &= \frac{1}{(4\pi)^2} \frac{1}{8} \bigg( g^2_3 \Big( 8 m_W M_\phi + 18 M^2_\phi - 3 m^2_W \Big) 
  - 4 g^2_3 (m^2_W - 4 M^2_\phi) \ln(\frac{\Lambda_3}{m_W + 2 M_\phi})   \nonumber \\
& - (4 a^2_{2,3} + 15 g^4_3) x^2 
+ 8 ( g^2_3 M^2_\phi - a^2_{2,3} x^2 ) \ln\Big(\frac{\Lambda_3}{2 M_\phi}\Big) \nonumber \\ 
& + \frac{1}{2} g^2_3 (g^2_3 x^2 - m^2_W) \Big( 8\ln(2) - 9 \ln\Big(\frac{\Lambda_3}{m_W}\Big) \Big)  \bigg)  \nonumber \\
&+ \bigg( -\frac{1}{12\pi} \Big( m^3_{\Sigma_0} + 2 m^3_{\Sigma^\pm} \Big) \bigg)_{\rmii{soft EFT}},
\end{align}
(note that the penultimate line in fact vanishes identically) and
\begin{align}
V_5(x)  &= \frac{1}{(4\pi)^2} \Big( - \frac{2 \pi \mu^2_{h,3}}{M_\phi}   \Big) + \frac{1}{(4\pi)^2} \bigg(  (\sqrt{2} a^{\frac{3}{2}}_{2,3} + 4 g^{\frac{3}{2}}_3) (m_{\Sigma_0} + m_{\Sigma^\pm}) x \bigg)_{\rmii{soft EFT}},
\end{align}
at N4LO and N5LO respectively. In particular, the soft EFT pieces at N5LO result from resummations of $V_1$ using
\begin{align}
\widetilde{m}^2_{\Sigma_0} &= \frac{d^2}{dx^2} \Big( V_0 + V_1 \Big), \\
\widetilde{m}^2_{\Sigma^{\pm}} &= \frac{1}{x} \frac{d}{dx} \Big( V_0 + V_1 \Big). 
\end{align}
Again, our result is RG invariant at the order we truncate.

\section{Direct minimisation}
\label{sec:comparing-2steppers}

In the past few decades, direct minimisation of the real part of the thermal effective potential in Landau gauge has solidified itself as the standard meta in studies of cosmological phase transitions. Most studies resort to a one-loop approximation, as for generic models this has an explicit and relatively simple formula, however the downside is that it suffers from rather large theoretical uncertainties. When going beyond one-loop accuracy, the functional form of the effective potential becomes much more complicated and direct minimisation becomes numerically expensive. On the other hand, strict EFT expansions are numerically cheap to evaluate, once higher order corrections to the effective potential are known, as they can be obtained by straightforward Taylor expansions around leading order results. All higher order corrections are obtained simply by evaluating these expressions numerically. In order to provide a comparison of these approaches, in this appendix we present results obtained by direct minimisation of the real part of the Landau-gauge effective potential, both for the supersoft and soft scale EFTs. 

\begin{figure}
\centering
\begin{subfigure}{\textwidth}
    \centering
    \includegraphics[width=\textwidth]{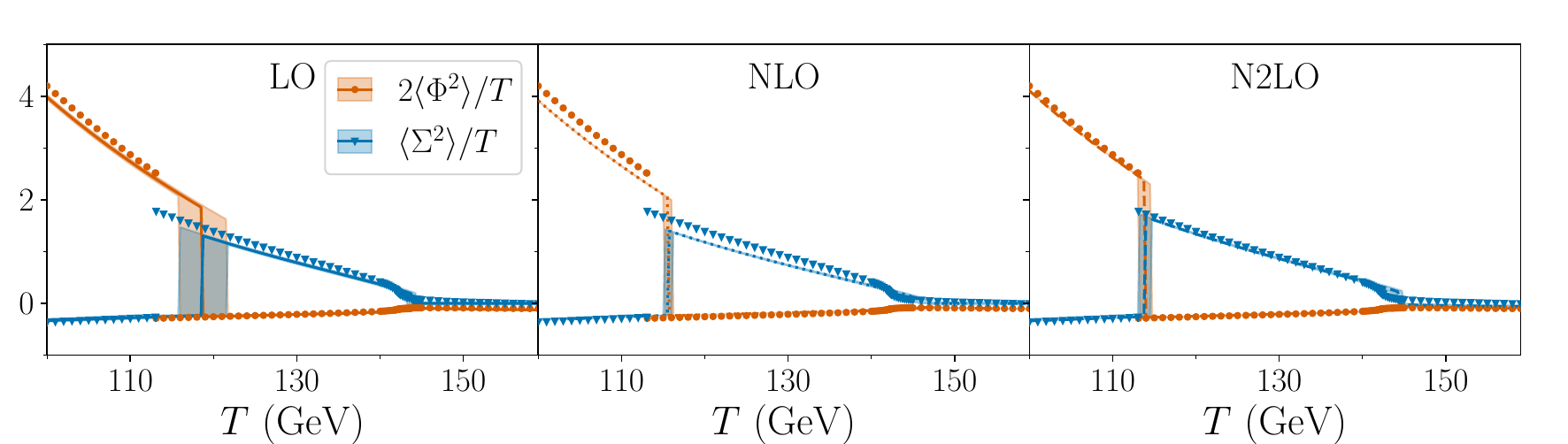}
    \caption{BM1}
    \label{fig:BM1_supersoft_strict_direct}
\end{subfigure}
\hfill
\begin{subfigure}{\textwidth}
    \centering
    \includegraphics[width=\textwidth]{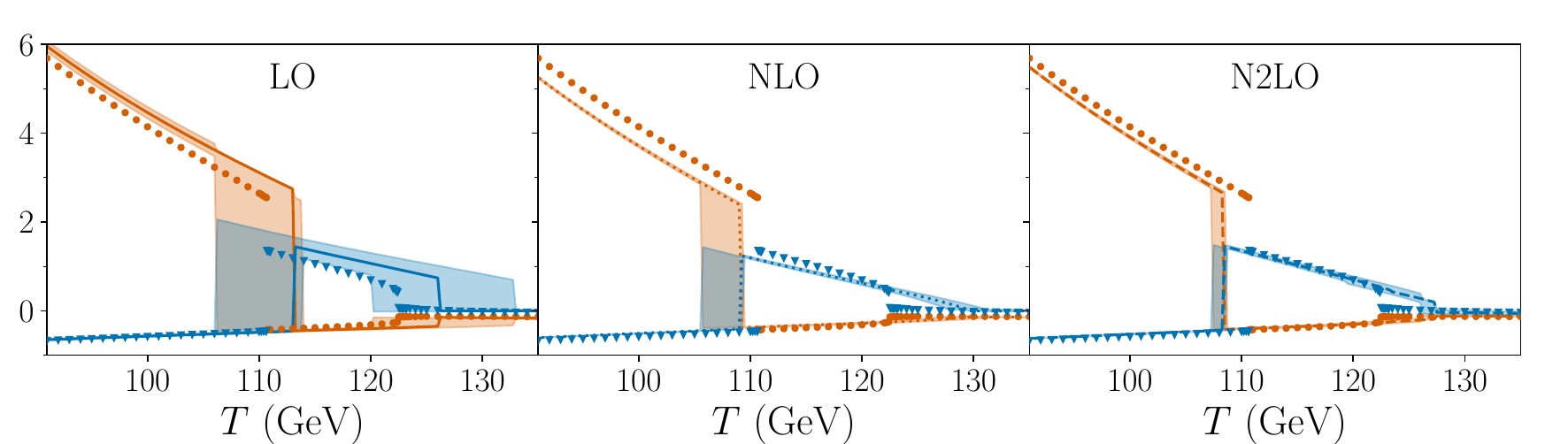} 
    \caption{BM2}
    \label{fig:BM2_supersoft_strict_direct}
\end{subfigure}
\caption{
Convergence of the direct minimisation method within the supersoft EFT for both transitions. Already lower order results are fairly close to lattice results, and convergence is clear. Notably in BM2, the first transition is seemingly weaker at higher orders compared to LO, as indicated by the size of the discontinuity in the triplet condensate.
}
\label{fig:BM-supersoft-direct}
\end{figure}

First, Fig.~\ref{fig:BM-supersoft-direct} showcases the supersoft EFT and the direct method for both transitions at both benchmark points.
Despite residual gauge dependence and the need to discard the imaginary part of the potential in minimisation, the result for the value of the condensate in the broken phase and for the critical temperature align with the lattice data fairly reliably.
Notably, already the leading order result is reasonably good, yet N2LO is even better, indicating convergence.
On the other hand, given the computational cost of direct minimisation, there should be no practical reason why not to upgrade this computation by a strict expansion, the results of which are shown in Fig.~\ref{fig:supersoft_strict_strict}.

\begin{figure}
\centering
\begin{subfigure}{\triptychwidth}
    \centering
    \includegraphics[width=\textwidth]{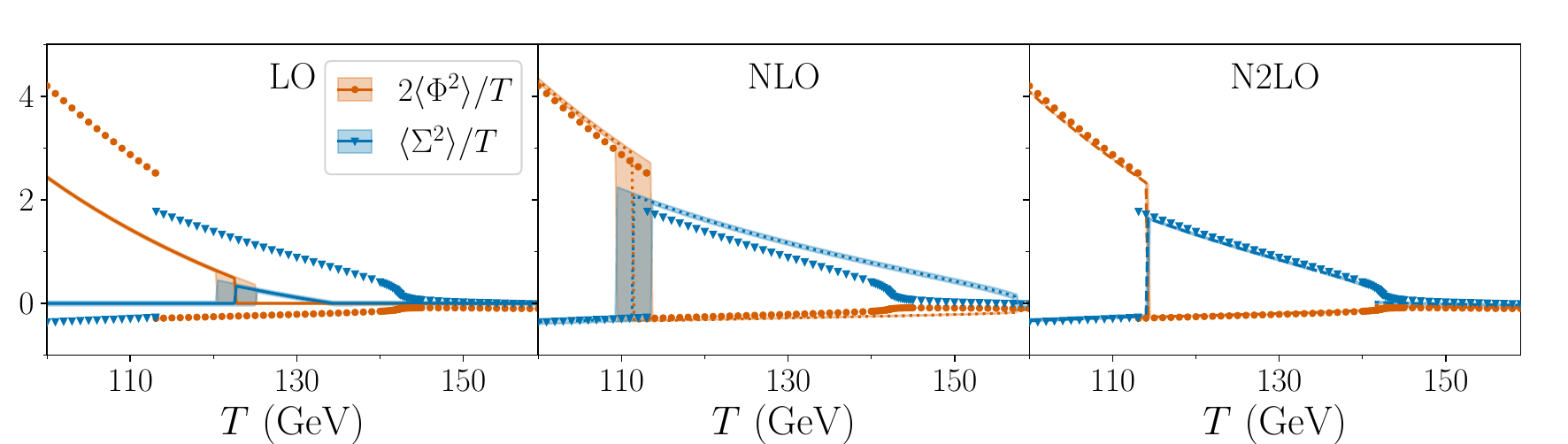}
    \caption{BM1}
    \label{fig:BM1_soft_direct_direct}
\end{subfigure}
\hfill
\begin{subfigure}{\triptychwidth}
    \centering
    \includegraphics[width=\textwidth]{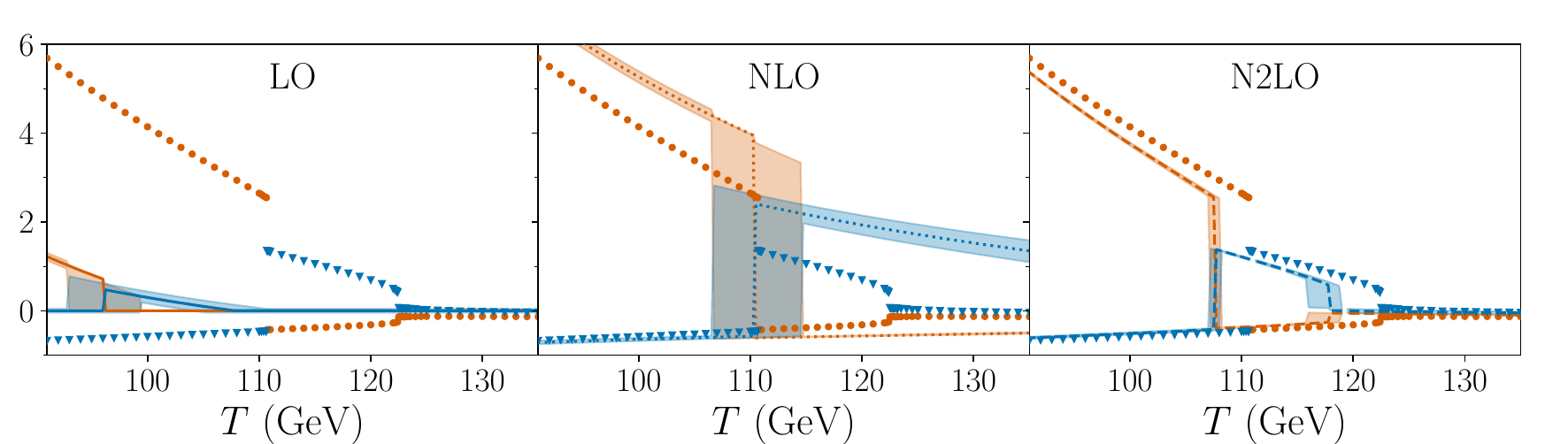} 
    \caption{BM2}
    \label{fig:BM2_soft_direct_direct}
\end{subfigure}
\caption{
Convergence of the direct minimisation method for the hard$\to$soft EFT. Results at lower orders are seemingly far off, while at N2LO, or two-loops, the perturbative result agrees fairly well with the lattice data despite its theoretical hiccups related to imaginary parts and gauge dependence. However, note that around the higher critical temperatures there are a number of perturbative data points missing. This is due to the numerical minimisation algorithm failing with default tolerance and method arguments, and is a common issue arising when directly minimising the real part of effective potentials at higher loop orders.
}
\label{fig:soft_direct_direct}
\end{figure}

Finally, we present triptychs of the convergence of direct minimisation for the hard$\to$soft EFT in Fig.~\ref{fig:soft_direct_direct} for both BM1 and BM2 and for both transitions.
We observe that while LO (tree-level) and NLO (one-loop) leave much to hope for, the result at N2LO (two-loop) agrees fairly well with the lattice results for both benchmark points, for both the value of the condensates and the critical temperatures. This observation provides some support for using the direct minimisation method within the hard$\to$soft EFT in parameter-space scans of BSM theories, as long as the computation is performed at \textit{two-loop} order. Despite the lack of theoretical robustness, this method allows one to scan wide regions of BSM theory parameter space in a single EFT, in contrast to EFT expansions which require delicate usage of chains of EFTs, potentially even in a single parameter point. Yet in practice, one major downside of the direct approach at two-loop order is the computational cost of minimising complicated multivariate functions. This motivates pursuing the automation of EFT expansions, which are numerically significantly less expensive and furthermore theoretically sound.

%
\bibliographystyle{JHEP}
\bibliography{refs}

\end{document}